\documentclass[referee]{aa} 
\usepackage{graphicx}
\usepackage{natbib}
\usepackage{txfonts}
\usepackage{ulem} 
\begin{document}
   \title{Formation rates of complex organics in UV irradiated CH$_3$OH-rich ices I: Experiments}

   \author{Karin I. \"Oberg\inst{1}
   		\and
		Robin T. Garrod\inst{2}
		  \and
          	Ewine F. van Dishoeck\inst{3,4}
		  \and
		  Harold Linnartz\inst{1}
          }

   \institute{Raymond and Beverly Sackler Laboratory for Astrophysics, Leiden Observatory, Leiden University, P.O. Box 9513, NL 2300 RA Leiden, The Netherlands.
              \email{oberg@strw.leidenuniv.nl}
         \and
         	Department of Astronomy, Cornell University, Ithaca, NY 14853, USA. \email{rgarrod@astro.cornell.edu}
         \and
             Leiden Observatory, Leiden University, P.O. Box 9513, NL 2300 RA Leiden, The Netherlands
         \and
             Max-Planck-Institut f\"ur extraterrestrische Physik (MPE), Giessenbachstraat 1, D 85748 Garching, Germany\\
             }

   \date{}

 
  \abstract
{Gas-phase complex organic molecules are commonly detected in the warm inner regions of protostellar envelopes, so-called hot cores. Recent models show that photochemistry in ices followed by desorption may explain the observed abundances. There is, however, a general lack of quantitative data on UV-induced complex chemistry in ices.}
{This study aims to experimentally quantify the UV-induced production rates of complex organics in CH$_3$OH-rich ices under a variety of astrophysically relevant conditions.}
{The ices are irradiated with a broad-band UV hydrogen microwave-discharge lamp under ultra-high vacuum conditions, at 20--70~K, and then heated to 200~K. The reaction products are identified by reflection-absorption infrared spectroscopy (RAIRS) and temperature programmed desorption (TPD), through comparison with RAIRS and TPD curves of pure complex species, and through the observed effects of isotopic substitution and enhancement of specific functional groups, such as CH$_3$, in the ice.}
{Complex organics are readily formed in all experiments, both during irradiation and during the slow warm-up of the ices after the UV lamp is turned off. The relative abundances of photoproducts depend on the UV fluence, the ice temperature, and whether pure CH$_3$OH ice or CH$_3$OH:CH$_4$/CO ice mixtures are used. C$_2$H$_6$, CH$_3$CHO, CH$_3$CH$_2$OH, CH$_3$OCH$_3$, HCOOCH$_3$, HOCH$_2$CHO and (CH$_2$OH)$_2$ are all detected in at least one experiment. Varying the ice thickness and the UV flux does not affect the chemistry. The derived product-formation yields and their dependences on different experimental parameters, such as the initial ice composition, are used to estimate the CH$_3$OH photodissociation branching ratios in ice and the relative diffusion barriers of the formed radicals. At 20~K, the pure CH$_3$OH photodesorption yield is $2.1(\pm1.0)\times10^{-3}$ per incident UV photon, the photo-destruction cross section $2.6(\pm0.9)\times10^{-18}$ cm$^2$. }
   {Ice photochemistry in CH$_3$OH ices is efficient enough to explain the observed abundances of complex organics around protostars. Some complex molecules, such as CH$_3$CH$_2$OH and CH$_3$OCH$_3$, form with a constant ratio in our ices and this can can be used to test whether complex gas-phase molecules in astrophysical settings have an ice-photochemistry origin. Other molecular ratios, e.g. HCO-bearing molecules versus (CH$_2$OH)$_2$, depend on the initial ice composition and temperature and can thus be used to investigate when and where complex ice molecules form.}

  \titlerunning{Formation of complex organics in CH$_3$OH ice}
   \maketitle
%

\section{Introduction\label{sec:intro}}

Organic molecules of increasing complexity are being detected in star-forming regions \citep{Blake87, Nummelin00, Bisschop07, vanDishoeck95, Cazaux03, Bottinelli04, Bottinelli07, Belloche09}; however, the origins of these complex molecules are the subject of debate. Commonly-suggested formation routes include various gas-phase reactions involving evaporated CH$_3$OH ices, atom-addition reactions on dust grains, and UV- and cosmic ray-induced chemistry in the granular ices \citep[][Herbst \& Dishoeck 2009, AR\&A in press]{Charnley92, Nomura04}. Recently, the focus has shifted to an ice formation pathway \citep[e.g.][]{Garrod08}, but due to the lack of quantitative experimental data, it is still not clear whether these molecules form in granular ices during the colder stages of star formation or in the warm gas close to the protostar. Nor has the relative importance of different grain formation routes been resolved. Establishing the main formation route is needed to predict the continued chemical evolution during star- and planet-formation and also to predict the amount of complex organics incorporated into comets and other planetesimals. In light of this, and the recent failures of gas phase chemistry to explain the observed complex molecules, we aim to quantify  the formation of complex molecules through UV-induced chemistry in CH$_3$OH-rich ices. 

Simple ices, such as solid H$_2$O, CO, CO$_2$, CH$_4$ and NH$_3$, are among the most common species found in dark cloud cores and towards protostars. The ices form sequentially in the cloud, resulting in a bi-layered structure dominated by H$_2$O and and CO, respectively \citep{Bergin05, Knez05, Pontoppidan06}. Laboratory experiments suggest that CH$_3$OH also forms in the ice during the pre-stellar phase, through hydrogenation of CO \citep{Watanabe03}, although, as yet, CH$_3$OH ice has only been detected toward protostars. Based on this formation route, CH$_3$OH is probably present in a CO-rich phase during most of its lifetime. More complex ices have been tentatively detected towards a few high-mass protostars \citep{Schutte99,Gibb04}, though specific band assignments are uncertain. Towards most other objects, derived upper limits on complex ices are too high to be conclusive.

Indirect evidence of complex molecule formation in ice mantles exists from millimeter observations of shocked regions and the innermost parts of low- and high-mass protostellar envelopes, so called hot cores and corinos.  The observations by \cite{Arce08} of HCOOCH$_3$, HCOOH and CH$_3$CH$_2$OH at abundances of $\sim$10$^{-2}$ with respect to CH$_3$OH towards the low-mass molecular outflow L1157 are especially compelling; this outflow has been above 100~K for a period that is an order of magnitude shorter than is required for the gas-phase production of such complex molecules. Ice evaporation through sputtering is, in contrast, efficient in shocks \citep{Jones96}. The observed abundances towards L1157 are remarkably similar to those observed in galactic-center clouds and high-mass protostars,  suggesting a common formation route.

In hot cores and corinos, gas-phase production may still be a viable alternative for some of the detected molecules, because of the longer time scales compared to outflows. However, recent calculations and experiments suggest that some key gas-phase reactions are less efficient than previously thought; for example, the gas-phase formation of HCOOCH$_3$ was found to be prohibitively inefficient \citep{Horn04}. Furthermore, \citet{Bisschop07} recently showed that complex oxygen-bearing molecules and H$_2$CO are equally well correlated with CH$_3$OH. Since CH$_3$OH and H$_2$CO are proposed to form together in the ice, this suggests that those complex oxygen-bearing molecules are also `first-generation' ice products. 

One possible ice formation route for complex species is atomic accretion and recombination on grain surfaces, which appears efficient for smaller species. \citet{Charnley04} suggested that the hot-core molecules may form either through hydrogenation of molecules and radicals, such as CO and HCCO, or through a combination of hydrogenation and oxidation starting with C$_2$H$_2$. Similar reaction schemes are suggested to explain observations by e.g. \citet{Bisschop08} and \citet{Requena-Torres08}. Such formation mechanisms have not been comprehensively tested under astrophysically relevant conditions, although the models of \cite{Belloche09} found them to be ineffective in the formation of nitriles up to ethyl cyanide, under hot core conditions. Experimental studies show that dissociative reactions may be the favored outcome of hydrogenating larger molecules and fragments \citep[e.g.][]{Bisschop07b}, hampering the build-up of large quantities of complex molecules. Quantitative experiments are, however, still lacking for most reactions. 

The alternative grain-surface formation route, which is investigated in this study, is the energetic destruction of the observed simple ices and subsequent diffusion and recombination of the radicals into more complex species. Within this framework, \citet{Garrod06} and \citet{Garrod08} modeled the formation of complex molecules during the slow warm-up of ices in an in-falling envelope, followed by ice evaporation in the hot core region. In the model, photodissociation of simple ices, especially CH$_3$OH, produces radicals in the ice. The luke-warm ices in the envelope (20--100~K) allow for the diffusion of `heavy' radicals like CH$_3$ and CH$_2$OH, which recombine to form complex molecules. The model continues until all the ice is evaporated and the resulting gas phase abundances reproduce some of the abundance ratios and temperature structures seen in the galactic center and towards hot cores.  Improvement of these model predictions is mainly limited by lack of quantitative experimental data on CH$_3$OH photodissociation branching ratios in ices, diffusion barriers of the formed radicals and binding energies of most complex molecules. The ultimate objective of the present study is to provide these numbers by experimental investigation of the photochemistry in CH$_3$OH-rich ices,  followed by quantitative modeling (paper II, Garrod \& \"Oberg, in preparation).

There have been multiple studies of photochemistry in ices containing organic molecules, stretching back to the 1960s \citep[e.g][]{Stief65}. Most studies provide only a limited amount of the kind of quantitative data needed for astrochemical models and instead focus on the qualitative assignment of final photochemistry products, following irradiation of ice mixtures that are proposed to mimic ice compositions in star forming regions. To produce enough detectable products, deposition and irradiation were typically simultaneous in the early experiments, rather than a sequential deposition, irradiation and warm-up scheme. This was the approach of, for example, \cite{Hagen79} and \cite{Dhendecourt82}. \citet{Allamandola88} included CH$_3$OH in ice mixtures in similar experiments and found that CH$_3$OH mainly photodissociates into smaller fragments at 10~K, while several new unidentified features appear following the warm-up of the irradiated ice, indicating an efficient diffusion of radicals. \citet{Gerakines96} investigated the photochemistry of pure CH$_3$OH ice more quantitatively at 10~K, and determined the CH$_3$OH photolysis cross section, and the H$_2$CO and CH$_4$ UV formation cross sections averaged over the lamp wavelength range at a specific flux setting. \citet{Gerakines96} also detected HCOOCH$_3$, and several other complex species have also been identified in CH$_3$OH-rich ices following UV-irradiation, though some peaks have been assigned to different carriers in different studies. The difficulty in identifying most complex products is discussed by e.g. \citet{Hudson00} who investigated the production of complex molecules in CH$_3$OH:H$_2$O and CH$_3$OH:CO ice mixtures both following UV irradiation and ion bombardment in a number of studies, most recently in \citet{Moore05} and \citet{Hudson05}. 

A few studies exist on the formation of complex molecules from ion bombardment of pure CH$_3$OH ice, though similarly to the UV photolysis experiments,  the formation of more complex molecules than CH$_3$OH is in general not quantified \citep{Baratta02, Bennet07a}.  \citet{Bennet07a} identified HCOOCH$_3$, HOCH$_2$CHO and (CH$_2$OH)$_2$, mainly based on comparison with calculated spectra and desorption patterns following irradiation, but provided only upper limits of their formation rates. The quantitative data from both studies include formation rates of small molecules and CH$_3$OH and H$_2$CO ion-bombardment dissociation rates. In a separate study \citet{Bennet07b}  determined the formation rate of HCOOCH$_3$ and HOCH$_2$CHO in a CO:CH$_3$OH ice mixture for the ion-bombardment flux used in their experiment. 

The aim of this study and its follow-up paper is to combine experiments with kinetic modeling to completely quantify the photochemistry of CH$_3$OH rich ices. This includes determining the CH$_3$OH photodesorption yield, the CH$_3$OH dissociation branching ratios upon UV irradiation, the diffusion barriers of the formed radicals and reaction barriers to form more complex molecules, where present. Here, we present the experiments on CH$_3$OH ice chemistry under a large range of astrophysically relevant conditions and quantify the formation of all possible first generation complex molecules. To ensure that reaction products are correctly assigned, we also present RAIR spectra and temperature-programmed desorption experiments of all stable expected complex products, together with their derived binding energies. 

The paper is organized as follows. Section 2 presents the experimental and data analysis methods. Section 3 reports on both qualitative and quantitative results of the experiments. Discussion follows in section 4, and includes estimates of photodissociation branching ratios and diffusion barriers. Preliminary astrophysical implications are discussed in section 5. A summary of results and  concluding remarks are given in section 6.

\section{Experiments and analysis\label{sec:exps}}

\begin{table}
\begin{center}
\caption{Experimental parameters for UV-irradiation experiments.}             
\label{tab:ph_exps}      
\centering                          
\begin{tabular}{l c c c c c c }        
\hline\hline                 
Exp. &Species &Temp.$^{\rm a}$  & Thick. &UV flux\\
 & &(K) & (ML) &(10$^{13}$ cm$^{-2}$ s$^{-1}$)\\
\hline                
1 &CH$_3$OH &20 &21 &1.1\\
2 &CH$_3$OH &30 &19 &1.1\\
3 &CH$_3$OH &50 &20 &1.1\\
4 &CH$_3$OH &70 &22 &1.1\\
5 &CH$_3$OH &20 &19 &4.3\\
6 &CH$_3$OH &50 &15 &4.3\\
7&CH$_3$OH:CO &20 &12:17 &1.1\\
8&CH$_3$OH:CO &30 &12:11 &1.1\\
9&CH$_3$OH:CO &50 &16:7 &1.1\\
10&CH$_3$OH:CH$_4$ &30 &11:27 &1.1\\
11&CH$_3$OH:CH$_4$ &50 &11:6 &1.1\\
12&CH$_3$OH:CO &20 &6:60 &1.1\\
13 &CH$_3$OH &20 &6 &1.1\\%
14 &CH$_3$OH &50 &6 &1.1\\
15 &CH$_3$OH &20 &4 &4.3\\
16 &CH$_3$OH &50 &8 &4.3\\
17$^{\rm b}$  &CH$_3$OH &20 &18 &1.1\\
18$^{\rm c}$ &CH$_3$OH &20 &20 &1.1\\
19 &CH$_3$OD &20 &$\sim$20 &1.1\\
20 &CH$_3$OD &50 &$\sim$20 &1.1\\
21 &CD$_3$OH &20 &$\sim$20 &1.1\\
22 &CD$_3$OH &50 &$\sim$20 &1.1\\
\hline
\end{tabular}
\end{center}
$^{\rm a}$ The ice-deposition and -irradiation temperature.\\
$^{\rm b}$ Following irradiation the ice is quickly heated to 50~K for 2 hours.\\
$^{\rm c}$ Following irradiation the ice is quickly heated to 70~K for 2 hours.
\end{table}

\begin{table*}
\begin{center}
\caption{Pure ice spectroscopy and TPD experiments.}             
\label{tab:sup_exps}      
\centering                          
\begin{tabular}{l c c c c c c }        
Species & Formula&Mass (amu) &Thick.$^{\rm a}$ (ML) &$E_{\rm des}^{\rm a}$ (K)\\
\hline\hline                 
Ethane &C$_2$H$_6$ &30 &5 [3] &2300 [300]\\
Methanol &CH$_3$OH &32 &25 [10] & 4700 [500]\\
Acetaldehyde &CH$_3$CHO  &44 &4 [2] &3800 [400]\\
Dimethyl ether &CH$_3$OCH$_3$  &46 &$\sim$4 & 3300 [400]\\
Ethanol &CH$_3$CH$_2$OH  &46 &5 [2]&5200 [500]\\
Formic acid &HCOOH  &46& 5 [2] &5000 [500]\\
Methyl formate &HCOOCH$_3$ &60 &3 [1] & 4000 [400]\\
Acetic acid &CH$_3$COOH &60 &3 [1] &6300 [700]\\
Glycolaldehyde &HOCH$_2$CHO &60 &3 [1] &5900 [600]\\
Ethylene glycol &(CH$_2$OH)$_2$ &62 &7 [3] & 7500 [800]\\
\hline
\end{tabular}
\end{center}
$^{\rm a}$Values in brackets indicate uncertainties.\\
\end{table*}

All experiments are carried out under ultra-high vacuum conditions ($\sim$10$^{-10}$~mbar) in the CRYOPAD set-up, which is described in detail in \citet{Fuchs06}  and \citet{Oberg09b}. The ices are grown \textit{in situ} with monolayer precision at thicknesses between 3 and 66 ML, by exposing a cold substrate at the center of the vacuum chamber to a steady flow of gas, directed along the surface normal. The substrate is temperature controlled between 20 and 200~K. The relative temperature uncertainty is less than a degree, while the absolute uncertainty is about two degrees. All UV-irradiation experiments are performed with CH$_3$OH from Sigma-Aldrich with a minimum purity of 99.8\%. The mixed ice experiments contain CH$_4$ or CO gas of 99\% purity (Indogas). Pure, complex ice experiments with C$_2$H$_6$, CH$_3$CHO, CH$_3$OCH$_3$, CH$_3$CH$_2$OH, HCOOH, HCOOCH$_3$, CH$_3$COOH, HOCH$_2$CHO and (CH$_2$OH)$_2$ are carried out with chemicals of 99--99.9\% purity from Sigma-Aldrich. All liquid samples are further purified with several freeze-thaw cycles to remove any volatile gas from the sample. The dominant source of contaminants is from the vacuum inside of the chamber once the ice is deposited; during each experiment, up to 0.5 ML of H$_2$O adsorbs onto the substrate from the small H$_2$O contamination always present in the chamber. This has no measurable impact on the photochemistry from test experiments with CH$_3$OH isotopologues. 

The set-up is equipped with a Fourier transform infrared (FTIR) spectrometer in reflection-absorption mode (Reflection-Absorption InfraRed Spectroscopy or RAIRS). The FTIR covers 750 -- 4000 cm$^{-1}$, which includes vibrational bands of all investigated molecules, and is operated with a spectral resolution of 1 cm$^{-1}$. To increase the signal to noise the spectra are frequently binned when this can be done without reducing the absorbance of sharp features. RAIRS is employed both to acquire infrared spectra of complex molecules and to quantify the changing ice composition during UV irradiation of CH$_3$OH-rich ices. All spectra are corrected with a linear baseline alone, to avoid distorting any spectral profiles.

Temperature Programmed Desorption (TPD) is another analytical tool, which is employed in this study to identify ice photoproducts. In a TPD experiment, ice evaporation is induced by linear heating of the ice, here with a heating rate of 1 K min$^{-1}$. The evaporated gas phase molecules are detected by a Quadrupole Mass Spectrometer (QMS). The resulting TPD curves depend on the evaporation energy of the ice, which can be uniquely identified for most of the investigated species. For mixed ices the TPD curve also depends on such quantities as ice trapping, mixing and segregation. The QMS software allows for the simultaneous detection of up to 60 different $m/z$ values (the molecular mass divided by the charge). Hence in the TPD experiments of irradiated ices, all possible reaction-product masses, which contain at most two oxygen and two carbon atoms,  are monitored.

In the CH$_3$OH photochemistry experiments, the ice films are irradiated at normal or 45$^{\circ}$ incidence with UV light from a broadband hydrogen microwave-discharge lamp, which peaks around Ly $\alpha$ at 121~nm and covers 115--170~nm or 7--10.5~eV \citep{Munozcaro03}.  The lamp flux was calibrated against a NIST calibrated silicate photodiode prior to the experimental series and is monitored  during each experiment using the photoelectric effect in a gold wire in front of the lamp. The lamp emission resembles the spectral distribution of the UV interstellar radiation field that impinges externally on all clouds. It is also consistent with the UV radiation produced locally inside clouds by the decay of electronic states of H$_2$, following excitation by energetic electrons resulting from cosmic-ray induced ionization of hydrogen, see e.g. \cite{Sternberg87}. Each irradiation experiment is followed by a TPD experiment, where RAIR spectra are acquired every 10~K up to 200~K.

Supporting experiments consist of RAIR spectra and TPD curves of nine complex organic ices, which are potential photoproducts of CH$_3$OH ices. Spectra of these ices have been reported previously in the literature in transmission, but because of known band shifts in RAIRS compared to transmission spectroscopy, their RAIR spectra are also presented here.

The identification process is complicated by spectral overlaps of most of the potential photochemistry products. Thus great care is taken in securing each assignment, especially where they disagree with previous work or where disagreements between previous studies exist. To call an identification secure we test it to be consistent with up to seven criteria. These identification tools are described in detail in section 3.5.

Following the spectral band identification, RAIR spectroscopy is used to determine the initial CH$_3$OH ice abundance and the formed simple and complex ice abundances as a function of fluence during each photochemistry experiment. This requires known band strengths. The absolute RAIRS band strengths have been estimated previously in our set-up for CO and CO$_2$ ice \citep{Oberg09a, Oberg09b}. Using the same method, new measurements on CH$_3$OH are consistent with the CO and CO$_2$ results; i.e. the determined band strengths have the same relative values compared to the transmission band strength ratios reported in the literature, within 20\%. The \textit{relative} ice band strengths from transmission are thus still valid, with some exceptions, and most experimental objectives only require knowing the ice fraction that has been converted into products. The main caveat is that ices thicker than a few monolayers are not guaranteed to have a linear relationship between the absorbance of strong bands and ice thickness due to RAIRS effects \citep{Teolis07}. This is circumvented by selecting weak enough bands, especially for CH$_3$OH, whose absorbance remains linear with respect to the amount of deposited ice at all the investigated ice thicknesses. The transmission band strength of CH$_3$OCH$_3$ is not present in the literature and its band strength is estimated by deposition of a dilute CH$_3$OCH$_3$:CH$_3$OH 1:10  mixture and assuming a constant sticking coefficient and that the CH$_3$OCH$_3$ is 'dragged' along with the CH$_3$OH to reach the substrate at a similar deposition rate. 

Table \ref{tab:ph_exps} lists the CH$_3$OH, CH$_3$OH:CO and CH$_3$OH:CH$_4$ photochemistry experiments. The experiments are designed to study the impact of ice temperature, ice thickness, UV flux, UV fluence and mixed-in CO and CH$_4$ on the reaction-product abundances. Each ice is irradiated for $\sim$6 hours at the reported flux. The high flux/fluence experiments are also used to determine the CH$_3$OH photodesorption rate using the same procedure as reported by \citet{Oberg09b}. In two experiments (17 and 18) the irradiation is followed by fast heating to a specified temperature to investigate the impact of the heating rate for radical diffusion. Table \ref{tab:sup_exps} lists the investigated complex organics for which RAIR spectra and TPD experiments have been acquired.

In addition to the experiments listed in Table \ref{tab:ph_exps} and \ref{tab:sup_exps}, two experiments were performed to test the CO accretion rate due to UV-induced out-gassing from chamber walls and the life time of spectral features in the ice when diffusion is slow. In the first experiment a blank substrate was irradiated for the typical experiment time of six hours and a build-up of 0.2 CO~ML was recorded. In the second experiment a photolyzed CH$_3$OH ice was monitored for five hours at 17~K after the UV lamp was turned off; the spectra of complex products did not change measurably during this period.

In all experiments, systematic uncertainties dominate and include the absolute calibration of the temperature ($\sim$2 K), the UV flux ($\sim$30\%) and the conversion between transmission and RAIRS band strengths used to determine the absolute ice abundances ($\sim$50\%), while the relative RAIRS band strengths are more accurate ($\sim$20\% uncertainty from comparison between different trannsmission spectroscopy studies). The conversion between transmission and RAIRS band strengths will not affect the uncertainty of the photochemistry rates, since these depend only on the fraction of the ice that is converted into products. The determined diffusion barriers in Paper II are furthermore not affected by the uncertainty in UV flux because these only depend on which species the produced radicals react with, not on how many of them are produced per UV photon. Another source of error is the local baseline determination, which results in relative abundance uncertainties of up to 30\% for a couple of the detected products, which is reported in detail in \S \ref{sec:res_quant} for each species. Thus the formation yield of products relative to the original ice abundance has a total uncertainty of 35--50\%.

\section{Experimental results\label{sec:res}}

This section begins with experimental results that quantify CH$_3$OH bulk photolysis (\S\ref{sec:res_cs}) and surface photodesorption (\S\ref{sec:res_pd}). \S \ref{sec:res_depend} qualitatively describes how the CH$_3$OH photoproducts are affected by different experimental variables for the experiments listed in Table \ref{tab:ph_exps}. This information, together with RAIR spectra and TPD data on pure complex organics in \S \ref{sec:res_sup}, is used in \S \ref{sec:res_id} to identify the CH$_3$OH photoproducts. Following identification, the formation of all identified products from pure CH$_3$OH ice photochemistry are shown quantitatively in \S \ref{sec:res_quant}. Section \ref{sec:res_warmup} describes quantitatively the formation and desorption of molecules during warm-up of the irradiated ices. Finally, \S \ref{sec:res_sum} summarizes the effects of different experimental parameters on the final ice composition after irradiation and during warm-up.

\subsection{The CH$_3$OH UV photolysis cross-section\label{sec:res_cs}}

\begin{figure}
\resizebox{\hsize}{!}{\includegraphics{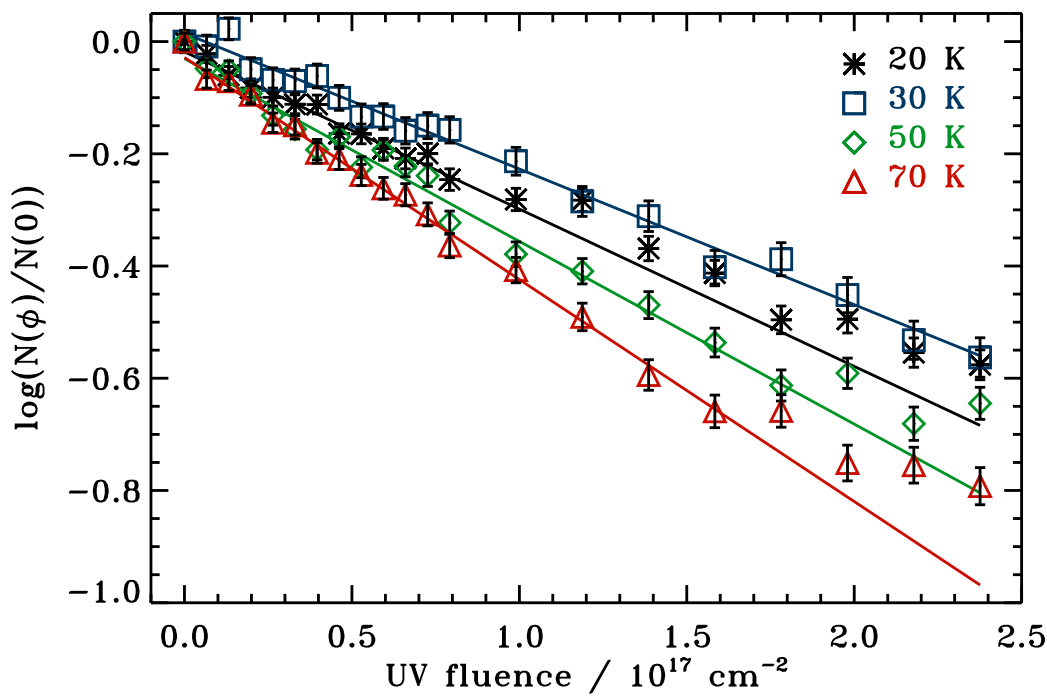}}
\caption{The logarithm of the normalized CH$_3$OH abundance as a function of UV fluence at 20, 30, 50 and 70~K. The lines are exponential fits to the first 10$^{17}$ photons in each experiment.}
\label{fig:ch3oh_cs}
\end{figure}

The UV-destruction cross-section of CH$_3$OH ice, averaged over the lamp spectrum, determines the total amount of radicals available for diffusion and subsequent reaction. The cross section is calculated from the measured loss of CH$_3$OH ice band intensity with fluence. The initial UV destruction of CH$_3$OH, i.e. before back-reactions to reform CH$_3$OH become important,  in an optically thin ice is given by

\begin{equation}
	\label{eq:cs}
	N(\phi )=N(0){\rm exp}(-\phi \times \sigma_{\rm ph}),
\end{equation}

\noindent where $N$ is the CH$_3$OH column density in cm$^{-2}$, $\phi$ is the UV fluence in photons cm$^{-2}$ and $\sigma_{\rm ph}$ is the UV-photolysis cross section in cm$^{2}$. Figure \ref{fig:ch3oh_cs} shows that photodestruction during the first 10$^{17}$ photons is well described by this equation for $\sim$20~ML thick ices at different temperatures. The loss of CH$_3$OH is calculated from the combination band around 2550 cm$^{-1}$ and the resulting photodissociation cross sections are 2.6[0.9], 2.4[0.8], 3.3[1.1] and 3.9[1.3] $\times10^{-18}$ cm$^2$ at 20, 30, 50 and 70~K respectively. The uncertainties in brackets are the absolute errors; the relative uncertainties are 10--20\%. The increasing cross section with temperature is indicative of significant immediate recombination of dissociated CH$_3$OH at low temperatures when diffusion is slow. The measured effective CH$_3$OH-ice cross sections thus underestimate the actual photodissociation rate. This is consistent with the higher gas-phase CH$_3$OH UV-absorption cross section; convolving the absorption spectra from \citet{Nee85} with our lamp spectra results in a factor of three higher absorption rate than the observed photodissociation rate in the ice at 20~K. 

The measured photolysis cross sections also depend on whether a `clean' CH$_3$OH band is used to calculate the CH$_3$OH loss. The $\nu_{11}$ CH$_3$OH band around 1050 cm$^{-1}$ is commonly used in the literature \citep{Gerakines96,Cottin03}. This band overlaps with strong absorptions of several complex photoproducts and using it results in a 30\% underestimate of the CH$_3$OH destruction cross section at 20~K. This explains the higher cross-section value obtained in these experiments compared to \citet{Gerakines96} and \citet{Cottin03}, who recorded $1.6\times10^{-18}$ cm$^{2}$ and $6\times10^{-19}$ cm$^{2}$, respectively.  These destruction cross sections were also measured after greater fluences, $\sim1.8\times10^{17}$ and $\sim6\times10^{17}$ UV photons cm$^{-2}$, respectively, when back reactions to form CH$_3$OH confuse the measurements. The measurements in this paper thus demonstrates the importance of a high fluence resolution and of picking a `clean' band when determining the photodestruction cross section of an ice.

\subsection{CH$_3$OH photodesorption yields\label{sec:res_pd}}

\begin{figure}
\resizebox{\hsize}{!}{\includegraphics{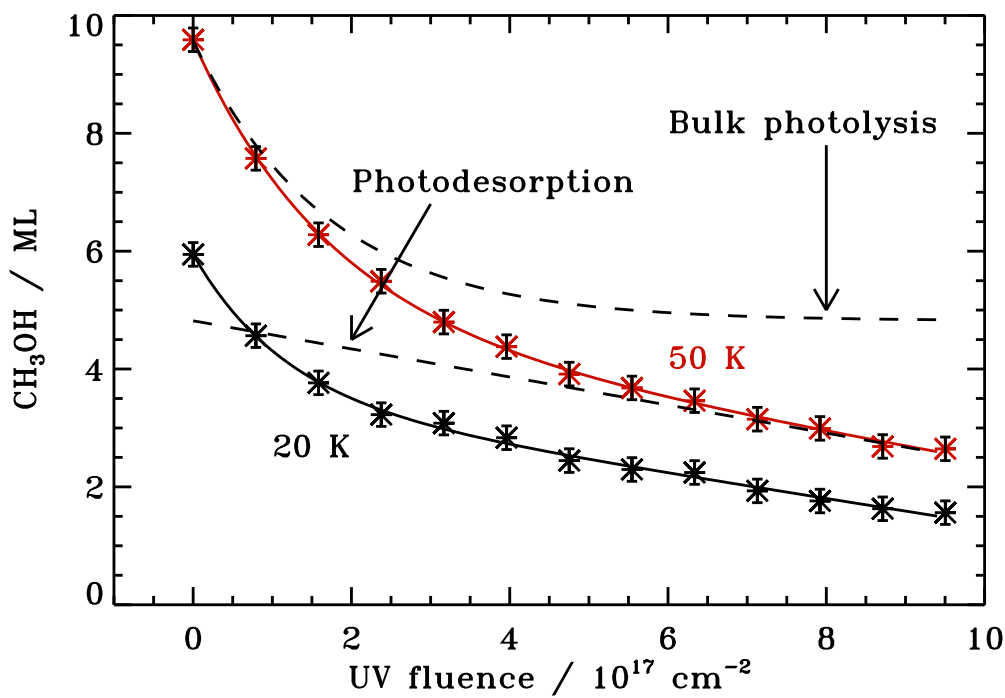}}
\caption{The loss of CH$_3$OH ice through photolysis and photodesorption at 20 and 50~K as a function of UV fluence. The curves are fitted with $A_0 + A_1\times\phi + A_2(1-{\rm exp}(-\phi\times \sigma))$. The thin dashed lines show the offset decomposition of the function belonging to the 50~K experiment into its exponential photolysis part and its linear photodesorption part.}
\label{fig:ch3oh_pd}
\end{figure}

Previous experiments show that several ices (pure CO, CO$_2$ and H$_2$O) are efficiently photodesorbed upon UV irradiation. To constrain the photodesorption of CH$_3$OH ice and thus determine the loss of CH$_3$OH molecules into the gas phase rather than into photoproducts in the ice, the same procedure is followed as reported by \citet{Oberg09b}. This method is based on the fact that photodesorption from a multilayer ice is a zeroth order process with respect to photon fluence, since it only depends on the amount of molecules in the surface layer. The photodesorption yield will thus not change with fluence as long as the original ice is sufficiently thick. In contrast ice photolysis is a first order process, since it depends on the total amount of ice. Through simultaneous modeling of the ice loss with an exponential decay and a linear function, these two processes can be separated and the photodesorption yield determined (Fig. \ref{fig:ch3oh_pd}). The resulting yields are 2.1[1.0]$\times10^{-3}$ and 2.4[1.2]$\times10^{-3}$ desorbed molecules per incident UV photon at 20 and 50~K, respectively. There is thus no evidence for a temperature dependence of the photodesorption yield within the investigated temperature range. 

These yields agree with previous photodesorption studies of other molecules \citep{Westley95, Oberg07b, Oberg09a, Oberg09b} and confirms the assumption in several observational and model papers that most ice molecules have similar photodesorption yields, around 10$^{-3}$ per incident UV photon. The CH$_3$OH photodesorption mechanism is suggested to be similar to H$_2$O and CO$_2$, i.e. a photodesorption event is initiated by photodissociation of a surface CH$_3$OH molecule. The fragments contain excess energy and either desorb directly or recombine and desorb. The insensitivity to temperature suggest that longer range diffusion is comparatively unimportant and that most molecules desorb through the escape of the produced photodissociation fragments or through immediate recombination, in the same site, and desorption of the fragments following photodissociation. A more complete study including different ice thicknesses and temperatures is however required to confirm the proposed photodesorption pathway. 

Another conceivable indirect photodesorption mechanism is desorption due to release of chemical heat following recombination of two thermalized radicals, first suggested by \citet{Williams68} and more recently investigated theoretically by \citet{Garrod07}. Its quantification requires more sensitive QMS measurements than is possible with this setup. The temperature independence suggests, however, that this is a minor desorption pathway in this setup, compared to direct photodesorption.

\subsection{Dependence of photo-product spectra on experimental variables\label{sec:res_depend}}

The influence, if any, of different experimental variables on the resulting infrared spectra of irradiated CH$_3$OH ice is investigated in detail below. These dependences are then used in the following sections to identify absorption bands and to subsequently quantify reaction rates, diffusion barriers and photodissociation branching ratios. The results are also independently valuable, since many of these experimental variables also vary between different astrophysical environments.

\subsubsection{UV fluence\label{sec:res_depend_fluence}}

\begin{figure}
\resizebox{\hsize}{!}{\includegraphics{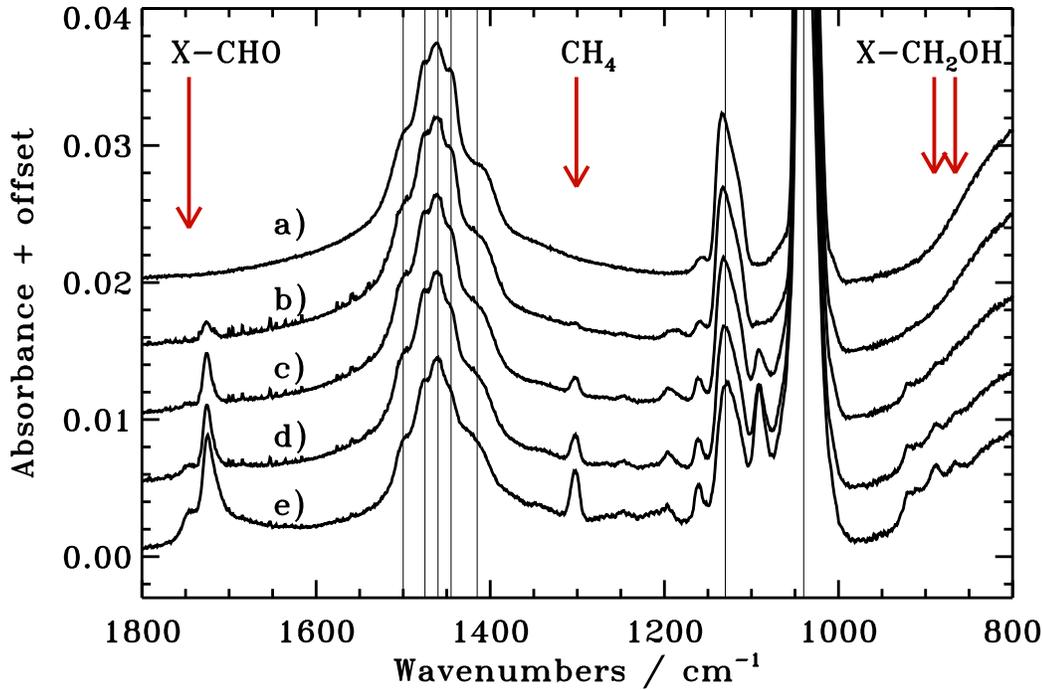}}
\caption{The different growth rates of spectral features in a CH$_3$OH ice at 50~K a) before UV irradiation and after a UV fluence of b) 7$\times10^{15}$, c) 4$\times10^{16}$, d), 8$\times10^{16}$ and e) 2.4$\times10^{17}$ cm$^{-2}$. The CH$_3$OH features are marked with thin lines. New features are present at 1750--1700, 1400--1150, 1100--1050 and 950--850 cm$^{-1}$, including the bands belonging to  CH$_4$, complex aldehydes and acids (X-CHO), and complex alcohols (X-CH$_2$OH), where X$\neq$H.} 
\label{fig:sp_fluence}
\end{figure}

In most experiments, the ices are exposed to a total UV fluence (i.e. total flux integrated over the time of the experiment) of $\sim2.4\times10^{17}$ cm$^{-2}$, which is comparable to the UV fluence in a cloud core after a million years with a UV flux of 10$^4$ cm$^{-2}$ s$^{-1}$ \citep{Shen04}. This agreement is important since the composition of photoproducts changes with UV fluence  in all experiments. This is illustrated in Fig. \ref{fig:sp_fluence}, which shows spectra of an originally 20~ML thick CH$_3$OH ice at 50~K after different fluences. This effect is demonstrated numerically below for three of the bands representing complex OH bearing molecules (X-CH$_2$OH) at 866/890 cm$^{-1}$, simple photoproducts (CH$_4$) at 1301 cm$^{-1}$, and  HCO/COOH bearing complex molecules (shortened to X-CHO in most figures for convenience) at 1747 cm$^{-1}$. These molecular class assignments agree with previous studies and are discussed specifically in \S \ref{sec:res_id}. After a fluence of $\sim7\times10^{16}$ cm$^{-2}$ the relative importance of the integrated bands at 866/890, 1301 and 1747 cm$^{-1}$ is 0:90:10. After a fluence of $\sim2.2\times10^{17}$ cm$^{-2}$ this has changed significantly to 35:34:31. The product composition after a particular fluence cannot therefore be linearly scaled to a lower or a higher fluence.

\subsubsection{UV flux\label{sec:res_depend_flux}}

\begin{figure}
\resizebox{\hsize}{!}{\includegraphics{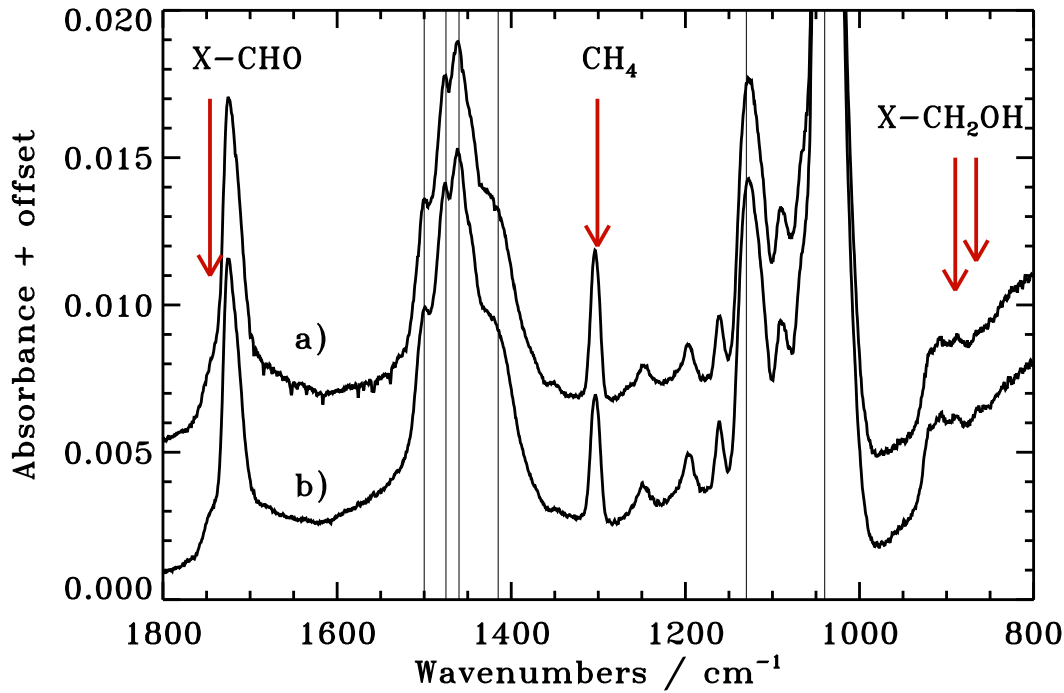}}
\caption{The spectra of originally (a) 21 and (b) 19 ML thick CH$_3$OH ice irradiated with a UV flux of (a) 1.1$\times10^{13}$ cm$^{-2}$ s$^{-1}$ and (b) 4.3$\times10^{13}$ cm$^{-2}$ s$^{-1}$ at 20~K achieve the same fluence of 2.2 $\times10^{17}$ cm$^{-2}$. The CH$_3$OH features and some product bands are marked as in Fig. \ref{fig:sp_fluence}.} 
\label{fig:sp_flux}
\end{figure}

The flux levels in the laboratory ($\sim$10$^{13}$ UV photons cm$^{-2}$ s$^{-1}$) are several orders of magnitude higher than those found in most astrophysical environments -- the interstellar irradiation field is $\sim$10$^{8}$ photons cm$^{-2}$ s$^{-1}$ \citep{Mathis83}. Hence the product dependence on flux, if any, is required before translating laboratory results into an astrophysical setting. Figure \ref{fig:sp_flux} shows that two spectra acquired after the same fluence, but irradiated with a factor of four different flux, are identical within the experimental uncertainties. Numerically, the relative importance of the integrated bands at 866/890, 1301 and the 1747 cm$^{-1}$ are 15:56:29 in the low flux experiment and 24:51:25 for the high flux experiment after a total fluence of 2.2 $\times10^{17}$ cm$^{-2}$. Including a 10--20\% uncertainty in the band intensities (the higher value for the 866/890 cm$^{-1}$ band), there is thus no significant dependence on flux within the explored flux range at 20~K. The same holds for similar experiments at 50~K (not shown). This does not exclude a flux dependence at astronomical time scales, but it does provide a benchmark for models aiming to translate laboratory results into astrophysical ones.

\subsubsection{Ice thickness\label{sec:res_depend_thick}}

\begin{figure}
\resizebox{\hsize}{!}{\includegraphics{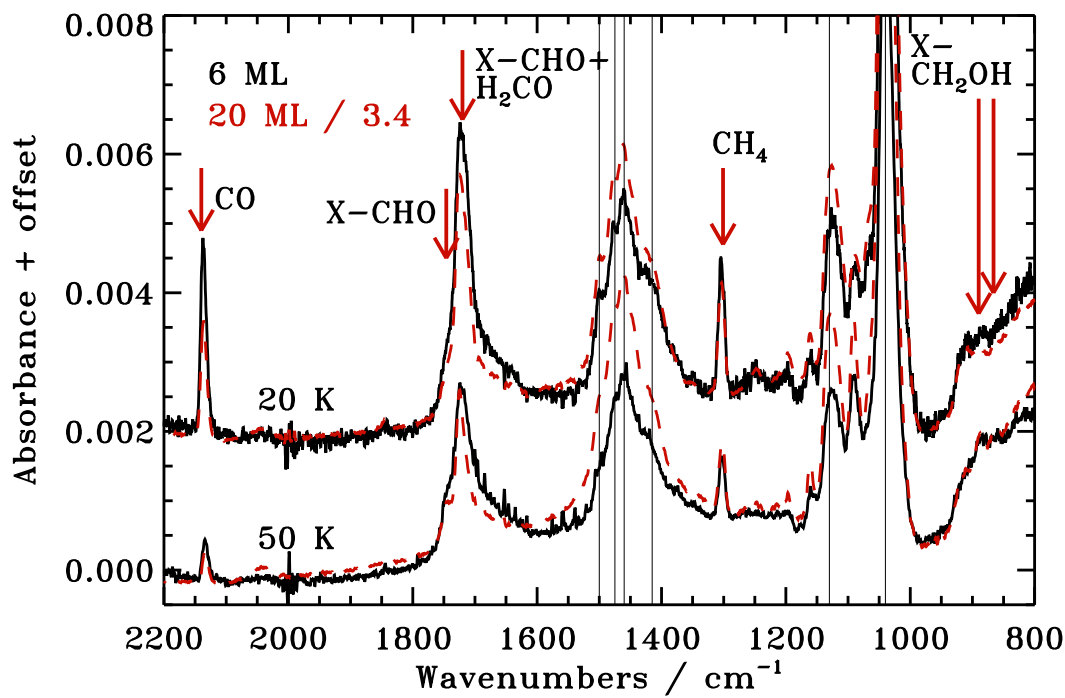}}
\caption{The differences in the photolyzed CH$_3$OH spectra of two different original thicknesses after the same fluence of $\sim$2.4$\times10^{17}$ cm$^{-2}$. The 20~ML (red dashed lines) ice spectrum is normalized to have the same CH$_3$OH absorbance as the 6~ML (black solid lines) ice experiment before irradiation to facilitate comparison of fractional photoproduct rates. CO and H$_2$CO bands are marked in addition to the spectral features focused on in previous figures.} 
\label{fig:sp_thick}
\end{figure}

Figure \ref{fig:sp_thick} shows that both the fractional CH$_3$OH destruction, as evidenced by e.g. the $\nu_7$ band intensity, decrease around 1130 cm$^{-1}$,  and the fractional formation of a few new spectral features are enhanced in thin ices ($\sim6$~ML) compared to the standard 20~ML experiment. However, this does not necessarily imply a different chemistry in thinner ices. Rather the difference in CH$_3$OH destruction may be explained by an increased escape probability of photoproducts in the thinner ice and by the greater importance of direct photodesorption. Similarly, the observed relative enhancements of the CO band at 2150 cm$^{-1}$ and the 1700 cm$^{-1}$ band in the 20~K experiment are probably due to the constant freeze-out of CO during the experiment, up to 0.2~ML out of 0.5~ML CO ice detected at the end of the 6~ML experiment, and its reactions to form more HCO-bearing carriers of the 1700 cm$^{-1}$ band. Therefore, despite the apparent dependence of the spectral features on ice thickness, there is no significant evidence for different formation yields in 6 and 20~ML thick ices. This means that bulk reactions still dominate over the potentially more efficient surface reactions in ices as thin as 6~ML, at these fluences.

\subsubsection{Ice temperature during irradiation\label{sec:res_depend_temp}}

\begin{figure}
\resizebox{\hsize}{!}{\includegraphics{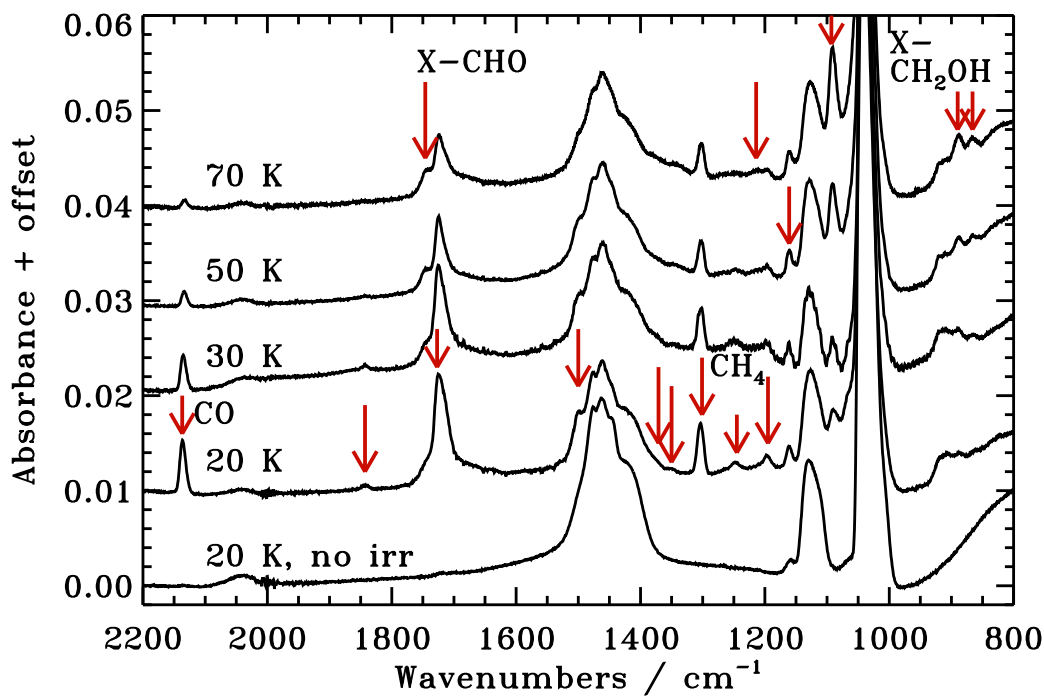}}
\caption{The photolyzed CH$_3$OH spectra at different temperatures after the same fluence of $\sim$2.4$\times10^{17}$ cm$^{-2}$ for 19--22~ML thick ices. The arrows mark new bands at the temperature at which they are most abundantly produced and some key features are also named. } 
\label{fig:sp_temp}
\end{figure}

The photolyzed ice spectra depend on the ice temperature, illustrating the different temperature dependencies of different photochemistry products (Fig. \ref{fig:sp_temp}). The 1727 (H$_2$CO + X-CHO) and 1300 (CH$_4$) cm$^{-1}$ features are most abundantly produced at the lowest investigated temperature of 20~K, while the 866/890  cm$^{-1}$ bands increase in strength with temperature and the 1747 (X-CHO) cm$^{-1}$ feature is barely affected by temperature changes. The different temperature dependencies can be used to infer the size of the main contributor to each band; photolysis fragments and molecules that form through hydrogenation of such fragments are expected to be most abundant at 20~K, while molecules that form from two larger fragments will be more efficiently produced at higher temperatures where diffusion is facilitated. This is complicated by competition between different reaction pathways, which may inhibit the formation of some complex molecules at higher temperatures where new reaction channels become possible. Nevertheless, the dependence on temperature of different bands can aid in identifying their molecular contributors. All formed bands are thus classified according to the temperature at which they are most abundantly produced (Fig. \ref{fig:sp_temp}), except for a few bands, where the dependence on temperature is too weak to assign them to a certain temperature bin. This information is summarized in Table \ref{tab:bands}.
  
\subsubsection{Pure CH$_3$OH ice versus CH$_3$OH:CO 1:1 and CH$_3$OH:CH$_4$ 1:2 ice mixtures\label{sec:res_depend_comp}}

\begin{figure}
\resizebox{\hsize}{!}{\includegraphics{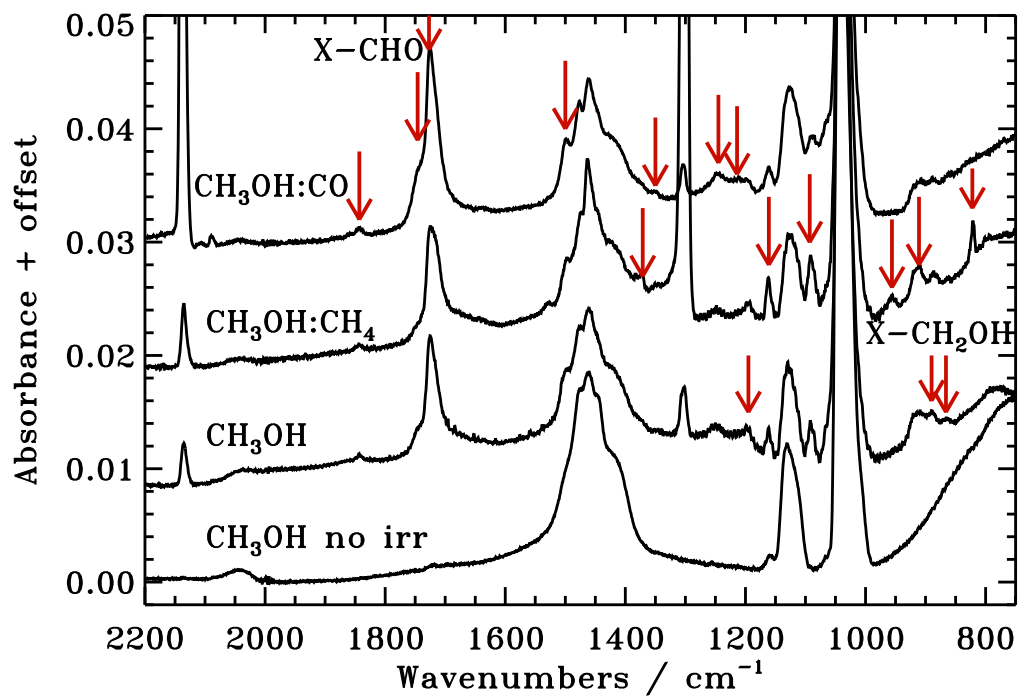}}
\caption{The changes in the photolyzed ice spectra for pure CH$_3$OH (19~ML), and 1:1 CH$_3$OH:CO and 1:2 CH$_3$OH:CH$_4$ ice mixtures at 30~K (23 and 38~ML, respectively) after the same fluence of $\sim$2.4$\times10^{17}$ cm$^{-2}$. The spectra have been scaled to correspond to the same initial CH$_3$OH abundance. The arrows mark new bands in the ice mixture where they are most abundantly produced. } 
\label{fig:sp_comp}
\end{figure}

In the set of experiments where CH$_3$OH is mixed with CH$_4$ or CO at $\sim$1:1 ratio, the resulting photoproduct compositions are significantly different compared to those obtained from pure CH$_3$OH ice experiments. This is illustrated in Fig. \ref{fig:sp_comp} for ices irradiated at 30~K. In the CH$_3$OH:CH$_4$ mixture, bands corresponding to complex molecules at 822, 890, 956, 1161, 1350 and 1382 cm$^{-1}$ are enhanced, while in the CH$_3$OH:CO mixtures, the 1214, 1245, 1350, 1498, 1727, 1746 and 1843 cm$^{-1}$ bands grow faster compared to pure CH$_3$OH ice. A few bands are most prominent when no other species is added to the CH$_3$OH ice, for example the 866 and 1195 cm$^{-1}$ bands. The 1726-1747 cm$^{-1}$ band excess in the CO-containing ices shows that molecules that contain an HCO group can be overproduced by adding CO. Similarly, the band enhancements in the CH$_4$-containing ices are expected to arise from overproduction of CH$_3$-containing molecules. These observations are used below for band identifications and later to explain variations in abundances of different complex molecules in star-forming regions.

\subsubsection{CH$_3$OH deuteration level\label{sec:res_depend_deut}}

\begin{figure}
\resizebox{\hsize}{!}{\includegraphics{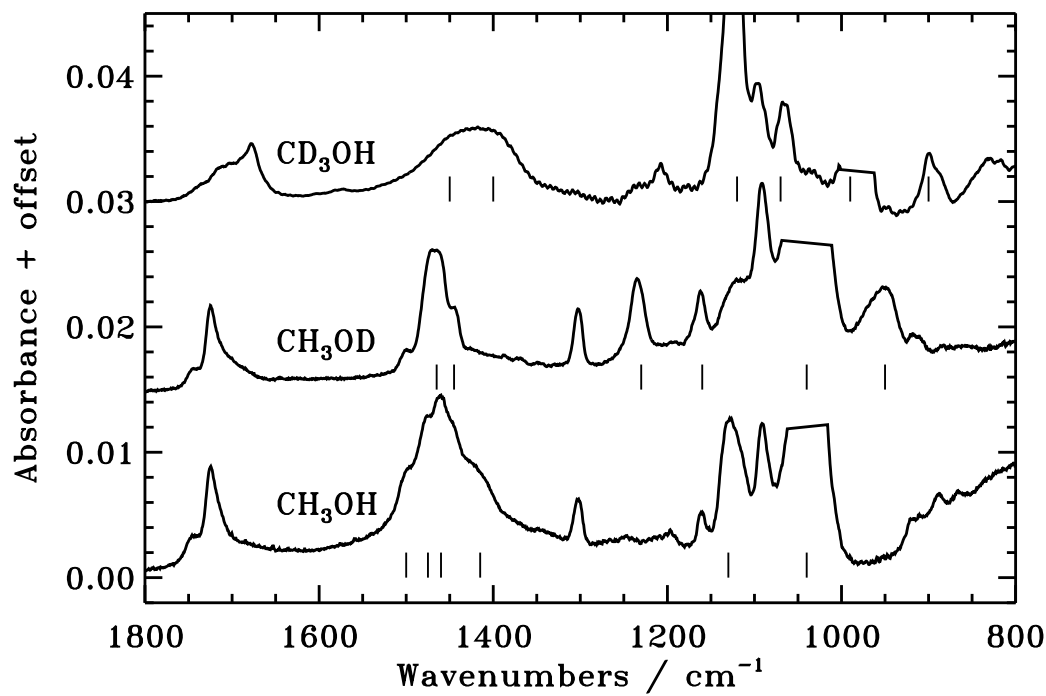}}
\caption{The resulting spectra of photolyzed pure CH$_3$OH , CH$_3$OD and CD$_3$OH ices at 50~K after the same fluence of $\sim$2.4$\times10^{17}$ cm$^{-2}$. The thin lines below each spectra mark the original CH$_3$OH, CH$_3$OD and CD$_3$OH features. The strongest band in each spectrum is blanked out for visibility.} 
\label{fig:sp_deut}
\end{figure}

In the partially deuterated ices (CH$_3$OD and CD$_3$OH) some band positions do not change compared to regular CH$_3$OH, while others are either shifted or completely missing (Fig. \ref{fig:sp_deut}). Bands that are present in the CH$_3$OH ice and missing in the CH$_3$OD experiments must originate from either OH(D)-containing molecules with the H involved in the vibrational mode in question or from simple hydrogenated species. These two groups of molecules are seldom confused and thus comparison between the photolyzed CH$_3$OH spectra and the photolyzed CH$_3$OD spectra can be used to assign some alcohol-bands. The band positions that are constant between the CH$_3$OH and CH$_3$OD do not however exclude the contributions of OH-containing molecules to these bands, since the OH group can be present in the molecule without involvement in the vibration in question. Comparing the CH$_3$OH and CH$_3$OD experiments, the bands at 866 and 890 cm$^{-1}$ are obviously affected. The 1700 cm$^{-1}$ band is somewhat reduced in the CH$_3$OD experiment, suggesting that one of the carriers is HOCH$_2$CHO. An underlying broad feature around 1600 cm$^{-1}$ is also reduced in the CH$_3$OD experiment.

 As expected, few of the bands in the CH$_3$OH experiments are still present in the UV-irradiated CD$_3$OH ice. The complex at 900-860 cm$^{-1}$and some of the X-CHO features are exceptions, though the bands are shifted. The only bands expected to appear at their normal positions come from H$_2$O$_2$ and possible H$_2$O dependent on the main source of hydrogen in the ice; neither species is obviously present in the ice from the photolyzed CD$_3$OH-ice spectra.

\subsubsection{Spectral changes during warm-up\label{sec:res_depend_warm}}

\begin{figure}
\resizebox{\hsize}{!}{\includegraphics{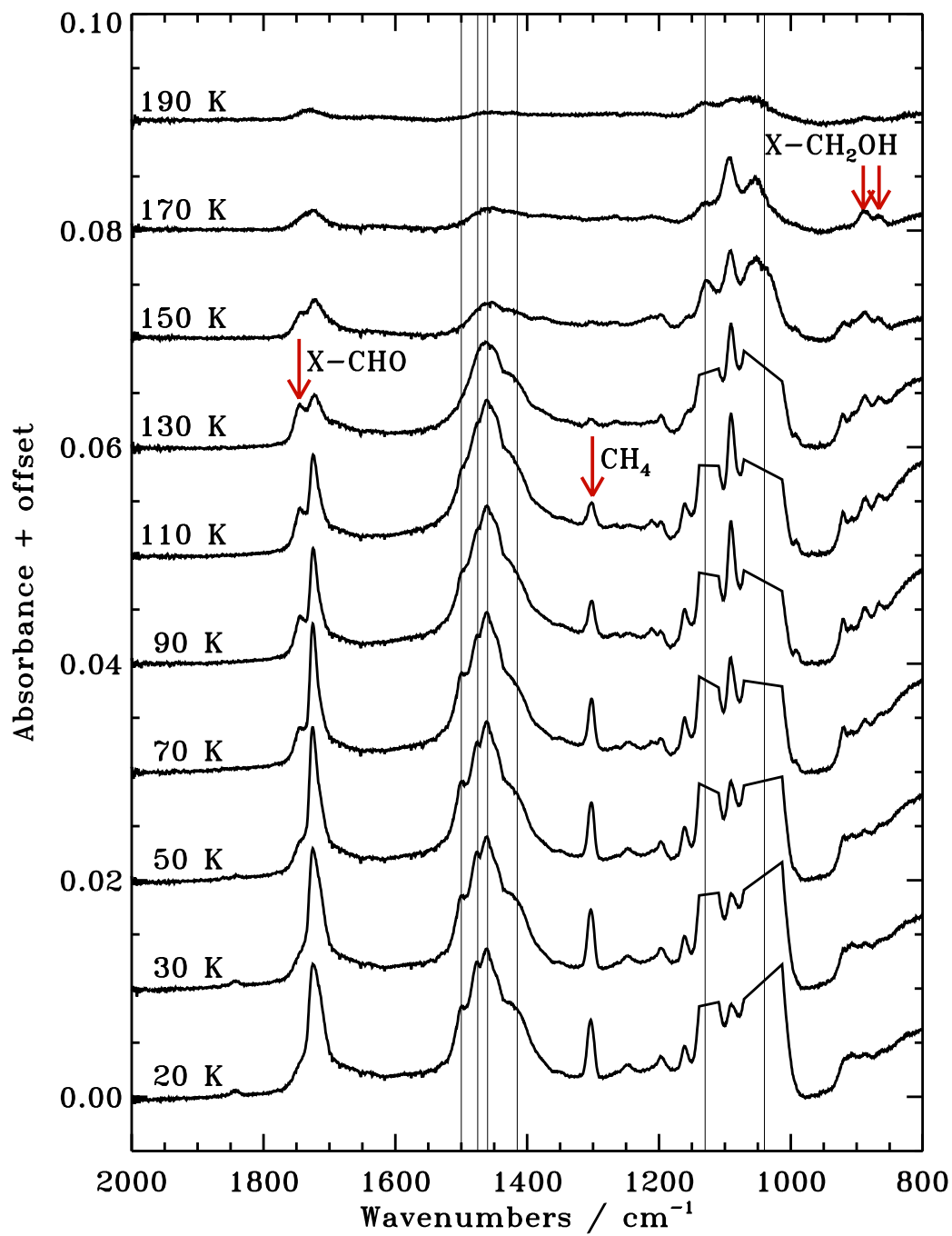}}
\caption{The photolyzed ice spectra during warm-up following irradiation at 20~K with a fluence of $\sim$2.4$\times10^{17}$ cm$^{-2}$ -- the UV lamp is turned off during the warm-up.  The thin lines mark CH$_3$OH features with the two strongest bands blanked out for visibility.} 
\label{fig:sp_warm}
\end{figure}

Following irradiation at the specified temperatures, the ices are heated by 1 K min$^{-1}$ to 200~K and spectra acquired every 10~min. Figure \ref{fig:sp_warm} shows the irradiated CH$_3$OH ice between 20 and 190~K. The UV lamp is turned off during the warm-up and thus the ice composition only depends on thermal desorption and reactions of previously produced radicals. As the ice is heated (Fig. \ref{fig:sp_warm}), several new spectral bands appear, while others increase or decrease in strength with temperature.

The 866, 890 and 1090 cm$^{-1}$ bands increase most dramatically in intensity during warm-up of the 20~K pure CH$_3$OH ice. Simultaneously the 1195  cm$^{-1}$ feature loses intensity. The 866 and 890 cm$^{-1}$ bands remain until 170~K and are the last sharp features to disappear. The 1747 and 1214 cm$^{-1}$ bands also increase in intensity with temperature. In contrast the bands at 1245, 1300, 1727 and 1850 cm$^{-1}$ only lose intensity during warm-up. Most of these bands only disappear completely at  the desorption temperature of CH$_3$OH, 120--130~K, indicating significant trapping of molecules inside the CH$_3$OH ice. Significant bulk chemistry is thus required to explain the results (see also \S \ref{sec:res_depend_thick}). At 190~K there are still some shallow bands left, which only disappear after the substrate has been heated to room temperature.

Experiments 17 and 18 show that the warm-up rate matters somewhat for the final ice composition. The quantification and discussion of this effect is saved for Paper II.

\subsection{Reference RAIR spectra and TPD experiments of pure complex ices\label{sec:res_sup}}

\begin{figure}
\resizebox{\hsize}{!}{\includegraphics{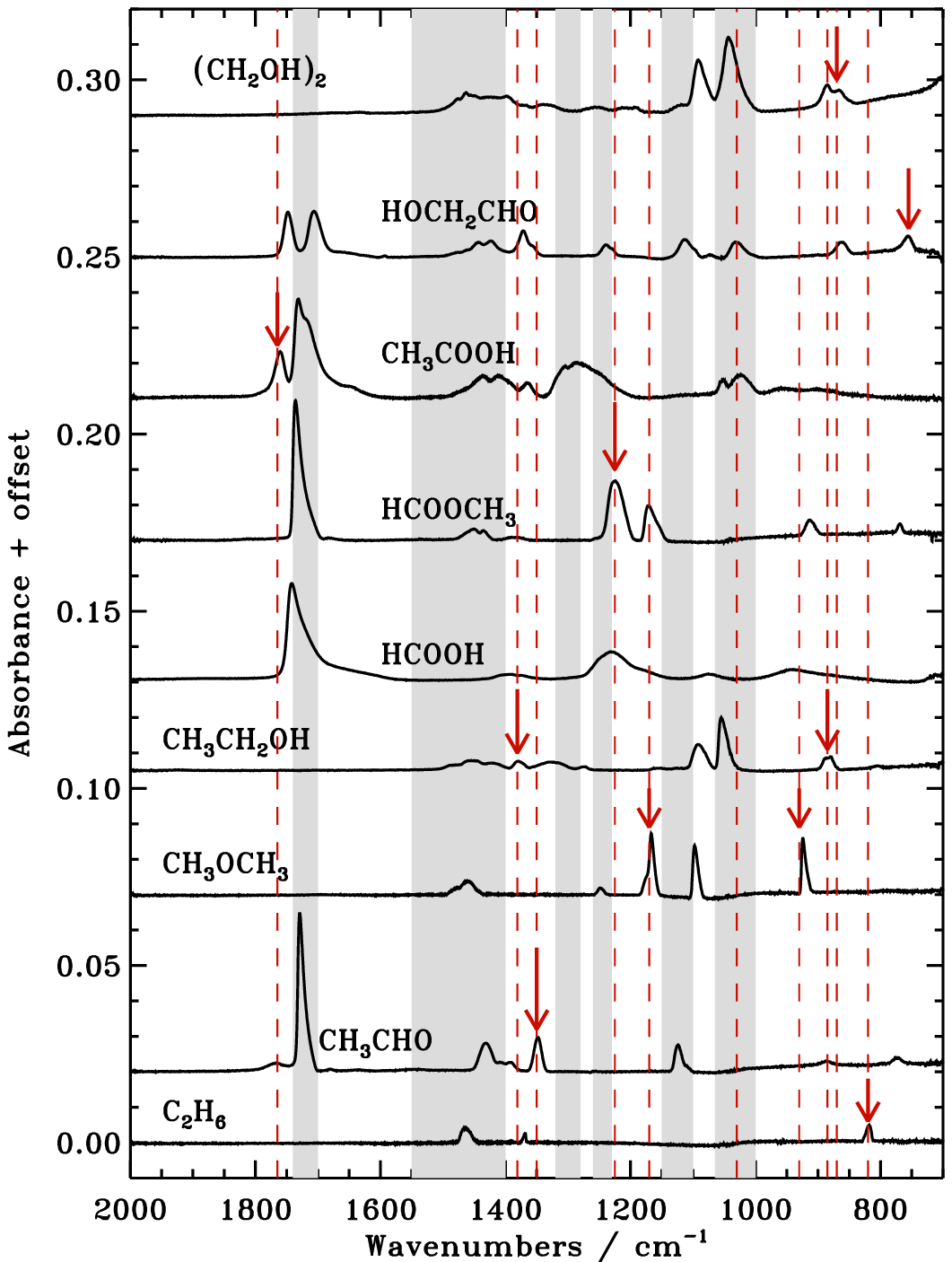}}
\caption{RAIR spectra of 3--9~ML thick pure complex organic ices at 20~K (except for (CH$_2$OH)$_2$ at 150~K) used to identify stable photoproducts. The red arrows indicate the bands mainly used for identification and abundance determinations -- HCOOH has no sharp isolated feature. The dashed lines follow the arrows through all spectra to visualize overlaps with other spectral bands. The shaded regions show the wavelength regions where fundamental modes of CH$_4$, H$_2$CO and CH$_3$OH absorb.}
\label{fig:comp_sp}
\end{figure}

\begin{figure}
\resizebox{\hsize}{!}{\includegraphics{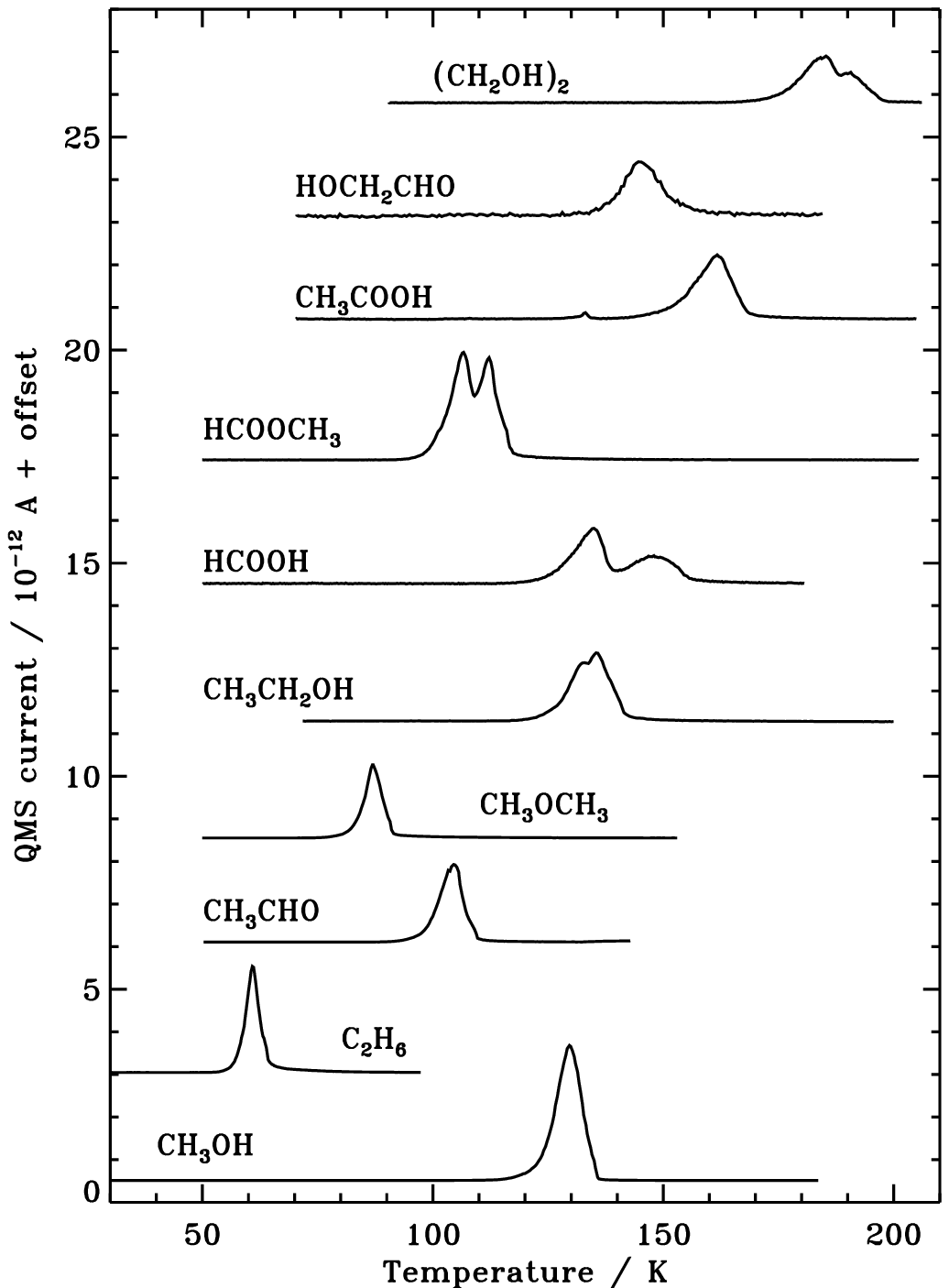}}
\caption{TPD spectra of pure complex organic ices used to identify photoproducts together with a CH$_3$OH TPD curve.}
\label{fig:comp_tpd}
\end{figure}

Photolysis of CH$_3$OH ice and recombination of the fragments can theoretically result in a large number of new species. To facilitate the identification of these species, this section briefly presents new RAIR spectra and TPD time series of all stable, complex photoproducts considered in this study. Radicals are expected to form in the photolyzed ice, but these cannot be produced in pure form in ices and thus comparison spectra are difficult to obtain.

Figure \ref{fig:comp_sp} shows the RAIR spectra of C$_2$H$_6$ (ethane), CH$_3$CHO (acetaldehyde), CH$_3$OCH$_3$ (dimethyl ether), CH$_3$CH$_2$OH (ethanol), HCOOH (formic acid), HCOOCH$_3$ (methyl formate), CH$_3$COOH (acetic acid) and HOCH$_2$CHO (glycolaldehyde) at 20~K and (CH$_2$OH)$_2$ (ethylene glycol) at 150~K. Below 150~K the (CH$_2$OH)$_2$ spectrum contains CH$_3$OH features despite the high stated purity (99\%) of the sample. The spectral bands above 2000 cm$^{-1}$ are not shown since all complex molecule spectral features in that region overlap with strong CH$_3$OH features and thus cannot be used for identification. The figure illustrates that most bands overlap with at least one band from another complex molecule or with bands of H$_2$CO and/or CH$_3$OH, which is expected for complex molecules with the same or similar functional groups (e.g. absorption by HCO/COOH stretches at 1700--1750 cm$^{-1}$). The red arrows indicate the bands mainly used for identification of each species. These bands are chosen to overlap as little as possible with strong absorption features of other species. For HCOOH, CH$_3$COOH, HCOOCH$_3$ and HOCH$_2$CHO there are no suitable bands for determining the produced abundances in most experiments, i.e. the isolated bands are too weak to provide detections or strict upper limits. For these molecules the red arrows indicate the bands used to derive upper limits, while the 1700 cm$^{-1}$ band is used to derive the sum of their abundances once the H$_2$CO contribution has been subtracted. Where bands are partly overlapping only the low- or high-frequency half of the band is used for identification. 

Figure \ref{fig:comp_tpd} displays the TPD curves for the same complex molecules as shown in Fig. \ref{fig:comp_sp}. The QMS signal belonging to the molecular mass is plotted for each TPD experiment. The $m/z$ values of all possible fragments have also been gathered, and are used to separate TPD curves of molecules of the same molecular mass that desorb in a similar temperature interval. For example, CH$_3$COOH and HCOOCH$_3$ both have a molecular mass of 60, but CH$_3$COOH frequently loses an OH group in the QMS, resulting in m/z = 17 and 43, while HCOOCH$_3$ does not. The TPD curves were modeled using the IDL routine MPFIT under the assumption of zeroth order desorption behavior, which is expected for multilayer ices. The desorption rate is then $\nu\times N_{\rm sites}\times{\rm exp}(-E_{\rm des}/T)$. The vibrational frequency $\nu$ is defined as a function of the desorption energy $E_{\rm des}$: $\nu = \sqrt(2k_{\rm B}N_{\rm sites}E_{\rm des}\pi^{-2}m^{-1})$, where $N_{\rm sites}$ is the number of molecular sites per cm$^{-2}$ and $m$ is the molecular mass -- $N_{\rm sites}$ is a constant and assumed to be 10$^15$ cm$^{-2}$ in line with previous studies. The resulting desorption energies are reported in Table \ref{tab:sup_exps}. The uncertainties include both model uncertainty and experimental errors. The values agree with those published by \citet{Garrod06}, based on experiments on water rich ice mixtures by \citet{Collings04}, within 20\%, except for HCOOCH$_3$ where the discrepancy is larger. This may however be due to experimental differences, i.e. pure ices here versus their ice mixtures, rather than experimental errors.

\subsection{Identification of CH$_3$OH ice UV photoproducts\label{sec:res_id}}

\begin{figure}
\resizebox{\hsize}{!}{\includegraphics{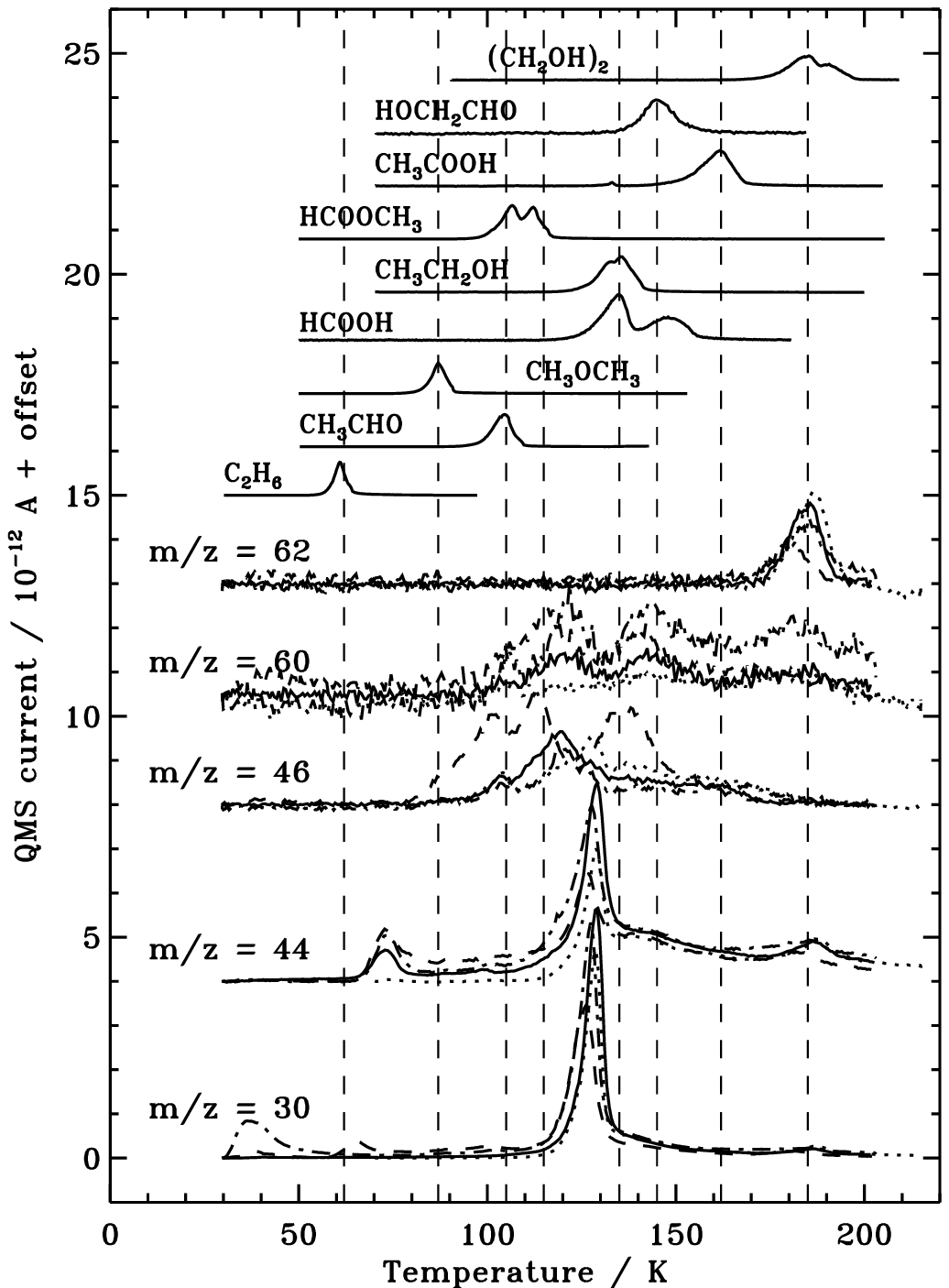}}
\caption{TPD experiments following UV irradiation of a pure CH$_3$OH ice at 30~K (black), 70~K (green), a CH$_3$OH:CH$_4$ 1:2 ice mixture at 30~K (red) and a CH$_3$OH:CO 1:1 mixture at 30~K (blue). $m/z=62$ can only contain contributions from (CH$_2$OH)$_2$, $m/z=60$ from (CH$_2$OH)$_2$, HOCH$_2$CHO, CH$_3$COOH and HCOOCH$_3$, $m/z=46$ from HCOOH, CH$_3$CH$_2$OH and CH$_3$OCH$_3$, and $m/z=44$ from CO$_2$, CH$_3$CHO and all heavier complex organics. Finally $m/z=30$ can contain contributions from C$_2$H$_6$, CH$_3$OH, H$_2$CO and several heavier organics compounds. All TPD series are scaled to the same initial CH$_3$OH abundance to facilitate comparison}
\label{fig:comp_tpd_irr}
\end{figure}

In addition to the complex molecules described in the previous section, smaller molecules and radicals that form via CH$_3$OH photodissociation are also considered when assigning spectral bands following CH$_3$OH ice photolysis. Identification of all photoproducts is a two-step process where the first step is the comparison between new band positions and experimentally or calculated band positions of molecules and fragments to establish a list of possible carriers to each observed band. In the second step the behavior of the band when changing experimental variables, as described in \S \ref{sec:res_depend}, is employed  together with QMS data (Fig. \ref{fig:comp_tpd_irr}) to determine which ones(s) of the possible candidates is the most important contributor to the formed band.  Combining all information, a band identification is considered secure when consistent with the list of criteria below.

\begin{enumerate}
\item The spectral band position in the photolysis spectra must agree within 15 cm$^{-1}$ of a measured pure ice band \citep[e.g.][]{Hudson05} or within 50 cm$^{-1}$ of a calculated band position for the species in question to be considered a candidate carrier of the observed band. In each case all spectral features within the spectrometer range are checked for consistency even if only one band is used for identification. 
\item The temperature at which the band starts to disappear during warm-up is compared for consistency with the observed desorption temperatures of different complex molecules in pure-ice TPD experiments. 
\item The mass signature in the TPD experiment following UV irradiation is checked at each temperature where a tentatively assigned band disappears during warm-up (Fig. \ref{fig:comp_tpd_irr}). 
\item The band positions of new carriers in UV-irradiated CH$_3$OH and partly-deuterated CH$_3$OH experiments are compared to check that the expected shifts occur as discussed in \S \ref{sec:res_depend_deut}. 
\item The irradiated spectra are examined for band enhancements and suppressions in CH$_3$OH mixtures with CO and CH$_4$. In CH$_4$ experiments, species containing CH$_3$ groups are expected to be over-produced compared to pure CH$_3$OH ice experiments. Similarly in CO containing experiments, HCO-group containing species should have enhanced abundances.
\item Irradiation experiments at different temperatures are compared to ensure consistency with the expected relative diffusion barriers of differently sized radicals (see also \S \ref{sec:res_depend_temp}). 
\item Finally, the temperatures at which radicals disappear during warm-up are compared with the spectroscopic appearance or increase of the molecular band in question; where radicals are detected, the loss of radical bands during warm-up should correspond to the enhancement of molecular bands formed from recombination of these same radicals.
\end{enumerate}

All observed bands and their inferred carrier properties are listed in Table \ref{tab:bands}, i.e. if a band is enhanced in the CO:CH$_3$OH mixture its main contributor contains a CO group, if it is enhanced in the CH$_4$:CH$_3$OH mixture the main contributor contains a CH$_3$ group and if the band disappears in the CH$_3$OD experiment the main contributor contains an OH group.

\begin{table*}
\begin{center}
\caption{Detected bands between 700 and 2200 cm$^{-1}$ and identifications.}             
\label{tab:bands}      
\centering                          
\begin{tabular}{l c c c c c c c}        
\hline\hline                 
Wavenumber (cm$^{-1}$) &$T_{\rm form}$ (K)$^{\rm a}$ & $T_{\rm des}$ (K)$^{\rm b}$ & CO & CH$_3$ & OH &Candidates \\
\hline                 
2135 & 20 & 30 & y &  & &CO \\
1843 & 20 & 30 & y & &  &HCO \\
1746 & 70 & 140--190 & y &  &  y/--&HOCH$_2$CHO + HCOOH + CH$_3$CHO \\
1727 & 20 & 70/110/150 & y &  & &H$_2$CO + CH$_3$CHO + HCOOCH$_3$\\
1498 & 20 & 70 & y &  &  & H$_2$CO\\
1382 & 20 & 140&  & y &  & CH$_3$CH$_2$OH\\
1372 & 30 & 40 &  & y &   & C$_2$H$_6$\\
1350 & 20 & 100 & y & y &  &CH$_3$CHO \\
1301 & 20 & 40 &  & y &  &CH$_4$ \\
1245 & 20 & 70 & y &  &  &H$_2$CO \\
1214 & 70 & 130 & y &  &  &HCOOCH$_3$ \\
1195 & 20 & 50 &  &  &  y & CH$_2$OH \\
1161 & 50 & 90 &  & y &  &CH$_3$OCH$_3$ \\
1093 & 70 & 110/130/180&  &  &  & (CH$_2$OH)$_2$ + CH$_3$OCH$_3$ + CH$_3$CH$_2$OH\\
956 & 30 & 70 &  & y &  & \\
921 & ? & 110 & & y & & CH$_3$OCH$_3$\\
911 & ? & 150 &  &  &  & HCOOCH$_3$ \\
890 & 70 & 180 &  &  & y & (CH$_2$OH)$_2$\\
885 & ? & 130 &  & y & y & CH$_3$CH$_2$OH \\
866 & 70 & 150/180 &  &  & y & (CH$_2$OH)$_2$ (+ HOCH$_2$CHO)$^{\rm c}$ \\
822 & 30 & 50 &  & y & &C$_2$H$_6$\\
\hline
\end{tabular}
\end{center}
$^{\rm a}$The temperature at which the band carrier is most efficiently produced.\\
$^{\rm b}$The temperature at which the band starts to disappear during warm-up with 1 K min$^{-1}$.\\
$^{\rm c}$ Minor contributor in most experiments.
\end{table*}

\subsubsection{Small molecules: H$_2$, H$_2$O, CH$_4$, CO, O$_2$, H$_2$CO, H$_2$O$_2$ and CO$_2$\label{sec:res_id_small}}

H$_2$ probably forms in the photolysis experiments, since CH$_2$OH is observed and thus H atoms must be produced in the ice (see below). Two H atoms can subsequently recombine to form H$_2$ as observed in D$_2$O photodesorption experiments \citep{Watanabe00} and following irradiation of H$_2$O:CH$_3$OH:NH$_3$:CO experiments \citep{Sandford93}. Detections or determinations of strict upper limits of H$_2$ are however not possible because of a lack of strong infrared transitions together with the expected fast desorption of any H$_2$ from ices above 20~K. 

H$_2$O (D$_2$O) is visible during warm-up following the desorption of CH$_3$OH (CH$_3$OD). However, its origin is unclear, since H$_2$O is the main contaminant in the chamber. During irradiation, the H$_2$O stretching and libration modes are hidden under strong CH$_3$OH bands and thus cannot be used for identification. The 1670 cm$^{-1}$ bending mode coincides with the broad wing of the 1727 cm$^{-1}$ band found in most irradiated CH$_3$OH spectra. The wing is probably reduced in the CH$_3$OD ice, depending on the baseline determination, but it is also not visible in the CD$_3$OH ice (Fig. \ref{fig:sp_deut}). Thus, there is no clear evidence for the amount that H$_2$O contributes to this feature. The best constraints on water formation come instead from the CH$_3$OD experiments where $\sim$0.5~ML D$_2$O is detected at the end of the experiment, corresponding to a few percent with respect to the initial CH$_3$OD amount.

CH$_4$ is formed in all experiments, identified by its relatively isolated $\nu_4$ band at 1301 cm$^{-1}$ \citep{dHendecourt86}. The CH$_4$ $\nu_3$ feature at 3008 cm$^{-1}$ is also one of the few bands that are clearly visible on top of the CH$_3$OH bands in that region. 

CO has a single strong feature at 2139 cm$^{-1}$ \citep{Gerakines95}, which is shifted to 2135 cm$^{-1}$ in the experiments here. The feature cannot be confused with those of any other species and the identification is thus clear. Of the observed CO ice amount $\sim$0.2~ML originates outside of the CH$_3$OH ice. This is however only significant for the thin ($\sim$6~ML) 20~K ice experiments. At 30~K and above the CO sticking coefficient is low \citep{Bisschop07}. 

O$_2$ has no strong infrared bands. During warm-up of the 20 and 30~K experiment there is no $m/z$ = 32 detected at the expected O$_2$ desorption temperature of $\sim$30~K \citep{Acharyya07}. In contrast there is a clear $m/z$=28 band from CO in this temperature region. If CO and O$_2$ can be assumed to behave similarly this puts an upper limit on the O$_2$ production to less than 5\% of that of CO at 20~K.

H$_2$CO has three strong bands between 2000 and 800 cm$^{-1}$ at 1723, 1494 and 1244  cm$^{-1}$. The 1723 cm$^{-1}$ band overlaps with several complex spectral features, and the 1494 cm$^{-1}$ band sits on the shoulder of a strong CH$_3$OH band. The 1244 cm$^{-1}$ band is relatively isolated and has been used previously to constrain the H$_2$CO production \citep{Bennet07a}. All three bands are readily observed in the irradiated ices between 20 and 50~K. The 1244 and 1494 cm$^{-1}$ bands are strongly correlated and are most abundantly formed at low temperatures, while the 1723 cm$^{-1}$ band is less dependent on temperature, as expected from its multiple carriers. All three bands are enhanced in all CH$_3$OH:CO ice-mixture experiments and these identifications are therefore considered secure. 

All fundamental H$_2$O$_2$ infrared bands completely overlap with strong CH$_3$OH bands in the investigated spectral region \citep[e.g.][]{Loeffler06,Ioppolo08}. A small $m/z$=34 peak was observed during CH$_3$OH desorption. Without comparison TPD this cannot be used quantitatively, however. Thus, while some h$_2$O$_2$ probably forms, its formation rate cannot be constrained experimentally. 

CO$_2$ has a strong band at $\sim$2340 cm$^{-1}$, which is seen in all experiments at high fluences. Because of purge problems this spectral region is somewhat polluted with CO$_2$ lines from gas outside of the vacuum chamber. This results in a larger uncertainty in the derived CO$_2$ abundance than would have been the case otherwise.

\subsubsection{Radicals: OH, CH$_3$, HCO, CH$_3$O, CH$_2$OH\label{sec:res_id_rad}}

OH absorbs at 3423 and 3458 cm$^{-1}$ from a study on pure H$_2$O ice photolysis by \citet{Gerakines96}, using transmission spectroscopy. \citet{Bennet07a} calculated the band position to be at 3594 cm$^{-1}$ using the hybrid density functional theory with the 6-311G(d,p) basis set and refining the output with the coupled cluster CCSD(T) method -- these calculations have typical gas-phase band-position uncertainties of 0.5--1\% or $<$50 cm$^{-1}$ \citep{Galabov02}. There are some shallow bands in the 3400-3600 cm$^{-1}$ wavelength region in our CH$_3$OH experiments, that disappear between 30 and 50~K, and that are also present in the CD$_3$OH ices but not in the CH$_3$OD ones. The bands are however an order of magnitude broader than observed for OH radicals in matrix studies \citep{Acquista68}. This may be due to their interactions with other H-bonding molecules, but with such a disagreement present, these bands cannot be assigned to OH radicals with any certainty. The presence of the broad features also prevents the determination of strict upper limits.

HCO can be identified from its $\nu_2$ band between 1840 and 1860 cm$^{-1}$ \citep{Gerakines96, Bennet07a}. No stable species considered in this study absorbs in this region. \citet{Bennet07a} calculated the band positions of a large number of unstable species that can theoretically form in an irradiated CH$_3$OH ice and none of these have absorption features within 50~cm$^{-1}$ of 1840 or 1860 cm$^{-1}$. Furthermore the band enhancement in the CO ice mixture and the low temperature at which the bands disappear all point to HCO. The assignment of this band to HCO is thus considered secure. 

Calculations predict that CH$_3$ absorb at 1361 and 3009 cm$^{-1}$ \citep{Bennet07a}. In the 30~K CH$_4$ ice mixture, two bands appear close to these wavelengths, at 1385 and 2965 cm$^{-1}$. Using the band width from the CH$_4$ ice mixture, this wavelength region could be used to constrain the CH$_3$ production in all ices. Unfortunately, the band strength is not known well enough to derive useful upper limits. 

\citet{Bennet07a} calculated that the CO stretching band around 1170 cm$^{-1}$ is the strongest CH$_2$OH band within our spectral range. From matrix isolation experiments CH$_2$OH has been found to have one strong band at 1183 cm$^{-1}$ in agreement with the calculations \citep{Jacox73}. Similarly to \citet{Gerakines96} and \citet{Bennet07a}, we detect a feature at 1195 cm$^{-1}$, which appears at the onset of irradiation in all pure CH$_3$OH experiments and is most abundant in the 20~K ice as would be expected for a radical (Fig. \ref{fig:sp_temp}).  The band is not enhanced in the CO- or CH$_4$-containing ice mixtures, which confirms that the band  forms from CH$_3$OH alone. It is also not present in the CH$_3$OD ice after irradiation. The band starts to disappear around 50~K during warm-up, at the same time other bands grow, which can be assigned to more complex CH$_2$OH bearing species. Thus we confirm the results of  \citet{Gerakines96} and \citet{Bennet07a} that the 1195 cm$^{-1}$ band observed at low temperatures is due to CH$_2$OH. Above 50~K, it is only possible to derive CH$_2$OH upper limits, because of overlap with other absorption features.

CH$_3$O is predicted to have two fundamental transitions, isolated from strong CH$_3$OH bands, at 1319 and 1329 cm$^{-1}$ \citep{Bennet07a}. The closest observed bands are the CH$_4$ band at 1300 cm$^{-1}$ and a weak band at 1350 cm$^{-1}$. The 1350 cm$^{-1}$ feature only starts to disappear at 70~K, which is a higher temperature than expected for a radical. It is also somewhat enhanced in CO- and CH$_4$-containing mixtures. Hence there is no evidence for abundant build-up of CH$_3$O in any of the experiments. This is consistent with the comparatively small formation of CH$_3$O-containing molecules during warm-up of the irradiated ice as is reported in more detail below.

\subsubsection{CH$_3$-bearing complex molecules: C$_2$H$_6$, CH$_3$CHO, CH$_3$CH$_2$OH and CH$_3$OCH$_3$\label{sec:res_id_ch3}}

\begin{figure}
\resizebox{\hsize}{!}{\includegraphics{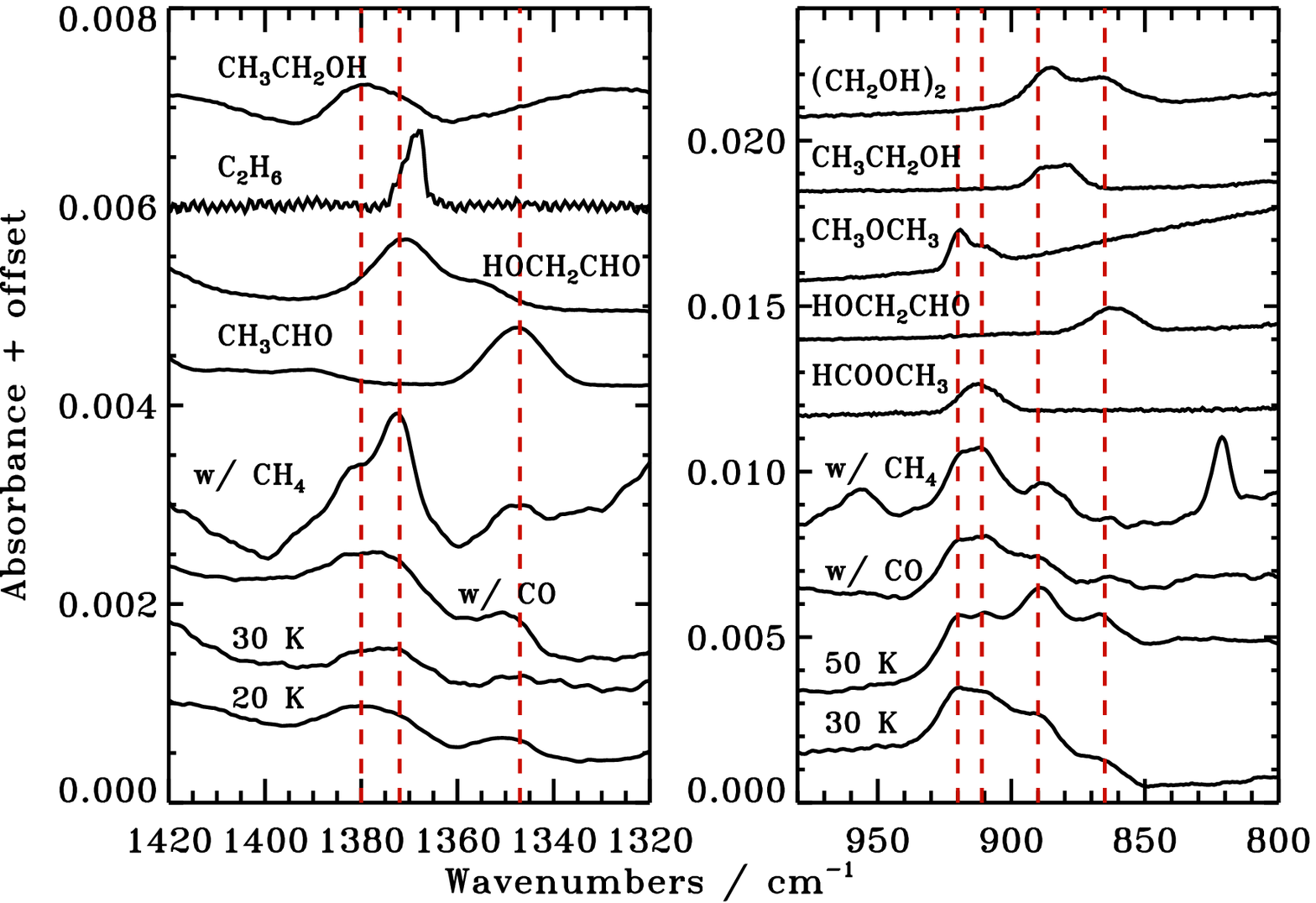}}
\caption{The left panel shows spectra of UV-irradiated pure CH$_3$OH ice at 20 and 30~K, and CH$_3$OH:CO 1:1 and CH$_3$OH:CH$_4$ 1:2 ice mixtures at 30~K after the same fluence of 2.4$\times10^{17}$ cm$^{-2}$, together with pure ice spectra of the complex species that absorb in this region. The right panel shows the complex absorption pattern between 980 and 820 cm$^{-1}$ in pure ices at 30 and 50~K and mixed ices at 30~K, together with the possible carriers of these bands. In both panels the CH$_3$OH bands have been subtracted from the spectra for visibility and the features have been scaled to the initial CH$_3$OH abundance in each experiment to facilitate comparison. The red dashed lines are present to guide the eye between band positions in the irradiated ices and in pure complex ice spectra. } 
\label{fig:sp_ch3}
\end{figure}

C$_2$H$_6$ is only detected in the UV-irradiated 30~K CH$_3$OH:CH$_4$ ice mixture experiment (Fig. \ref{fig:sp_comp}). This ice spectrum contains a clear feature at 822 cm$^{-1}$ with the expected band width of C$_2$H$_6$. The band starts to disappear at 50~K during warm-up and it can thus be securely assigned to C$_2$H$_6$. The same spectral region is used in the other experiments to derive strict upper limits on the C$_2$H$_6$ production.

Figure \ref{fig:sp_ch3} shows that there is a shallow band at the frequency of the CH$_3$CHO 1350 cm$^{-1}$ feature in most experiments and that CH$_3$CHO is the only molecule with a directly overlapping band, though the wing of the HOCH$_2$CHO band cannot be excluded as a carrier from spectral comparison alone. At 30~K, the feature is somewhat enhanced in both the CH$_4$- and CO-containing mixtures, when comparing ices at the same temperature. The temperature at which the feature starts to disappear (100~K) is close to the expected desorption temperature for CH$_3$CHO from the pure TPD experiments. The feature is completely gone before the onset of CH$_3$OH desorption which excludes significant contributions from less volatile molecules, like HOCH$_2$CHO. Together these observations are only consistent with CH$_3$CHO as the main carrier. 

Figure \ref{fig:sp_ch3} also shows that a band at 1380 cm$^{-1}$ in the photolysis spectra correlates with the pure CH$_3$CH$_2$OH ice feature, but HOCH$_2$CHO is also a potential carrier. The band is, however, enhanced in the CH$_4$-mixture experiments. Simultaneously a band at 885 cm$^{-1}$ is enhanced (Fig. \ref{fig:sp_ch3}), which can be attributed to CH$_3$CH$_2$OH as well. Furthermore, the 1380 cm$^{-1}$ band starts to desorb  at 120~K, with CH$_3$OH, while the complex HCO-bearing species only experience significant desorption at 150~K. This suggests that CH$_3$CH$_2$OH is the main carrier of the 1380 cm$^{-1}$ feature. The assignment is confirmed by a $m/z=46$ detection at the desorption temperature of CH$_3$CH$_2$OH; the QMS signal is enhanced by about a factor of three in the CH$_4$ mixture experiment, which is of the same order as the RAIRS band enhancement.  In the CO ice mixtures the contribution of HOCH$_2$CHO may however be significant and in these ices the 1380 cm$^{-1}$ feature is only used to provide upper limits on the CH$_3$CH$_2$OH production.  

\begin{figure}
\resizebox{\hsize}{!}{\includegraphics{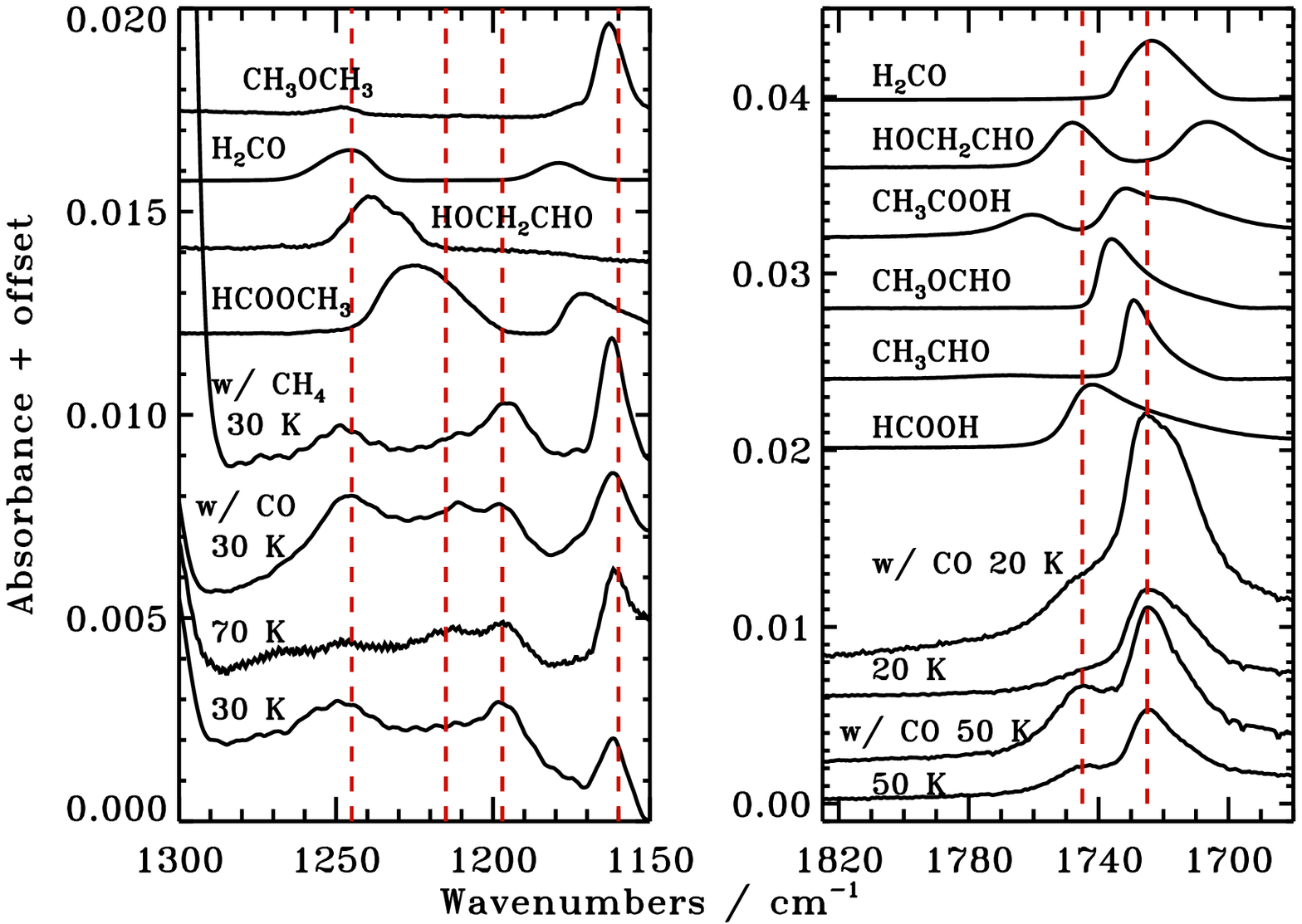}}
\caption{The left panel panel shows the observed absorption bands between 1140 and 1290 cm$^{-1}$ following UV irradiation of pure CH$_3$OH ice, CH$_3$OH:CO 1:1 and CH$_3$OH:CH$_4$ 1:2 ice mixtures, and the possible carriers of these features. The right panel shows the 1700-1770 cm$^{-1}$ HCO feature in irradiated pure CH$_3$OH ices at 20 and 50~K together with irradiated CH$_3$OH:CO $\sim$1:1 mixtures, and the possible carriers. In both panels the red dashed lines indicate some of the complex band positions.}
\label{fig:sp_hco}
\end{figure}

CH$_3$OCH$_3$ has not been generally considered as a photoproduct of CH$_3$OH ice in previous experimental studies \citep{Gerakines96, Bennet07a}. CH$_3$OCH$_3$ is however formed in the ice in our experiments; the feature at 1090 cm$^{-1}$, which can contain contributions from CH$_3$OCH$_3$, CH$_3$CH$_2$OH and (CH$_2$OH)$_2$ loses intensity at 90~K, before the onset of desorption of any of the other two species, which is only consistent with the presence of CH$_3$OCH$_3$ in the ice. Furthermore, Fig. \ref{fig:sp_hco} shows that the band seen around 1170 cm$^{-1}$ agrees better with CH$_3$OCH$_3$ than any other complex molecule and that it has no similarity with the nearby HCOOCH$_3$ band, with which it is usually identified \citep{Gerakines96, Bennet07a}. The band is not significantly affected by the ice temperature during irradiation and during warm-up it starts to lose intensity around 90~K, close to the CH$_3$OCH$_3$ desorption temperature (Fig. \ref{fig:comp_tpd}). At the same time a $m/z$ value of 46, characteristic of CH$_3$OCH$_3$, is detected by the QMS (Fig. \ref{fig:comp_tpd_irr}). The 1170 cm$^{-1}$ band is enhanced in the CH$_4$ mixture and so is a feature at 920 cm$^{-1}$, which can also be attributed to CH$_3$OCH$_3$. Neither feature is enhanced in the CO ice mixture, which confirms the identification of the band carrier with CH$_3$OCH$_3$.  The uncertainty of the integrated abundance of the CH$_3$OCH$_3$ feature is relatively large because of uncertainties in the baseline due to overlap with HCOOCH$_3$ and CH$_3$OH bands. 

\subsubsection{CHO-bearing complex molecules: HCOOH, HOCH$_2$CHO, HCOOCH$_3$ and CH$_3$COOH}

In the pure CH$_3$OH ice experiments, none of the CHO-bearing species, except H$_2$CO and CH$_3$CHO, are uniquely identified. Similarly HCOOH and CH$_3$COOH have no unique bands. A mixture of these compounds can, however, be identified from the growth of the 1700 cm$^{-1}$ complex (Fig. \ref{fig:sp_hco}). The relative importance of the different contributors in different mixtures and at different temperatures can also be assessed by spectral comparison. Figure \ref{fig:sp_hco} shows that the shape of the 1700 cm$^{-1}$ band depends on the ice temperature during irradiation. At lower temperatures, a broad feature, peaking at 1725 cm$^{-1}$ dominates, which can be identified with H$_2$CO. At higher temperatures the band is more similar in width to CH$_3$CHO and HCOOCH$_3$. This is true both when the ice is irradiated at higher temperatures and when an ice irradiated at 20~K is warmed up to $>$50~K. Simultaneously, the high frequency wing becomes more pronounced, suggesting an increasing importance of HOCH$_2$CHO, compared to HCOOH. This is the case in both pure and CH$_3$OH:CO ice mixtures. 

\begin{figure}
\resizebox{\hsize}{!}{\includegraphics{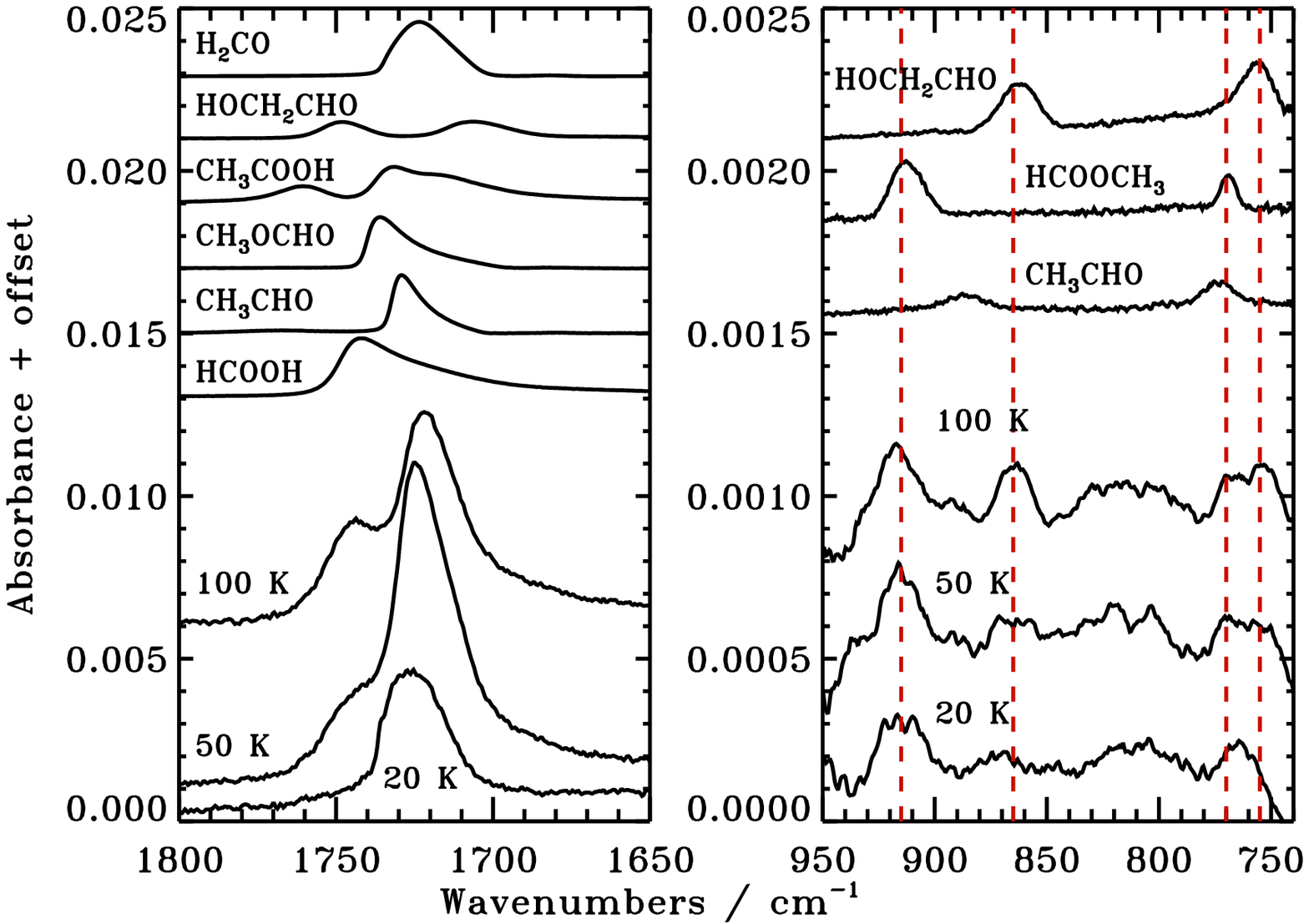}}
\caption{The left panel shows the 1700 cm$^{-1}$ feature during warm-up of a CH$_3$OH:CO 1:10 ice from 20 to 100~K, after irradiation at 20~K, together with six possible carriers of these bands. The right panel shows the 950-750 cm$^{-1}$ region in the same heated mixture together with possible carriers of the detected features. All spectra are scaled with the same fraction in both panels, so that the absorbance of each spectrum corresponds to $\sim$0.1~ML of each of the complex species.}
\label{sp_hco_2}
\end{figure}

In the CO-dominated CH$_3$OH:CO 1:10 mixtures, where the formation of many of the complex molecules is quenched, some of the weaker features of molecules like HOCH$_2$CHO and HCOOCH$_3$ are visible during warm-up of an ice irradiated at 20~K (Fig. \ref{sp_hco_2}). The final abundance ratio of HOCH$_2$CHO/HCOOCH$_3$ is approximately 1/2 at 100~K. However, HCOOCH$_3$ starts to form at a lower temperature; thus the ratio of the two species changes with temperature. The similar 1700 cm$^{-1}$ feature in this experiment and in the pure CH$_3$OH ice experiments suggests that these two molecules contribute the same fractions to the 1700 cm$^{-1}$ band in all mixtures, depending only on the temperatures at which they are irradiated, or the temperature to which they are
subsequently heated.

\subsubsection{(CH$_2$OH)$_2$}

(CH$_2$OH)$_2$ is the main carrier of the 866 cm$^{-1}$ feature in Fig. \ref{fig:sp_ch3}. This is inferred from the ice warm-up, where the band does not decrease by more than 10--20\% around the desorption temperature of HOCH$_2$CHO -- the only other possible carrier. The remainder of the band disappears around 180~K, the desorption temperature of (CH$_2$OH)$_2$. The QMS simultaneously detects a strong $m/z$=62 signal . The assignment of this band to mainly (CH$_2$OH)$_2$ is thus secure, but the uncertainty in the abundance is higher than for some other species because of a small contribution of a HOCH$_2$CHO feature to the band. In the CO ice mixtures the (CH$_2$OH)$_2$ abundance is instead calculated from the 890 cm$^{-1}$ band, which partially overlaps with a CH$_3$CH$_2$OH ice feature.


\subsection{Abundance determinations of photoproducts\label{sec:res_quant}}

\begin{table*}
\begin{center}
\caption{Potential photoproducts with infrared transitions.}             
\label{products}      
\centering                          
\begin{tabular}{l c c c c c c c c c}        
\hline\hline                 
Species & Band & Band position & Fitting region &$\sigma$ band area & Band strength$^{\rm a}$\\
& & (cm$^{-1}$)& (cm$^{-1}$)& (cm$^{-1}$)& (cm$^{-1}$)\\
\hline                 
H$_2$O & $\nu_2$ & 1670 & 1600--1670 & ul$^{\rm b}$ & 1.2$\times10^{-17}$ (1)\\
CH$_3$ &$\nu_2$ & 1385 & 1380--1390 & ul & 6.9$\times10^{-19}$ (2)\\
CH$_4$ & $\nu_4$ & 1301 & 1290--1310 & 0.002 & 1.2$\times10^{-17}$ (3)\\
CO & $\nu_1$ & 2135 & 2125--2145 & 0.001 & 1.1$\times10^{-17}$ (1)\\
HCO & $\nu_1$ & 1850 & 1840--1860 & 0.0005 & 2.1$\times10^{-17}$ (2)\\
H$_2$CO & $\nu_2$ & 1245 & 1228--1265 & 0.01 & 1.0$\times10^{-18}$ (4)\\
CH$_2$OH & $\nu_4$ & 1195 & 1185--1195 & 0.01 & 1.6$\times10^{-17}$ (2)\\
C$_2$H$_6$ & $\nu_{12}$ & 822 & 809--832 & 0.001 & 1.6$\times10^{-17}$ (5)\\
CO$_2$ & $\nu_1$ & 2340 & 2330--2350 & 0.005 & 1.2$\times10^{-17}$ (1)\\
CH$_3$CHO & $\nu_7$ & 1350 & 1337--1350 & 0.003 & 6.1$\times10^{-18}$ (5)\\
CH$_3$CH$_2$OH & $\nu_{12}$ & 1380 & 1380--1395 & 0.002 & 1.9$\times10^{-18}$ (5)\\
HCOOH & $\nu_3$ & 1740 & 1739--1773 & ul & 6.7$\times10^{-17}$ (6)\\
CH$_3$OCH$_3$& $\nu_{10}$ & 1161 & 1151--1171 & 0.003 & 1.2$\times10^{-17}$ (7)\\
HCOOCH$_3$& $\nu_9$ & 1214 & 1216--1233 & ul & 2.1$\times10^{-17}$ (2)\\
HOCH$_2$CHO& $\nu_{6}$ & 861 & 851--871 & ul & 3.4$\times10^{-18}$ (8)\\
CH$_3$COOH& $\nu_{14}$ & 1746/1770 & 1760--1772 & ul & 1.1$\times10^{-16}$ (9)\\
(CH$_2$OH)$_2$& $\nu_{14}$ & 866/890 & 851--876 & 0.006 & 3.4$\times10^{-18}$ (8)\\
X-CHO/X-COOH$^{\rm c}$& CO stretch & 1680--1780 & -- & 0.01 & 5.4$\times10^{-17}$ (2,6,8)\\
\hline
\end{tabular}
\end{center}
$^{\rm a}$The listed values are transmission band strengths, for our RAIRS experiments they are scaled by a factor of 5.5, determined empirically. \\
$^{\rm b}$Only an upper limit could be determined. \\
$^{\rm c}$An average of the HCOOH, HCOOCH$_3$ and HOCH$_2$CHO band strengths of 6.7, 4.8 and 4.6$\times10^{-17}$ cm$^{-1}$ is used.\\
(1) \citet{Gerakines95}, (2) calculated band strengths from \citet{Bennet07a},   (3) \citet{dHendecourt86},  (4) \citet{Schutte93},  (5) \citet{Moore98},  (6) \citet{Hudson99},  (7) from CH$_3$OH:CH$_3$OCH$_3$ ice mixture spectroscopy,  (8) \citet{Hudson05}, (9) \citet{Marechal87}.
\end{table*}

Table \ref{products} lists the possible photoproducts considered in the previous section that have strong infrared transitions, together with the bands used for quantification, the fitting regions, the band strengths and the estimated uncertainties of the integrated absorbance. The integrated absorbance of each molecular band is determined by first automatically scaling the previously acquired complex molecule spectra to the band over a defined fitting region, using a personal IDL routine. The scaled band in the complex molecule spectrum is then integrated to determine the molecular abundance rather than directly integrating the band in the irradiated CH$_3$OH spectra. The complex ice spectra used of comparison are pure ice spectra. Mixing these complex ices with CH$_3$OH may cause some changes in the band shape, position and strength of the complex features. The change in band shapes for some key species, such as CH$_3$OCH$_3$, between pure ice and ice mixtures with CH$_3$OH have therefore been investigated to ensure that they are not substantial enough to affect band assignments. The changes in band strengths remains unknown but are unlikely to exceed 20\% from previous studies on organic species in pure ices and in hydrogen-bonding ices \citep{Boogert97}. The fitting regions are chosen for each species to ensure that the template spectra are fitted to the part of the band that has no other possible contributors than the species in question. This is especially important in spectrally crowded regions, where the absorption features of different potential photoproducts partially overlap. For example, the CH$_3$CH$_2$OH spectrum is only fitted to the high frequency half of the 1360 cm$^{-1}$ band when determining the CH$_3$CH$_2$OH abundance to avoid contributions to the band intensity from HOCH$_2$CHO. The integrated absorbance uncertainty varies by an order of magnitude between the different species, mainly dependent on how crowded the region is, since a major source of error is the choice of the local baseline.  

\begin{figure}
\resizebox{\hsize}{!}{\includegraphics{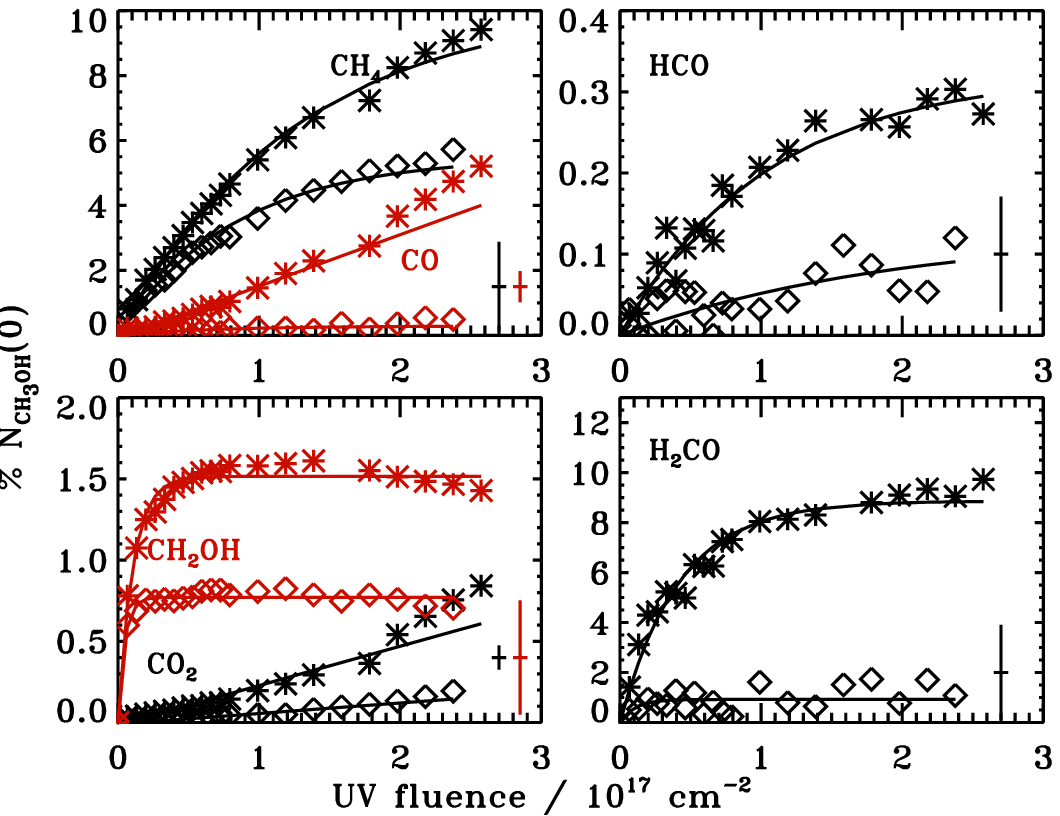}}
\caption{The evolution of small CH$_3$OH photo-products with respect to UV fluence in \% of the initial CH$_3$OH ice abundance (CH$_3$OH(0)) in each experiment during six hours of irradiation at 20~K (stars) and at 70~K (diamonds). The average abundance uncertainties are indicated in the bottom right corner of each panel. The lines are exponential fits to the abundance growths.}
\label{fig:quant_simp}
\end{figure}

\begin{figure}
\resizebox{\hsize}{!}{\includegraphics{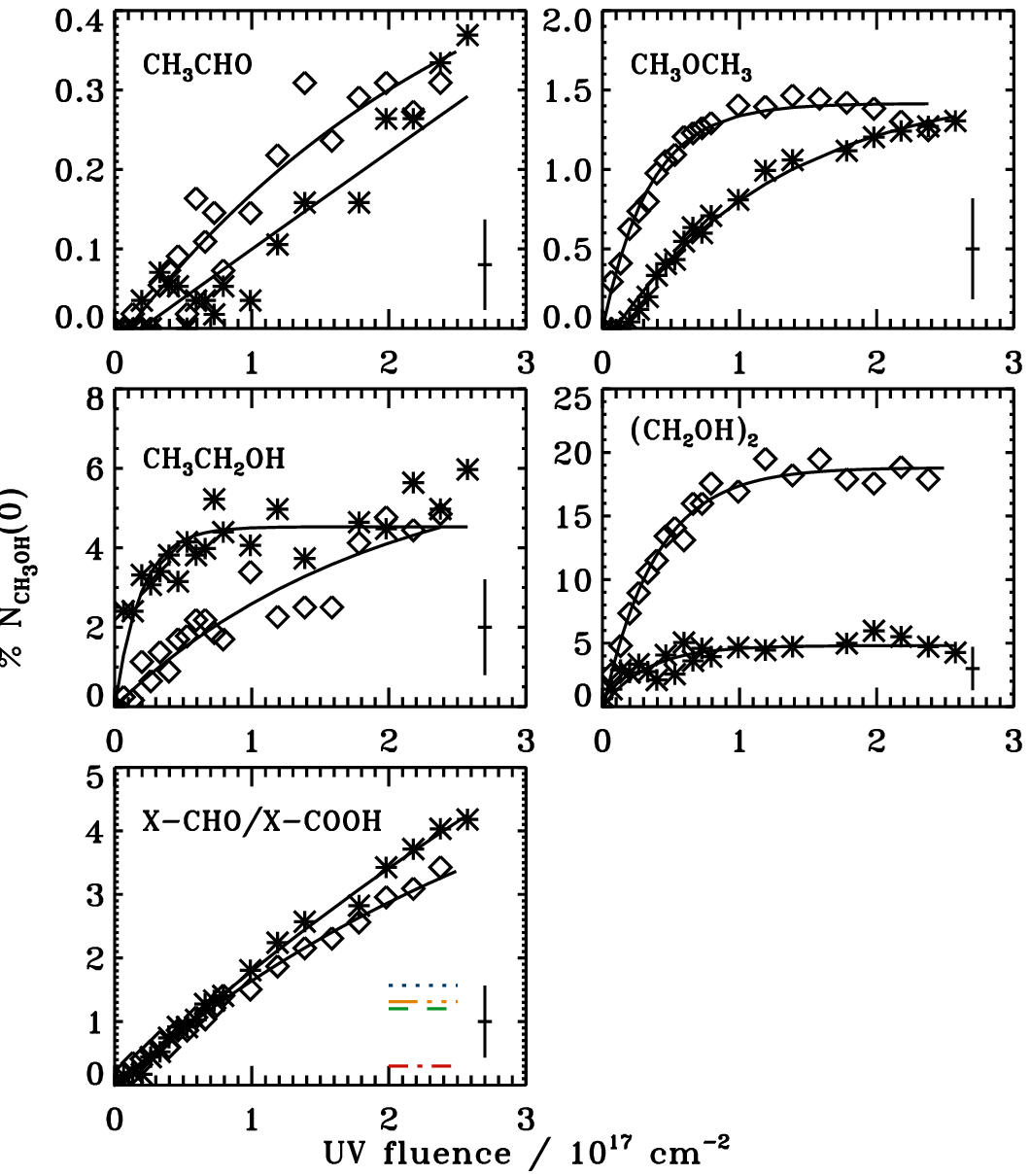}}
\caption{The evolution of complex CH$_3$OH photo-products with respect to UV fluence in \% of the initial CH$_3$OH ice abundance in each experiment during six hours of irradiation at 20~K (stars) and 70~K (diamonds). In the bottom panel the final upper limit abundances in the 20~K ice of HCOOH (blue dotted), HCOOCH$_3$ (green dashed), CH$_3$COOH (red dashed-dotted) and HOCH$_2$CHO (yellow dashed-triple-dotted) are also shown. The average uncertainty in each abundance is indicated in the bottom right corner. The lines are exponential fits to the abundance growths.}
\label{fig:quant_comp}
\end{figure}

The integrated absorbance of each species is converted to an abundance using literature band strengths for pure ices, with a $\sim$20\% uncertainty. Calculated band strengths  are less well defined for solid-state features, resulting in at least a factor of five uncertainty for the affected species (when comparing calculated values to experimental ones). All abundances are reported with respect to the initial CH$_3$OH abundance in each experiment to remove the uncertainty in the band strength conversion factor between transmission and reflection-absorption spectroscopy as well as to cancel the variations in the initial ice thickness between the different experiments. \S \ref{sec:res_depend_thick} showed that the ice chemistry does not depend significantly on ice thickness or irradiation flux and therefore only the results from the $\sim$20~ML thick ices with the low irradiation flux are shown explicitly. 

The sum of all CHO- and COOH-containing complex species, except for H$_2$CO and CH$_3$CHO, are also reported since these species have no uniquely detected features in most experiments. This sum is calculated by subtracting the H$_2$CO and CH$_3$CHO contributions from the 1700 cm$^{-1}$ feature and calculating the remaining integrating absorbance. The integrated absorbance is converted into a molecular abundance by using an averaged band strength, which is within 20\% of the reported band strengths of the three suspected main contributors, HCOOCH$_3$, HOCH$_2$CHO and HCOOH. The fourth potential major contributor CH$_3$COOH has a factor of two larger band strength, but as shown below the CH$_3$COOH upper limits are strict in most experiments and thus it contributes at most 10\% to the 1700 cm$^{-1}$ feature. 

Figures \ref{fig:quant_simp} and \ref{fig:quant_comp} show the photoproduct abundances with respect to the initial CH$_3$OH abundance as a function of UV fluence in 20 and 70~K ices (growth curves for the other experiments are shown in the appendices (\ref{sec:app_irr}1--3). The abundances of all fragments and molecules that can form directly from CH$_3$OH photodissociation or from hydrogenation of a dissociation fragment have a clear inverse dependence on ice temperature, i.e., the abundances of CO, CH$_4$, HCO and H$_2$CO are significantly higher at low temperatures than at high temperatures, where recombination of large radicals dominates the chemistry. For the more complex products the temperature dependence varies. Abundances of molecules that contain an HCO- or a CH$_3$-group do not depend on temperature within the experimental uncertainties, while the (CH$_2$OH)$_2$ abundance is strongly temperature dependent. 

Initial formation cross sections can be derived for the photochemistry products using

\begin{equation}
\label{eq:form_cs}
\frac{{\rm d}n}{{\rm \phi}t}=\sigma,
\end{equation}

\noindent where $n$ is the fractional abundance of the product with respect to CH$_3$OH,  $\phi$ the fluence in cm$^{-2}$ and $\sigma$ the formation cross section. A formation cross section only makes strict physical sense for fragments that form directly from CH$_3$OH photodissociation; however, it is also a relevant number whenever photodissociation is the formation-limiting step. Therefore, formation cross sections for all simple and complex products forming in the pure CH$_3$OH ice experiments between 20~K and 70~K are presented in Table \ref{tab:cs}. The cross sections are calculated from the first $5\times10^{16}$ UV-photons cm$^{-2}$, which is within the linear growth regime for most molecules.

\begin{table}
\begin{center}
\caption{Formation cross sections in 10$^{-19}$ cm$^2$ during pure CH$_3$OH ice photolysis.}             
\label{tab:cs}      
\centering                          
\begin{tabular}{l c c c c }        
\hline\hline                 
Species & 20~K &30~K& 50~K& 70~K\\
\hline                 
         CH$_4$&  7.0[$    1.2$]&  7.1[$    1.3$]&  5.8[$    1.3$]&  4.9[$    1.2$]\\
             CO&  1.0[$    0.5$]&  1.0[$    0.6$]&  0.5[$    0.5$]&  0.6[$    0.5$]\\
            HCO$^{\rm a}$&  0.4[$    0.2$]&  0.3[$    0.2$]&  0.3[$    0.2$]&  $ <0.2$\\
        H$_2$CO& 15.7[$    3.4$]& 15.4[$    3.8$]&  7.4[$    3.6$]&  $<3.3$\\
       CH$_2$OH$^{\rm a}$&  3.7[$    1.0$]&  2.9[$    1.1$]&  2.8[$    1.1$]&  1.9[$    1.0$]\\
         CO$_2$&  0.2[$    0.2$]&  $    <0.2$&  $ <   0.2$]&  $  <  0.2$\\
      CH$_3$CHO&  $    <0.1$&  $  <  0.1$& $  <  0.1$&  0.2[$    0.1$]\\
  CH$_3$OCH$_3$&  0.6[$    0.7$]&  1.1[$    0.8$]&  2.2[$    0.7$]&  2.4[$    0.7$]\\
 CH$_3$CH$_2$OH&  8.6[$    3.4$]&  8.8[$    3.8$]&  2.3[$    3.6$]&  3.9[$    3.3$]\\
 (CH$_2$OH)$_2$&  8.4[$    5.1$]& 10.4[$    5.7$]& 21.8[$    5.4$]& 32.2[$    5.0$]\\
   X-CHO/X-COOH&  1.5[$    1.8$]&  1.7[$    1.9$]&  2.3[$    1.8$]&  2.0[$    1.7$]\\
\hline
\end{tabular}
\end{center}
$^{\rm a}$The absolute uncertainty is a factor of 5 for HCO and CH$_2$OH.
\end{table}

The complete shapes of the ice growth curves belonging to the small products, CH$_4$, CH$_2$OH, HCO and H$_2$CO, are proportional to $A_1(1-e^{-A_2\phi})$, where $\phi$ is the fluence, $A_1$ the steady-state abundance, and $A_2$ combines information about the formation and destruction of the molecules in question in cm$^2$. At 50 and 70~K, this type of equation also fits the growth in abundances of complex species. Several complex production curves are however better fitted with $A_1(1-e^{-A_2(\phi-A_3)})$ at 20--30~K, where $A_3$ is a delay, in fluence units, before the production of the species in question starts. In other words, at low temperatures a certain build-up of radicals is required before the production of complex molecules becomes efficient (see also Fig. \ref{fig:sp_fluence}) and therefore a more general equation $A_1(1-e^{-A_2(\phi-A_3)})$, is used to fit their abundances curves. The measured delay is typically $(1-2)\times10^{16}$ photons cm$^{-2}$: significant for  CH$_3$CHO and CH$_3$OCH$_3$  abundances and for the upper limits of HCOOCH$_3$ and HCOOH, but not for the CH$_2$OH containing molecules. CO and CO$_2$ cannot be fitted with either equation; the CO$_2$ growth rate even increases with fluence. This behavior is explained below when the complete reaction scheme for the ice is discussed in detail. 

The exponential fits for all experiments are tabulated in the appendices (\ref{sec:app_coeff}). It should be noted however, that while this is a convenient way of describing the experimental outcomes, these equations cannot be directly translated to an astrophysical setting without intermediate modeling as will be discussed in \S \ref{sec:astro}.

\subsection{Ice formation and destruction during warm-up following irradiation\label{sec:res_warmup}}

\begin{figure}
\resizebox{\hsize}{!}{\includegraphics{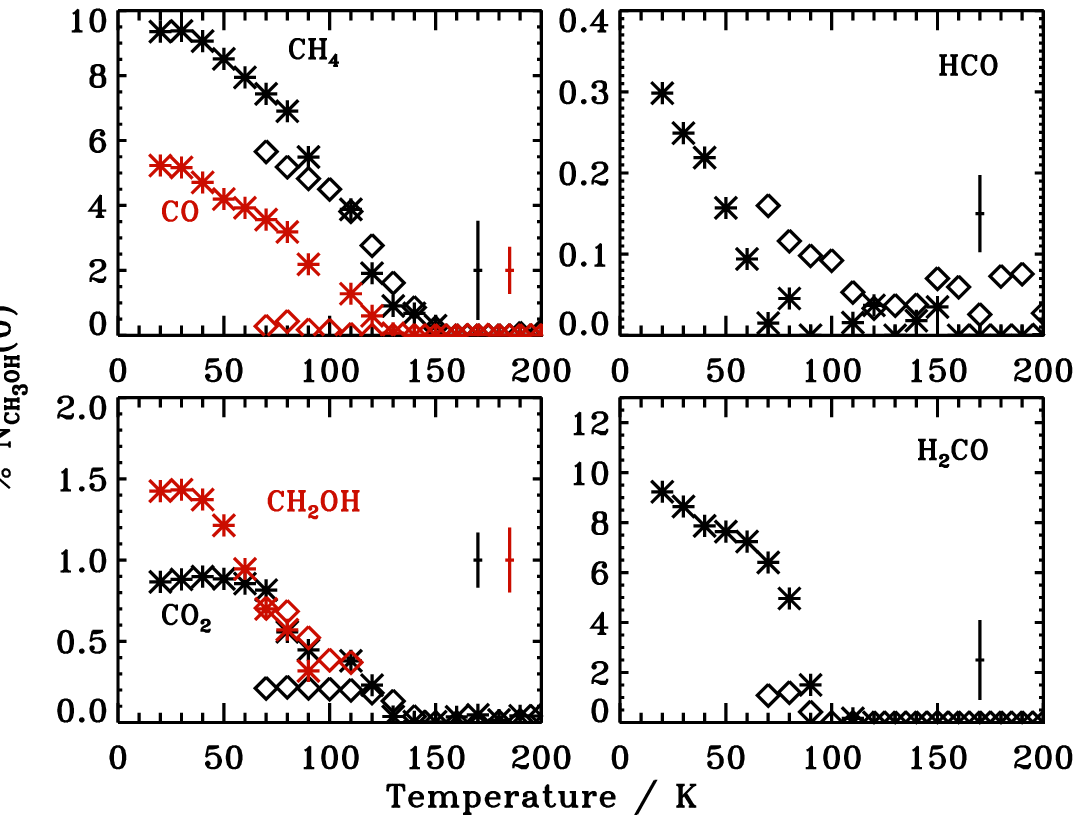}}
\caption{The evolution of small CH$_3$OH photo-products, in  \% of the initial CH$_3$OH ice abundance in each experiment, during 1 K min$^{-1}$ warm-up following UV irradiation at 20~K (stars)  and 70~K (diamonds). The average uncertainty in each abundance is indicated to the right in each panel.}
\label{fig:quant_oh_tpd}
\end{figure}

\begin{figure}
\resizebox{\hsize}{!}{\includegraphics{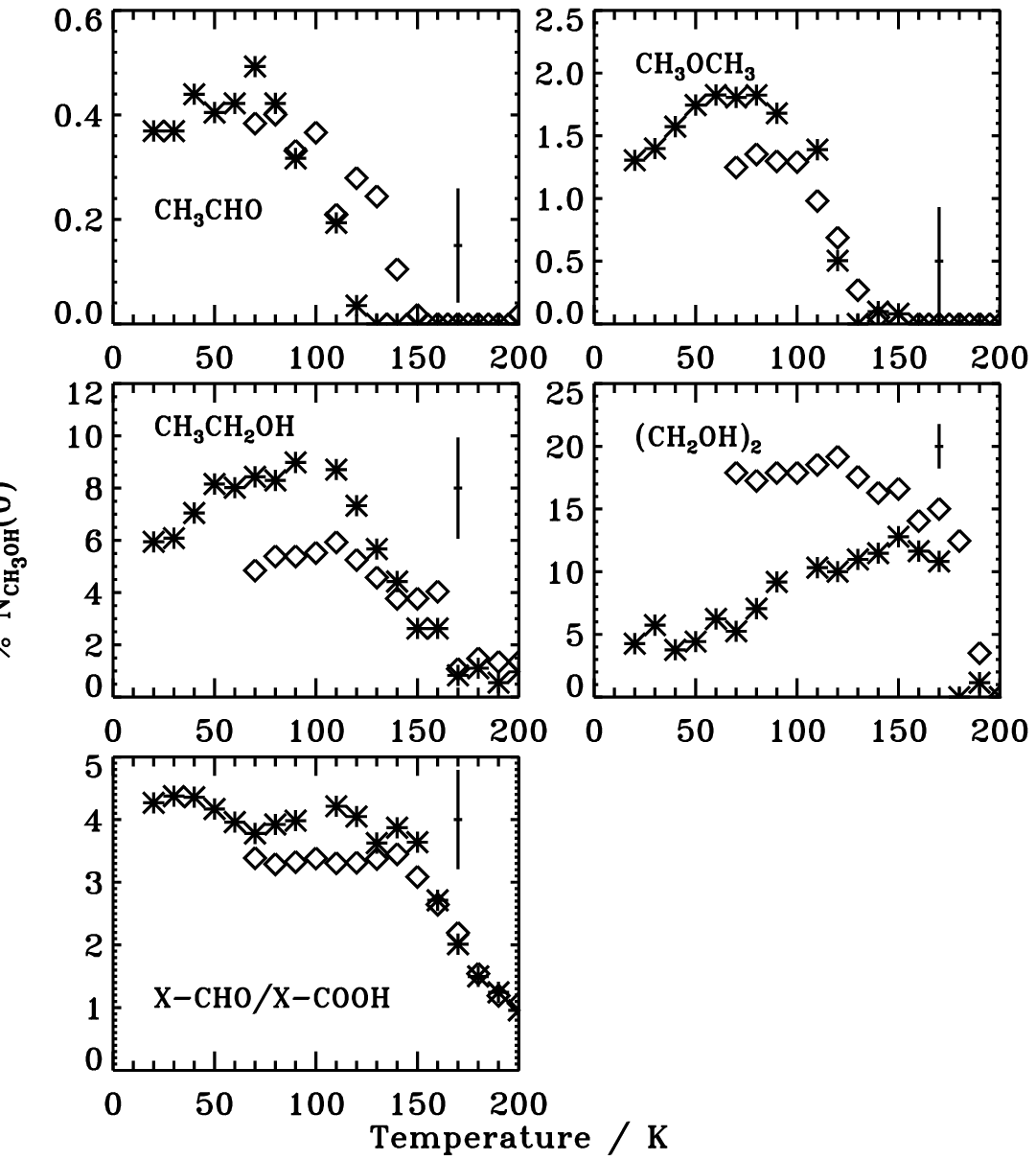}}
\caption{The evolution of complex CH$_3$OH photo-products, in  \% of the initial CH$_3$OH ice abundance in each experiment, during 1 K min$^{-1}$ warm-up following UV irradiation at 20~K (stars)  and 70~K (diamonds). The average uncertainty in each abundance is indicated to the right in each panel.}
\label{fig:quant_comp_tpd}
\end{figure}

During warm-up after the completion of the UV-irradiation experiment, the ice composition changes due to recombination of diffusing radicals and sequential thermal desorption. This results in the depletion of volatile molecules and radicals from the ice as the temperature increases, while more complex molecules increase in abundance before they also start to desorb. The temperature at which a species displays a maximum abundance depends both on the diffusion barriers of the radicals it forms from and the desorption temperature of the complex molecule in question. Figure \ref{fig:quant_oh_tpd} shows that the small molecules and radicals follow the expected behavior. HCO disappears the fastest, though it is not below the noise level until 90~K. The other small molecules all desorb slowly, and only completely disappear at 120~K, the temperature at which CH$_3$OH starts to desorb. This is indicative of significant trapping of volatiles inside the CH$_3$OH ice. 

The more complex molecules also behave as expected following the assumption that they form from recombining radicals; all complex ice abundances initially increase before desorption sets in (Fig. \ref{fig:quant_comp_tpd}). The temperature at which a maximum abundance is reached during warm-up varies with molecular species and also somewhat with the ice temperature during irradiation; for example, it appears that CH$_3$CH$_2$OH formed in the warmer ices experiences somewhat less co-desorption with CH$_3$OH, and thus it desorbs mainly at its pure-ice desorption temperature. This can be understood from recent experiments that show that segregation is a general feature of mixed ices when kept at elevated temperatures; the ice irradiated at 50 and 70~K may thus be partially segregated before the onset of CH$_3$OH desorption, while the ices irradiated at 20--30~K do not have time to segregate before reaching the CH$_3$OH desorption temperature, at which they co-desorb. Co-desorption with CH$_3$OH around 120~K is especially important for CH$_3$CH$_2$OH and CH$_3$OCH$_3$. CH$_3$CHO, CH$_3$OCH$_3$ and CH$_3$CH$_2$OH reach maximum abundances at $\sim$70,  80 and $\sim$110~K, respectively.  (CH$_2$OH)$_2$ only reaches a maximum at 120--150~K (Fig. \ref{fig:quant_comp_tpd}). The formation and desorption pattern of X-CHO and X-COOH is, as expected, complicated with several peaks, corresponding to formation maxima of the different contributors to the band. The initial decrease suggests that the subtraction of H$_2$CO is imperfect and that up to 20\% of the X-CHO/X-COOH abundance at low temperatures is due to H$_2$CO. 

\subsection{Dependence of ice products on physical conditions\label{sec:res_sum}}

\begin{figure}
\resizebox{\hsize}{!}{\includegraphics{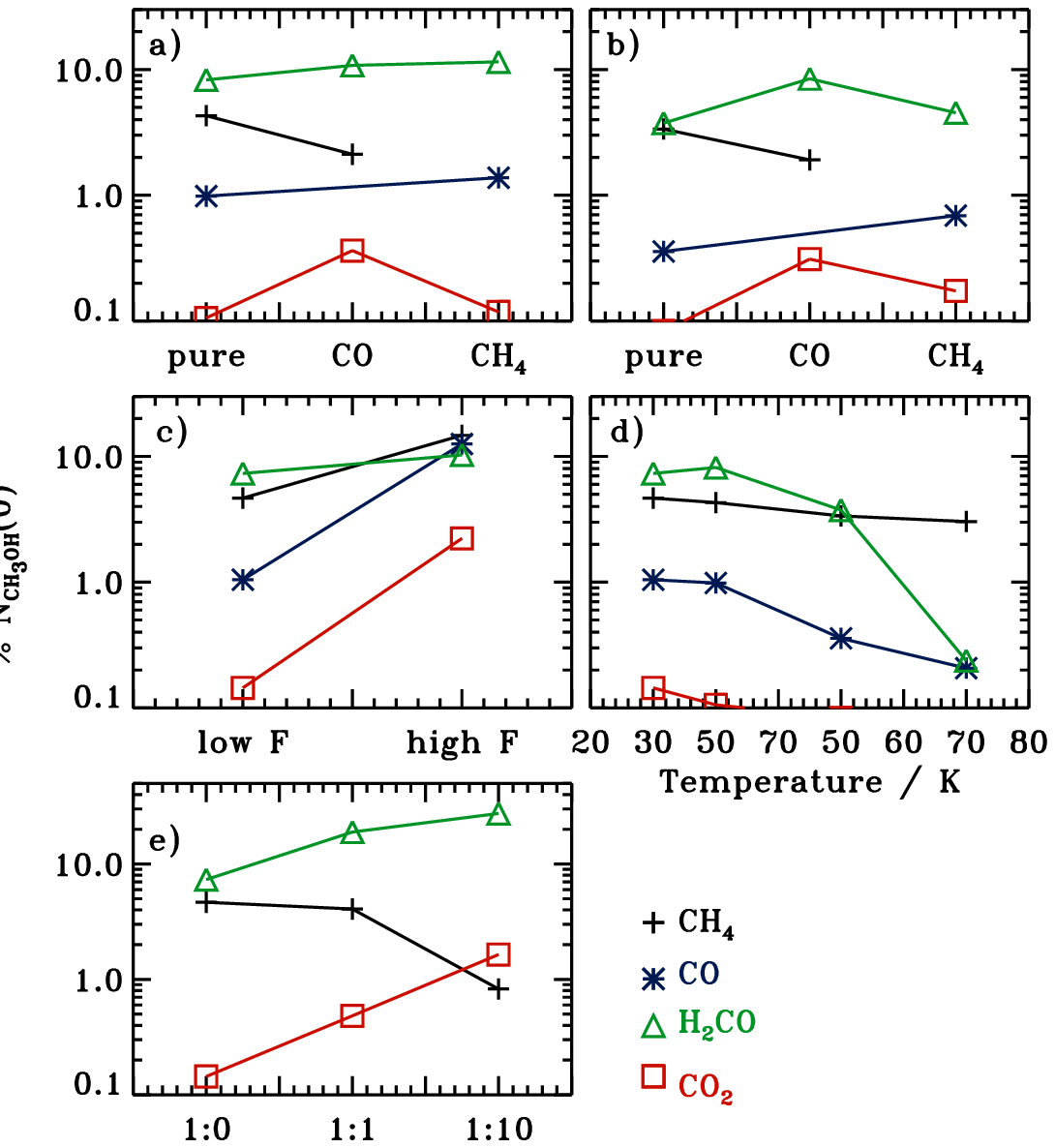}}
\caption{ 
The final simple ice abundances, with respect to initial CH$_3$OH abundance, after the completion of irradiation, plotted as functions of: a) and b) composition (pure CH$_3$OH vs CH$_3$OH:CH$_4$ 1:2 and CH$_3$OH:CO 1:1 ice mixtures) at 30 and 50~K, respectively; c) fluence; d) ice temperature; e) amount of CO mixed into the ice. All ices were irradiated with $2.4\times10^{17}$ photons cm$^{-2}$, except for the the high flux/fluence case in c), in which a final UV fluence four-times greater was attained. In d) and e) the ice temperature is  20~K. The uncertainties are as in Fig. \ref{fig:quant_simp}. 
}
\label{fig:fin_abund_simp}
\end{figure}

\begin{figure}
\resizebox{\hsize}{!}{\includegraphics{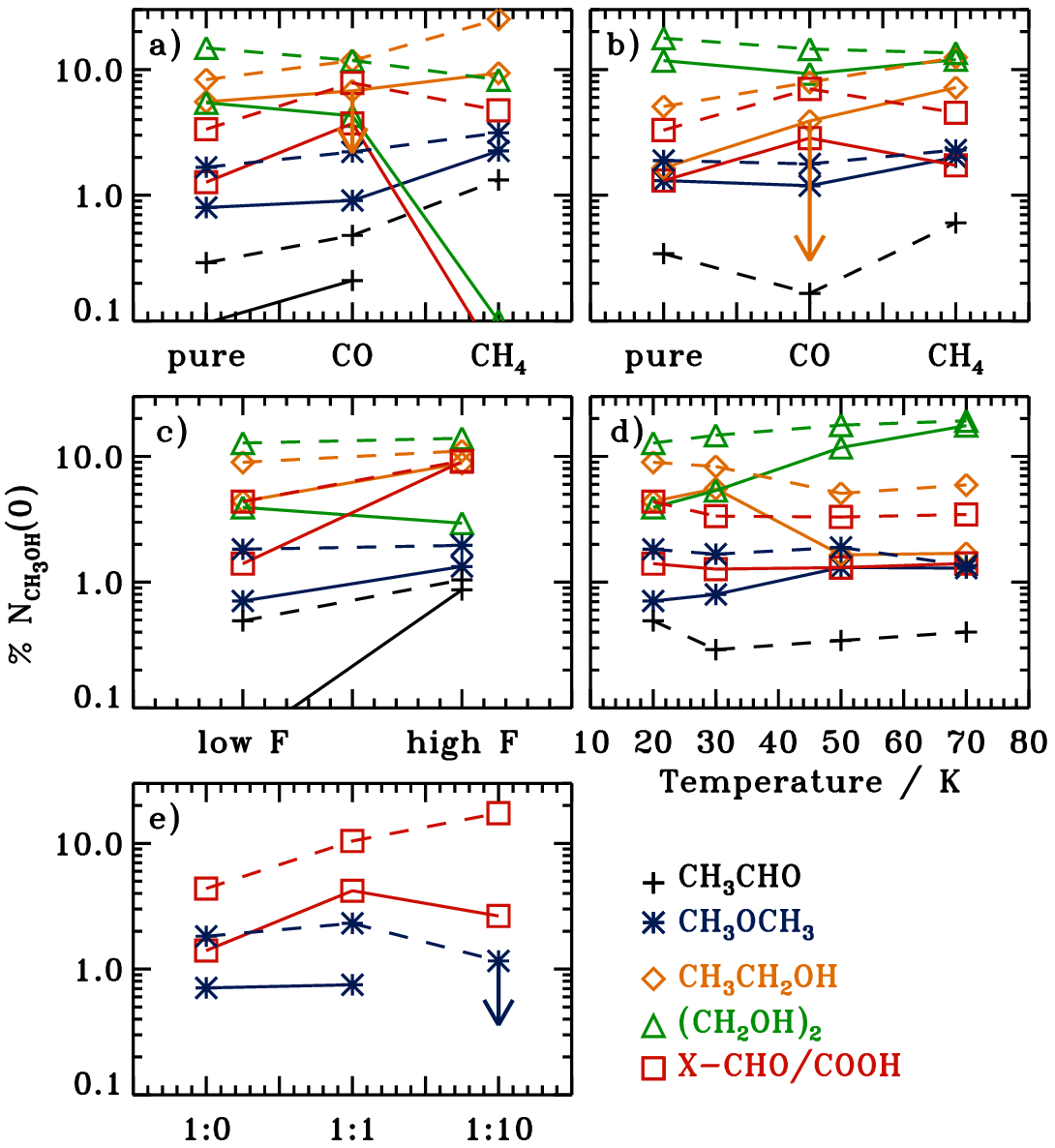}}
\caption{The equivalent of  Fig. \ref{fig:fin_abund_simp} for the complex ice products, with the addition that the product abundances are shown both at the completion of irradiation (solid lines) and the maximum abundance reached during warm-up (dashed lines). The uncertainties are as in Fig. \ref{fig:quant_comp}.}
\label{fig:fin_abund_comp}
\end{figure}

The previous two sections quantified the growth of complex ices in detail as a function of fluence and temperature under specific conditions. In this section, the focus is on the final complex ice abundances in each experiment after completion of irradiation at a low temperature and the maximum abundance reached during warm-up. This is used to determine trends in the final complex abundances as a function of the experimental conditions. Figures \ref{fig:fin_abund_simp} and \ref{fig:fin_abund_comp} show how the final irradiated ice composition changes with ice composition (pure vs. 1(2):1 mixtures with CO and CH$_4$), fluence, ice temperature during irradiation and the amount of CO mixed in at 20~K. The small photoproducts in Fig. \ref{fig:fin_abund_simp} are all affected by temperature, though the CO-containing ones, much more severly than CH$_4$. Both CO$_2$ and H$_2$CO are enhanced in the CO-containing mixtures, while CO$_2$ and CO are the species most affected by increases in fluence.

Among the complex products in Fig. \ref{fig:fin_abund_comp} CHO- and COOH-containing species are enhanced in the CO ice mixtures regardless of temperature, though CH$_3$CHO, is enhanced in the CH$_4$-containing ices as well. The CHO-containing molecules, except for CH$_3$CHO, thus trace CO-rich CH$_3$OH ices. In the CH$_3$OH:CO 1:10 mixture this enhancement of CHO- and COOH-bearing species is more extreme; only CHO- and COOH-bearing complex species are detected and these have a total abundance of  maximum $\sim$20\% during warm-up. CH$_3$OCH$_3$, CH$_3$CH$_2$OH and CH$_3$CHO are enhanced in the CH$_4$ ice mixture at 30~K, but this enhancement almost disappears at 50~K, except for the case of CH$_3$CH$_2$OH. Thus, CH$_3$CH$_2$OH is the most sensitive tracer of the addition of CH$_4$ to the ice. Overall, the complex-product abundances do not change by more than a factor of 2--3 in 1(2):1 mixtures compared to pure CH$_3$OH ice, nor between ices at different temperatures. However, the X-CHO abundance increases by more than a factor of five in the CH$_3$OH:CO 1:10 mixture.

Of all detected complex molecules, (CH$_2$OH)$_2$ depends most steeply on ice temperature, especially in the ice mixtures; its abundance varies by almost an order of magnitude between experiments of different temperatures after UV irradiation of the ice is completed. The dependence is, however, significantly weakened by the time of desorption. Increasing the UV fluence increases the importance of CH$_3$CHO and the other CHO- and COOH-bearing species. In contrast, the CH$_3$-containing species are not affected by fluence, while the (CH$_2$OH)$_2$ abundance decreases somewhat, for a fluence increased from 2.4 to 9.6$\times10^{17}$ cm$^{-2}$.

Different molecular production rates are thus affected differently by changes in ice composition, temperature and UV fluence. These varied responses of molecular production rates to changes in the experimental conditions can be used both to derive chemical properties and to make astrophysical predictions on complex molecule production. This is one of the topics in \S\ref{sec:disc} and \ref{sec:astro} below.

\section{Discussion\label{sec:disc}}

\subsection{Comparison with previous experiments\label{sec:disc_prev}}

The simple ice products found in this study at 20~K are qualitatively comparable to those found previously in UV irradiated pure CH$_3$OH ices at 10--15~K. Quantitatively, the relative amount of formed CH$_4$ and H$_2$CO are the same compared to \citet{Baratta02}, while an order of magnitude more CO and CO$_2$ is reported by the end of their experiment. The CO and CO$_2$ abundances are enhanced by a similar factor in \citet{Gerakines96} compared to the the experiments presented here. The enhancement may be due to  the higher UV fluence used in the previous studies, since the relative CO content increases in our ices with fluence. The high CO abundance may also be partly a result of investigating thick ices ($\sim$0.1--1 $\mu$m) in high vacuum chambers ($\sim$10$^{-7}$ mbar); the ice thickness reduces the escape probability of CO through photodesorption and the high vacuum{ (as opposed to ultra-high vacuum) increases the chance of contaminations, which may change the final ice composition.  The formation cross sections of CH$_4$ and H$_2$CO presented in \S \ref{sec:res_quant} agree with those reported by \citet{Gerakines96} within the experimental uncertainties of 50\%. Thus, the overall agreement is good, between previous studies and the pure CH$_3$OH photochemistry experiment at 20~K presented here.

Complex products are difficult to identify in both UV irradiated and ion-bombarded CH$_3$OH-rich ices, which has resulted in different assignments in the literature to the same bands, or simply no presentation of absolute complex product abundances. As discussed in the band-assignment section, more stringent criteria for band identification allow for both more secure and more numerous band identifications. The most important disagreements between this and previous studies are the assignments of HCOOCH$_3$ and CH$_3$OCH$_3$ bands \citep{Gerakines96,Bennet07a}. We agree with \citet{Bennet07a} on the assignments of bands to (CH$_2$OH)$_2$, but are hesitant with HOCH$_2$CHO assignments in the CH$_3$OH-dominated ices, because of overlap with features from e.g. HCOOH. However, HOCH$_2$CHO may be produced more efficiently during ion bombardment than during UV irradiation, facilitating band identification; we do not find so sharp a band at the position of one of the HOCH$_2$CHO bands as observed by \citet{Bennet07a} in their ion-bombardment study. No other complex molecules were identified in either study. \citet{Hudson00} also tentatively detected (CH$_2$OH)$_2$ following proton bombardment of a CH$_3$OH:H$_2$O mixture. Its formation appears prominent under a range of conditions. 

Similarly to \citet{Bennet07b}, we do observe bands of HCOOCH$_3$ and HOCH$_2$CHO in a CO dominated CH$_3$OH ice mixture, indicating that overall the chemistry induced by UV radiation and ion bombardment is similar. HOCH$_2$CHO was also tentatively detected by \citep{Hudson05} following UV irradiation of a CO:CH$_2$OH 100:1 ice mixture, though more work is required to confirm its formation under those conditions.

\subsection{Dependence of complex chemistry on experimental variables\label{sec:disc_depend}}

Section \ref{sec:res} showed that the final results of CH$_3$OH photochemistry depend on the initial ice composition and the ice temperature during irradiation, but not on UV flux or ice thickness. The independence of ice thickness for the final abundances of complex species between 6 and 20~ML puts an upper limit on the efficiency of surface photochemistry versus bulk photochemistry. Employing a typical relative product abundance uncertainty of 10\%, the upper limit on a relative enhancement in the 6~ML ice compared to the 20~ML one is $\sqrt{2\times10^2}\sim14$\%. This puts an upper limit on excess production in the 6 top ML combined, while the upper limit on excess production in the very top layer is $6\times14\sim84$\%. Surface reactions are therefore at most twice as efficient as bulk reactions in producing complex molecules. This does not directly limit surface diffusion, since photodesorption is efficient enough that many of the radicals and recombined products in the surface layers may escape, thus lowering the efficient surface yield. In fact the initial rates in the 6~ML ices seem somewhat higher, though this difference is barely significant (Appendix \ref{sec:app_coeff}).

The independence of the ice chemistry on UV flux, i.e. the ice composition depends only on the total UV fluence at any time, suggests that the experiments operate in a regime where photodissociation is the rate-limiting step for production of complex organics. If the opposite were true, more complex molecules should form at the lower flux level after the same fluence, when radicals have had more time to diffuse and recombine. This may be counter-intuitive since diffusion is expected to be slow at 20~K. It suggests that  the chemistry, at least at low temperatures, is dominated by fast non-thermalized diffusion following photodissociation. This is consistent with the small differences in formation rates between 20 and 70~K -- in cross-section terms, the formation rate increases by a factor of four or less for all complex organics. In contrast, formation rates dependent on thermal diffusion should change dramatically within this temperature regime since quantified segregation studies show that the thermal diffusion rate of molecules, such as CO$_2$, changes by an order of magnitude between only 50 and 60~K (\"Oberg et al., submitted). The low production rate of complex organics during irradiation at 20~K in the CH$_3$OH:CO 1:10 ice mixture is also consistent with the expected fast thermalization, and thus short diffusion range, of photo-fragments with excess energy \citep{Andersson08}. In this experiment it is instead thermal diffusion during warm-up that dominate the complex organic production. 

Whether thermalized or non-thermalized diffusion drives the chemistry thus depends on a number of factors, including the concentration of radicals in the ice and the temperature during irradiation. To quantify the relative importance of thermalized and non-thermalized diffusion for specific radicals at different temperatures requires quantitative modeling of the complex molecule production during both the irradiation and the warm-up phases. Qualitatively, an increase in thermal diffusion with temperature is required to understand the observed four times higher (CH$_2$OH)$_2$ abundance and formation rate during irradiation at 70~K compared to 20~K. Faster diffusion is also required to explain the immediate onset in formation of most molecules at high temperatures, compared to the short delay at 20~K. Some molecular abundances, such as CH$_3$CH$_2$OH, seem hardly affected by a change in ice temperature. Still it must form at least partly through diffusion of radicals since there is an abundance increase during warm-up. The insensitivity to ice temperature during irradiation may instead be due to increased competition between other reaction pathways once the CH$_2$OH radical becomes mobile, especially to form (CH$_2$OH)$_2$, as is discussed further below. The complexity of these interactions makes it impossible to better quantify the dependence on ice temperature for ice photochemistry until a complete grid of models has been run, which is the topic of Paper II.

CH$_4$ and CO are small molecules with a simple photodissociation chemistry: CH$_4$ mainly loses an H to form CH$_3$, though direct dissociation to CH$_2$ is also possible \citep{Romanzin08}, while CO does not photodissociate measurably with the UV lamp in this experiment \citep{Oberg07b}. Previous experiments show that CO can react with hydrogen to form HCO, and with OH to form CO$_2$ \citep{Watanabe03,Hudson99}. Thus adding CO or CH$_4$ to CH$_3$OH ice should as a first approximation not add any reaction pathways, but only provide excess functional group radicals. Figure \ref{fig:fin_abund_comp} shows that enriching the ice with these molecules also increases the importance of ice temperature on the final ice composition. (CH$_2$OH)$_2$ displays the most dramatic change; in the pure ice the final abundance changes by a factor of two between 30 and 50~K, in the CH$_4$ mixture the final abundance is an order of magnitude different at the two temperatures. This effect is mainly because less CH$_4$ is retained in the 2:1 ice mixture at 50~K compared to 30~K even though the same mixtures were deposited. The same is true for CO, where less than half of the originally deposited CO was trapped in the 50~K ice, resulting in less building material for HCO-containing species. This will be true in astrophysical ices as well and is therefore important to keep in mind, i.e. the building material of complex molecules will change with ice temperature also when the starting point is a typical interstellar ice mixture.

\subsection{CH$_3$OH photochemistry reaction scheme\label{sec:disc_scheme}}

\begin{figure}
\resizebox{\hsize}{!}{\includegraphics{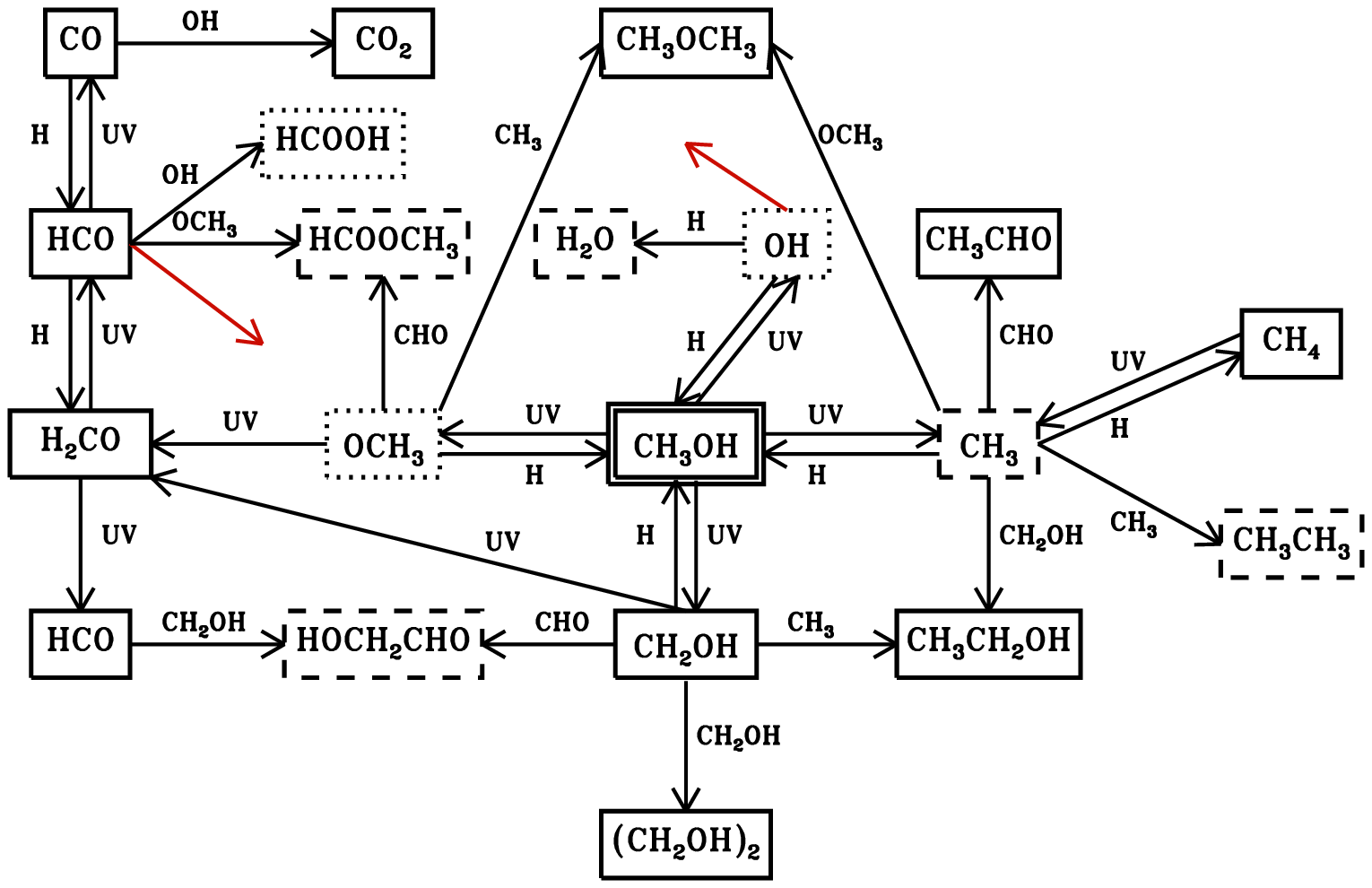}}
\caption{The proposed reaction scheme to form the observed products following UV-irradiation of pure CH$_3$OH ice (solid boxes), of CO or CH$_4$:CH$_3$OH ice mixtures (dashed boxes) and products whose production could only be constrained with upper limits (dotted boxes). Though not shown for clarity, all reactions can be reversed. The two red arrows indicate that the radicals in question take part in more reactions in other parts of the reaction scheme.}
\label{fig:reaction_scheme}
\end{figure}

When considering the possible chemistry induced in the CH$_3$OH ice through UV irradiation it is necessary to chose a level of complexity to investigate, since theoretically the products can continue to dissociate and recombine into ever more complex species. In this qualitative analysis we choose to only consider the dissociation and recombination of first generation radicals from CH$_3$OH dissociation, with one exception -- the dissociation of H$_2$CO into HCO and CO and its reaction to CO$_2$. This path was included because it only results in smaller molecules that are easily detected and thus the kinetics can be modeled. In fact, the photodissociation track from CH$_2$OH/CH$_3$O to H$_2$CO and further to HCO and CO is an important strand of reactions to understand, since these are the only reactions that do not depend on diffusion and thus their production yields provide a clean measure of radical production due to photodissociation of small molecules and radicals. Qualitatively, the relative production rates of CH$_2$OH, H$_2$CO, HCO, CO and CO$_2$ agree well with the prediction of the reaction scheme that the observed normalized formation rates should decrease from one generation of species to the next (Fig. \ref{fig:quant_simp}.)

Within this framework it is assumed that  the vast majority of reactions consists of radical recombination without breaking of bonds i.e. no abstraction. This needs to be further tested through experiments and modeling, but so far there has been no reported evidence for abstraction in ice chemistry experiments.

Taking all this into account, Fig. \ref{fig:reaction_scheme} shows the proposed reaction scheme describing the formation of the observed species from photoproduced radicals. The scheme shares many features with previous ones considered by e.g. \citet{Garrod06}. It starts with four possible photodissociation products of CH$_3$OH that form from breaking one bond i.e. CH$_2$OH, OCH$_3$, CH$_3$, OH and H (the evidence for no direct dissociation to H$_2$CO is discussed in \S\ref{sec:disc_branch}). The formed radicals are then allowed to recombine into stable species or photo-dissociate further.  This scheme should reproduce the chemistry in the ice at low fluences well. As the chemistry proceeds the destruction of complex molecules becomes important to obtain the observed equilibrium conditions and then a more elaborate reaction scheme is required to analyze the outcome. 

Within the framework of Fig. \ref{fig:reaction_scheme}, the final irradiated ice composition depends on both the production rate of each radical and the probability of two specific radicals finding each other and recombining in the ice. The radical production rate depends on the UV flux, the photodissociation branching ratio and photodissociation cross section of CH$_3$OH, and the relative photodissociation rates of CH$_2$OH, H$_2$CO and HCO into smaller species. Once produced, the radicals diffuse through the ice. The scheme shows that all complex molecules are formed in competition with several others that are formed from the same radicals. The recombination branching ratio of any one radical therefore depends both on the amount of other radicals and on their relative diffusion rates. As an example CH$_2$OH can react with H, CH$_3$, HCO, OCH$_3$ and CH$_2$OH as well as further dissociate to H$_2$CO. Whether dissociation or recombination occurs depends on both the UV flux and the diffusion rates of CH$_2$OH and other radicals, and the amount of other radicals in the ice. At high UV fluxes and low temperatures further dissociation to H$_2$CO and recombination with the volatile H to reform CH$_3$OH should dominate. At slightly higher ice temperatures recombination with less volatile radicals will become competitive and finally, once CH$_2$OH can diffuse through the ice, the CH$_2$OH+CH$_2$OH pathway will dominate, since CH$_3$OH ice photodissociation favors CH$_2$OH production as shown in \S \ref{sec:disc_branch}. 

To quantify which reaction pathway dominates at which temperature and flux, requires a large self-consistent model (see Paper II). Already this simplified analysis shows that all observed molecules can be formed straightforwardly with a simple reaction scheme involving recombination of small radicals. This is promising for  quantification of complex ice chemistry, both in the laboratory and in space. 

\subsection{CH$_3$OH photo-dissociation branching ratios\label{sec:disc_branch}}

UV photodissociation of CH$_3$OH ice can theoretically result in several different products. Experiments and calculations have shown that CH$_3$OH has multiple absorption bands in the range of our UV lamp, which are associated with fission of different bonds, producing the different radicals shown in Fig. \ref{fig:reaction_scheme} \citep[e.g.][]{Nee85,Cheng02}. With a broadband UV lamp it is thus difficult to predict which dissociation fragments are formed and their relative importance. 

While it is therefore not possible \textit{a priori} to exclude any photodissociation pathways, the direct dissociation to form H$_2$CO is however eliminated as an important dissociation pathway by the experiments. H$_2$CO production depends dramatically on ice temperature, such that an order of magnitude less H$_2$CO forms at 70~K compared to 20~K. This is consistent with a two-step process where CH$_3$OH is first photodissociated into CH$_3$O or CH$_2$OH followed by further photodissociation into H$_2$CO dependent on the relative time scales of diffusion and UV absorption at a certain temperature. The strong temperature dependence is inconsistent with H$_2$CO forming directly from CH$_3$OH since then a large amount of H$_2$CO should  form at higher temperatures,  similarly to CH$_4$, which only requires a single photodissociation event  followed by diffusion of H to form. The considered photodissociation channels for this discussion are then CH$_2$OH + H, OCH$_3$ + H and CH$_3$ + OH.

Without a complete model, the branching ratios cannot be calculated accurately, but inspection of the experimental results allows for some conclusions. Under the assumption that both the photodissociation cross sections of the formed CH$_3$OH fragments and the rate at which they recombine with hydrogen are approximately the same, the amount of formed complex molecules can be used to assess the importance of different photodissociation pathways. First CH$_3$CH$_2$OH and CH$_3$OCH$_3$ both form through recombination of CH$_3$ with OCH$_3$ and CH$_2$OH, respectively. CH$_3$ is predicted to be most mobile of the three radicals and thus its diffusion should determine the recombination rates of both complex molecules. The relative production rates of CH$_3$CH$_2$OH and CH$_3$OCH$_3$ should then to a first approximation only depend on the OCH$_3$:CH$_2$OH branching ratio during CH$_3$OH photodissociation. The CH$_3$CH$_2$OH/CH$_3$OCH$_3$ abundance ratio is consistently $4\pm2$ at all fluences, temperatures and ice compositions.  The value derived from the CH$_4$ ice mixture experiment at 30~K should describe the photodissociation branching ratio most accurate, since the enhancement of CH$_3$ radicals minimizes the effect of different diffusion properties of OCH$_3$ and CH$_2$OH. There the ratio is $5\pm1$.

The importance of the CH$_3$+OH dissociation pathway is more difficult to asses. An upper limit can be estimated by comparing the detected (CH$_2$OH)$_2$ abundances and the upper limits on C$_2$H$_6$ at 30~K following irradiation of pure CH$_3$OH ice. Since CH$_3$ is more volatile than CH$_2$OH and thus diffuses faster, more C$_2$H$_6$ will form per produced CH$_3$ radical in the ice than (CH$_2$OH)$_2$ per produced CH$_2$OH and thus the estimated upper limit will not be very strict. The observed (CH$_2$OH)$_2$ abundance to C$_2$H$_6$ upper limit is $\sim$40 in this experiment. The abundance of each complex molecule depends on the radical abundance squared, which results in a total branching ratio of CH$_2$OH:OCH$_3$:CH$_3$ of 5$\pm$1:1:$<$1. 

This effective photodissociation branching ratio of CH$_3$OH ice is not necessarily equal to the  branching ratio of gas phase CH$_3$OH molecules. The effective branching ratio may favor the CH$_3$ radical since hydrogenation of CH$_2$OH and OCH$_3$ results in CH$_3$OH, while CH$_3$ hydrogenation produces CH$_4$ that mainly photodissociates back into CH$_3$. This may however be compensated for by a higher recombination rate of CH$_3$ and OH than between the larger fragments and H, since H will diffuse away from its dissociation partner faster. It is interesting to note that the calculated branching ratio is consistent with a purely statistical one, i.e. three different bond breaks result in CH$_2$OH and one each in CH$_3$ and OCH$_3$. The model in Part II will further demonstrate whether this simple treatment of the branching ratio is valid.

\subsection{Diffusion of radicals}

The diffusion barriers of all radicals involved in complex molecule formation can only be properly quantified by modeling the entire chemical network self-consistenly under different irradiation and warm-up conditions. The relative heights of diffusion barriers can however be estimated from the decrease of radical abundances and the increase of molecular abundances during warm-up of irradiated ices. Inspection of the warm-up plots in Fig. \ref{fig:quant_comp_tpd} together with the reaction scheme in Fig. \ref{fig:reaction_scheme} qualitatively shows that the products of the radicals in question, H, OH, HCO, CH$_3$, OCH$_3$ and CH$_3$OH, depend differently on temperature. 

The only radicals that are detected are HCO and CH$_2$OH. HCO clearly disappears faster of the two and thus has a lower diffusion barrier than CH$_2$OH. This is also evident when comparing (CH$_2$OH)$_2$ and X-CHO production during warm-up; the X-CHO abundance does not increase beyond 80~K, while (CH$_2$OH)$_2$ is produced up to at least 110~K. CH$_3$OCH$_3$ and CH$_3$CH$_2$OH behave similarly during warm-up, though CH$_3$CH$_2$OH grows to slightly higher temperatures, suggesting that CH$_3$ diffusion is the most important limiting step, but that OCH$_3$ and CH$_2$OH mobility matters as well. The OCH$_3$ and CH$_2$OH diffusion is even more important in reactions with HCO; during warm-up of the CH$_3$OH:CO 10:1 ice, HCOOCH$_3$ forms at lower temperatures, and thus consumes most of the HCO, compared to HOCH$_2$CHO. The increased importance of the diffusion capabilities of the heavier radicals in reactions with HCO compared to CH$_3$ suggest that the diffusion barrier for HCO is higher than for CH$_3$. This is confirmed by the slightly lower formation temperature of CH$_3$OCH$_3$ and CH$_3$CH$_2$OH compared to X-CHO. The OH barrier is the only one that cannot be assessed since it is only involved in CO$_2$ and HCOOH production. HCOOH is not uniquely identified and CO$_2$ production may instead be limited by CO diffusion. Combining the above results the relative diffusion barriers increases as H$<$CH$_3<$HCO$<$OCH$_3<$CH$_2$OH. This is in qualitative agreement with the assumptions by \citet{Garrod08}.

These relative diffusion barriers, together with the dissociation branching ratios and the observed dependences on ice composition are used below to test a possibility of an ice origin of observed complex molecules in different astrophysical environments.

\section{Astrophysical implications\label{sec:astro}}

\subsection{Potential importance of photochemistry around protostars}

CH$_3$OH ice probably forms in dense cloud cores; it is absent from the cloud edges, but often abundant towards protostars. This is in contrast to for example CO$_2$ and H$_2$O ice, which are present already at a few A$_{\rm V}$ in dark clouds. Because of its formation deep into the cloud, the CH$_3$OH ice is shielded from external UV irradiation during most of its lifetime. A conservative test of whether enough radicals can be produced from irradiated CH$_3$OH ice to account for observations of complex molecules should then only include the locally produced UV field inside the cloud core from cosmic ray interactions with H$_2$. This cosmic-ray induced UV field results in an approximate flux of 10$^{4}$ cm$^{-2}$ s$^{-1}$ \citep{Shen04}. During a million years in the cloud core CH$_3$OH ice is thus exposed to a fluence of $3\times10^{17}$ cm$^{-2}$, which is the same as the final fluence in most of the experiments here, where more than 50\% of the CH$_3$OH is destroyed. Previous experiments at 10~K and the irradiated ice experiments at 20~K here show that of the radicals formed from photodissociation, a large fraction is either further dissociated or hydrogenated to form simpler species than CH$_3$OH. In this study $\sim$25\% of the destroyed CH$_3$OH is converted into simpler molecules at 20~K, but up to 50\% in previous studies \citep{Gerakines96}.  The amount of `frozen in' radicals may be further reduced in astrophysical environments, where hydrogen atoms accrete onto the ice surface and re-hydrogenate radicals. Hydrogenation studies of O$_2$ and CO have revealed that the hydrogen penetration depth at 10--15~K is limited to the top few monolayers \citep[][Fuchs et al. A\&A in press]{Ioppolo08} and re-hydrogenation will therefore mainly affect predictions of the complex chemistry on the ice surface. The radical formation rates in CH$_3$OH ices thicker than a few monolayers and in CH$_3$OH ices covered by CO ice should not be significantly perturbed by this effect.

More than 50\% of the photodissociated CH$_3$OH ice may thus result in `frozen in' radicals or complex molecule formation at the dark cloud stage. Once the radicals become mobile following cloud core collapse and the turn-on of the protostar, the approximate complex molecule to CH$_3$OH ice abundance ratio is 25\%, since each complex molecule forms from two CH$_3$OH fragments. This is assuming CH$_3$OH is the precursor of all complex O-bearing organics. The fraction of complex molecules in the ice will increase if CH$_4$, H$_2$O, CO and CO$_2$ ice  take part in the photochemistry, which will depend on the structure of the ice. The ratio will also increase in protostellar envelopes if the effects of the enhanced UV-radiation field from the star are included for ices in the inner parts.

Laboratory data on ice photochemistry thus predict that a large fraction of CH$_3$OH will be converted into more complex molecules during the pre- and proto-stellar stages. This can be compared with observations. The sum of detected oxygen-rich complex molecule abundances is approximately 50\% with respect to the CH$_3$OH abundances in hot cores \citep{Bisschop07} and closer to 10-20\% in other sources rich in complex molecules. Photochemistry in ices thus provide the right order of magnitude of complex ice species compared to what is observed in regions where ice desorption has occurred. This does not prove that the observed complex molecules have an ice origin, but quantified experimental results do provide a way to test this.

\subsection{Abundance ratios as formation condition diagnostics}

The analysis of the complex ice composition following irradiation in the laboratory can be used to test an ice formation scenario for complex species in star-forming regions. The first piece of information from the warm-up plots is that the ice temperature during irradiation only has a limited effect on most complex ice abundances at the time of desorption. Thus in protostellar envelopes with large CH$_3$OH ice fractions, where the complex ice chemistry is dominated by pure CH$_3$OH chemistry, similar relative fractions of complex ices would be expected in the gas phase over a large range of objects if only thermal desorption is assumed. If on the other hand the entire lifetime of the ice can be sampled by non-thermal desorption, temperature effects on the ice composition may be observed. This temperature effect will be especially clear if the coldest parts are still dominated by a CO-rich CH$_3$OH ice, which is predicted to favor an HCO-rich complex chemistry. However, parts of the complex ice product composition seem robust to a range of physical conditions and can therefore be used to test formation scenarios without making assumptions about the original ice composition or the ice-desorption mechanism.

Regardless of mixture composition and ice temperature, the CH$_3$CH$_2$OH and CH$_3$OCH$_3$ ratio is constant in the experiments and it is thus expected to be constant in astrophysical environments as well, even though the ratio of 5 to 1 may not be reproduced in interstellar regions because of different time scales, desorption temperatures and gas phase destruction rates. The ratio may also be affected by hot gas-phase chemistry around high mass protostars. Indeed, the detected abundances of CH$_3$OCH$_3$ may be strongly affected by gas-phase processes; Garrod et al. (2008) found that gas-phase formation of this molecule, following the evaporation of methanol, is efficient, even with conservative rate estimates. Thus the CH$_3$CH$_2$OH and CH$_3$OCH$_3$ ratio can only be used directly as an ice chemistry test where the gas phase processing is negligible. 

Similarly, the HCOOCH$_3$ and HOCH$_2$CHO ratio changes little with ice composition, though it does depend on ice temperature. Thus, while detailed modeling of absolute ratios of this pair of molecules is saved for Paper II, strong correlations between such pairs of molecules towards different astrophysical objects would support the idea that ice photochemistry followed by desorption is responsible for  gas phase organics. 

In contrast, the CH$_3$CH$_2$OH, HOCH$_2$CHO and (CH$_2$OH)$_2$ relative abundances at the time of ice desorption vary with ice temperature during irradiation and composition even on laboratory time scales. They range between 1:$<$1:1 (CH$_3$OH:CO 1:1, 30~K), 2:$<$1:10 (pure CH$_3$OH 70~K), 10:$<$1:4 (CH$_3$OH:CH$_4$ 1:2, 30~K)  and $<$1:8:$<$1 (CH$_3$OH:CO 1:10 20~K)  (Fig. \ref{fig:fin_abund_comp}). Of these product compositions, the pure CH$_3$OH and the different CO mixtures are perfectly plausible astrophysical compositions since CH$_3$OH is proposed to form from CO ice. The relative abundances of these three complex molecules can thus potentially be used to investigate when and under which conditions complex molecules form in different astrophysical objects, when an ice formation route has been established.  This is pursued qualitatively below though there is a general lack of statistical samples that contain the diagnostically most valuable abundances. 

The experiments did not consider the photochemistry of H$_2$O:CH$_3$OH ice mixtures because the expected formation path of CH$_3$OH from CO  in astrophysical environments. Once the original ice heats up, mixing between the H$_2$O and CO-rich ice phase may however occur at a similar time scale as radical diffusion within the CO:CH$_3$OH ice, resulting in a different complex product composition than expected from radical reactions within the CO:CH$_3$OH phase. In a H$_2$O-rich environment, some of the CH$_3$OH photodissociation fragments should react with OH rather than other CH$_3$OH fragments, forming species such as HOCH$_2$OH. Observations of OH-rich complex species together with a quantification of the mixed H$_2$O:CH$_3$OH and the layered CH$_3$OH/H$_2$O chemistry may therefore provide constraints both on the initial ice composition and the efficiency of ice mixing in the protostellar stage. 

\subsection{Comparison with astrophysical sources}

\begin{table*}
\begin{center}
\caption{Abundances of complex molecules relative to CH$_3$OH.}             
\label{tab:observations}      
\centering                          
\begin{tabular}{l c c c c c c c c c c}        
\hline\hline                 
&IRAS 16293-2422/A$^{\rm a,b}$ & Hot cores$^{\rm c}$ & L1157$^{\rm d}$ & MC G-0.02$^{\rm e}$ &Hale-Bopp$^{\rm f}$ & CH$_3$OH$^{\rm g}$& CH$_3$OH:CO$^{\rm g}$\\
\hline                
CH$_3$OH & 1/1& 1 & 1& 1&1& 1 &1\\
CH$_3$CHO &0.038/$<$0.0016 &2.9[3.1]$\times10^{-5}$ &--&0.033 &0.010 &0.01 &$<$0.04\\
CH$_3$CH$_2$OH&--/0.031 &0.019[0.012]&0.007 & 0.040&$<$0.042 &0.1&$<$0.01\\
CH$_3$OCH$_3$&0.20/0.013&0.41[0.51]&--& 0.050&--&0.04 &$<$0.01\\
HCOOCH$_3$& 0.30/0.0084&0.089[0.084]&0.019 &0.037 &0.033 &$<$0.03&$>$0.08\\
HOCH$_2$CHO& --/--& --& --&0.01 &$<$0.017&$<$0.04 & $>$0.04\\
(CH$_2$OH)$_2$&--/-- &-- &--&0.01 &0.10 &0.4&$<$0.01\\
\hline
\end{tabular}
\end{center}
$^{\rm a}$The first value is from single dish data, the second from interferometric studies of the A core. $^{\rm b}$\citet{Cazaux03,Huang05,Bisschop08} and van Dishoeck \& Herbst (2009).
$^{\rm c}$ Average data and standard deviations towards a sample of seven high-mass hot cores \citep{Bisschop08}. $^{\rm d}$\citep{Arce08}. $^{\rm e}$\citep{Requena-Torres06,Requena-Torres08}. $^{\rm f}$\citep{Crovisier04}. $^{\rm g}$The relative abundance with respect to CH$_3$OH during warm-up of an irradiated 20~K pure CH$_3$OH ice and a CH$_3$OH:CO 1:10 ice mixture.\\
\end{table*}

Complex molecules have been detected in the gas phase towards a variety of astrophysical environments. Table \ref{tab:observations} lists the detected abundances towards a low-mass protostar IRAS 16293-2422, a sample of high-mass protostars, a low-mass outflow L1157, a galactic-center cloud MC G-0.02 and the comet Hale-Bopp. Typical uncertainties are factors of a few, but can be larger along some lines of sight.

Overall, complex-molecule abundances vary by less than an order of magnitude, with respect to CH$_3$OH,  within these dramatically different astrophysical environments. For example the CH$_3$CH$_2$OH/CH$_3$OH ratio varies between 0.02 and 0.04 (Table \ref{tab:observations}). This suggests a formation scenario where reaction barriers are not rate determining, in agreement with an ice photochemistry scenario. 

Focusing on specific ratios, the HOCH$_2$CHO to HCOOCH$_3$ ratio is only available in one of these environments and therefore not possible to use as a test. The CH$_3$CH$_2$OH to CH$_3$OCH$_3$ ratio is unity within the observational uncertainties towards the IRAS 16293-2422 core and the galactic center sources, but it is an order of magnitude lower in the high-mass hot core sample. This may be due to a more extended temperature gradient in the hot core objects, which would increase the importance of CH$_3$OCH$_3$ release into the gas phase at lower temperatures and thus over a larger volume. It may also be the result of gas phase chemistry modifying the released ice abundances. The physics of these objects thus needs to be addressed and more molecules observed, to confirm an ice origin of these complex molecules. 

Another interesting ratio is the CH$_3$CHO to CH$_3$CH$_2$OH one, which varies by two orders of magnitude between the different objects (Table \ref{tab:observations}). In addition, an interferometric study has revealed spatial separations between these two species around the protostar IRAS 16293-2422 \citep{Bisschop08}. Part of this may be due to different destruction efficiencies of CH$_3$CHO towards different objects. It is however difficult to explain such a lack of correlation with a more traditional ice formation scenario of complex molecules, through successive hydrogenation and oxygenation of small carbon chains. In such a reaction network CH$_3$CHO and CH$_3$CH$_2$OH are formed under the same conditions. In contrast, photochemistry of ices does not predict a  correlation between CH$_3$CH$_2$OH and CH$_3$CHO since CH$_3$CHO  formation may not even require CH$_3$OH as a starting point \citep{Moore03}, while CH$_3$CH$_2$OH does.

The abundances of CHO-containing species are in general expected to vary between different astrophysical objects in the ice-photochemistry scenario because of their large enhancements in CO-rich and thus colder ices. Comparing the different types of sources in Table \ref{tab:observations}, only the single dish observations towards IRAS 16293 contain more HCOOCH$_3$ than CH$_3$OCH$_3$ or CH$_3$CH$_2$OH. This suggests that the complex molecules in the colder areas of IRAS 16293 are formed in a more CO-rich ice matrix than towards the warmer sources, which is consistent with the high volatility of CO. The uncertainties are yet too high for any conclusive comparison, however.

Quantitatively, the pure CH$_3$OH experimental results are in closest agreement with the ice desorption found in Hale-Bopp. This may not be too surprising, since these observations are less affected by gas-phase reactions that can both form and destroy complex molecules. It may also be a sign of a warmer formation path compared to most larger-scale astrophysical objects, and thus an insignificant CO ice content. The almost one-to-one correlation is tentative evidence that cometary ices have reached their current composition through ice photochemistry, though both more cometary observations and actual modeling of their chemical evolution are necessary to evaluate whether this holds in general. The formation and destruction mechanisms around massive protostars are clearly too complicated to evaluate the origins of complex molecules there directly, without a more complete gas-grain model. In addition, time-scale effects may be significant for all astrophysical abundances comapred to those found in the laboratory. It is, however, reassuring that the interferometric study towards the low mass protostar IRAS 16293, where high-temperature gas-phase reactions should be of less importance, agrees reasonably well with our experiments. Observations of HOCH$_2$CHO and (CH$_2$OH)$_2$ towards IRAS 16293 and other low-mass protostars would probably provide the strongest constraints available on the prevalence of UV-ice photochemistry in star-forming regions.

\section{Conclusions}

The major experimental and analytical results of this study are summarized below:

\begin{enumerate}

\item The CH$_3$OH ice photodissociation cross section increases from 2.6[0.9] to 3.9[1.3]$\times10^{-18}$ cm$^2$ between 20 and 70~K suggesting that a significant amount of the dissociated fragments recombines immediately to form CH$_3$OH at 20~K when radical diffusion in ices is slower.
\item CH$_3$OH ice photodesorbs with a yield of 2.1[1.0]$\times10^{-3}$ per incident UV photon at 20~K. The yield is independent of temperature and of the same order as the yields found previously for CO-, CO$_2$- and H$_2$O-ice photodesorption.
\item UV photolysis of 6--21~ML pure CH$_3$OH ice at 20--70~K results in a product mixture of simple and complex molecules, whose abundances have been quantified. The identified species are CO, CO$_2$, CH$_4$, HCO, H$_2$CO, CH$_2$OH, CH$_3$CHO, CH$_3$OCH$_3$, CH$_3$CH$_2$OH, (CH$_2$OH)$_2$ and a mixture of complex CHO- and COOH-containing molecules. The small temperature dependence may be explained by fast diffusion and recombination of non-thermalized radicals following CH$_3$OH photo-dissociation. 
\item The final product composition following CH$_3$OH photolysis depends on UV fluence and temperature, but not on the UV flux level or the ice thickness when both are varied by a factor 3--4.
\item In CH$_3$OH:CO and CH$_3$OH:CH$_4$ 1:1(2) mixtures the complex molecules containing HCO- and CH$_3$ groups are moderately enhanced following UV irradiation compared to pure CH$_3$OH ice photolysis. In an irradiated CO:CH$_3$OH 10:1 ice mixture, the HCO-containing products dominate and both HCOOCH$_3$ and HOCH$_2$CHO are detected as the originally 20~K ice is warmed up. With the exception of this ice the final complex product mixture is robust within a factor of few in all different experiments.
\item Additional formation of complex molecules occurs following irradiation at 20--70~K, when all ices are slowly heated with the UV lamp turned off; some abundances increase by up to a factor of ten between 20 and 100~K. Diffusion of thermalized radicals through the ice is thus important for complex molecule formation.
\item From abundance ratios of related products, formed during UV-irradiation of CH$_3$OH-rich ices, we infer an approximate CH$_3$OH photodissociation branching ratio into CH$_2$OH+H:OCH$_3$+H:CH$_3$+OH of 5:1:$<$1
\item From the peak formation temperature of related molecules during warm-up, we find that the relative radical diffusion barriers increase as H$<$CH$_3<$HCO$<$OCH$_3<$CH$_2$OH. While the mobility of the radical with the lowest barrier is most important in determining the formation temperature of the product, the diffusion barrier of the heavier radical matters as well. For example HCOOCH$_3$ forms at a lower temperature (from HCO and OCH$_3$) than HOCH$_2$CHO (from HCO and CH$_2$OH).
\item The predicted sum of oxygen-rich complex molecules compared to CH$_3$OH in ices is $>$25\% after a UV fluence corresponding to 6 million years in a cloud core, followed by moderate ice heating to 30--50~K during the protostellar stage, though quantitative modeling is required to test the competition between radical diffusion, desorption, dissociation and recombination on astrophysical time scales.
\item Some complex molecular ratios, especially  CH$_3$CH$_2$OH to CH$_3$OCH$_3$, do not depend significantly on experimental conditions and may thus to be constant in space as well, if gas-phase effects do not dominate. If quantitative modeling confirm this prediction, these ratios can be used to test the UV-induced ice formation scenario of complex molecules in astrophysical regions. Other ratios, such as (CH$_2$OH)$_2$/CH$_3$CH$_2$OH and HCOOCH$_3$/CH$_3$CH$_2$OH, depend both on the irradiated ice composition and the ice temperature and can therefore be used to investigate the formation conditions of observed complex molecules.
\item Comparison with astrophysical objects shows that the composition of complex ices in the Comet Hale-Bopp is readily explained by UV photolysis of of pure CH$_3$OH ice, while CH$_3$OH in a CO-dominated ice can explain the variations in the HCOOCH$_3$ and CH$_3$CHO abundances between cold and warmer regions of protostars.

\end{enumerate}

\noindent This study shows that complex ice chemistry can be quantified, though the process of doing so is more arduous than qualitative work. The discussed dependencies of molecule formation on different ice conditions show the impossibility of perfectly simulating the astrophysical ice evolution in the laboratory, especially since physical conditions vary in space as well. It is, however, possible to study ice processes under specific laboratory conditions, which through careful modeling (Paper II) can provide the energy barriers that govern these reactions both in the laboratory and in star forming regions. These can then be used to model ice chemistry in all possible astrophysical environments. 

\begin{acknowledgements}

The authors wish to thank Herma Cuppen for stimulating discussions and Lou Allamandola, Ted Bergin and Eric Herbst for valuable comments on the manuscript. Funding is provided by NOVA, the Netherlands Research School for Astronomy, a grant from the European Early Stage Training Network ('EARA' MEST-CT-2004-504604) and a Netherlands Organisation for Scientific Research (NWO) Spinoza grant.
\end{acknowledgements}

\bibliographystyle{aa}


\Online

\begin{appendix} 

\section{Photoproduct growth curves during UV-irradiation\label{sec:app_irr}}

\subsection{Pure CH$_3$OH ice at different temperatures}

Figures \ref{fig:app_quant_simp} and \ref{fig:app_quant_comp} show the increasing abundances of photoproducts during irradiation of pure CH$_3$OH ices at 30 and 50~K (experiments 2 and 3). The abundances are fitted as a function of UV fluence as described fully in \S\ref{sec:res_quant}  for 20~K and 70~K ices. The abundances follow the temperature trends suggested by the 20~K and 70~K ices (experiments 1 and 4), except possibly for CH$_3$CH$_2$OH, which seems to be enhanced at 30~K compared to the other ices. This enhancement is barely significant however. The fit coefficients and uncertainties for all irradiated $\sim$20~L, pure CH$_3$OH ices are reported in Table \ref{tab:fits_1}

\begin{figure}[ftp]
\resizebox{\hsize}{!}{\includegraphics{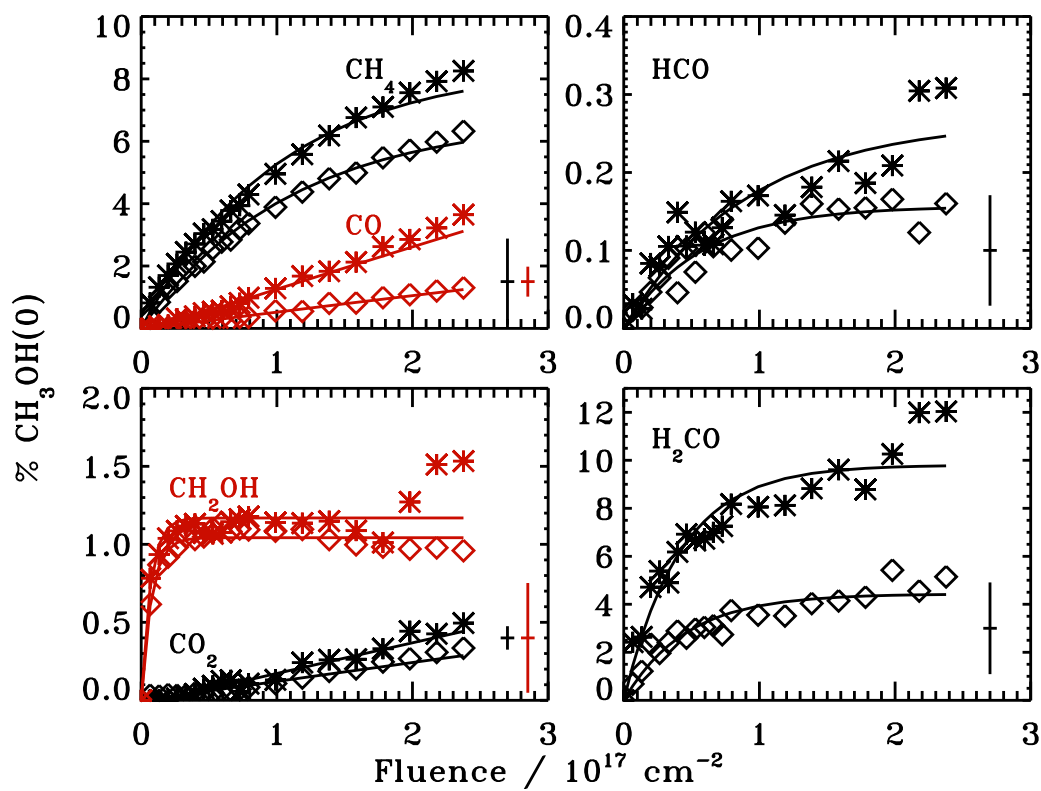}}
\caption{The evolution of small CH$_3$OH photo-products with respect to UV fluence in \% of the initial CH$_3$OH ice abundance, CH$_3$OH(0), in each experiment at 30~K (stars) and 50~K (diamonds). The relative uncertainty for each abundance is indicated in the bottom right corner. The lines are exponential fits to the abundance growths.}
\label{fig:app_quant_simp}
\end{figure}

\begin{figure}[ftp]
\resizebox{\hsize}{!}{\includegraphics{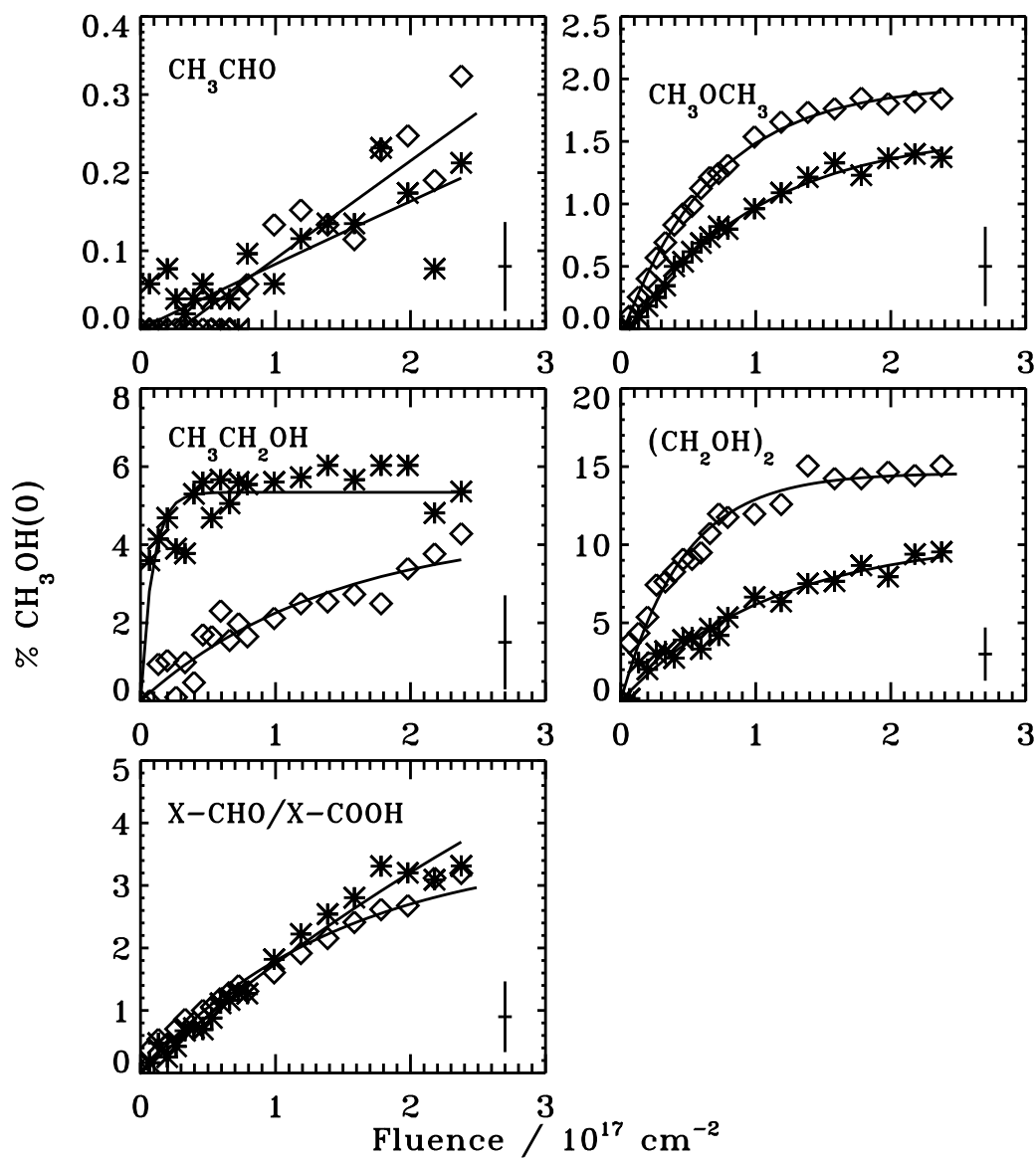}}
\caption{The evolution of complex CH$_3$OH photo-products with respect to UV fluence in \% of the initial CH$_3$OH ice abundance. Otherwise as Fig. \ref{fig:app_quant_simp}.}
\label{fig:app_quant_comp}
\end{figure}

\newpage

\subsection{High fluence experiments}

Figures \ref{fig:app_quant_simp_hf} and \ref{fig:app_quant_comp_hf} show the increasing abundances of photoproducts during irradiation at 20~K and 50~K with a high flux (experiments 5 and 6), together with the fitted growth curves. The first $2.5\times10^{17}$ photons cm$^{-2}$ are consistent with the low flux experiments within the experimental uncertainties. For molecules such as  CH$_3$CHO that forms slowly, these experiments provide better constraints on the production rates than the low flux experiments. In contrast, molecules and fragments with high production rates are better constrained by the lower flux experiments, since they have a abundance determinations at a higher fluence resolution. The production rates of molecules that are destroyed/photodesorbed faster than they are produced at high fluences cannot be fit in these experiments. The fit coefficients of these experiments are reported in Table \ref{tab:fits_1}.

\begin{figure}[ftp]
\resizebox{\hsize}{!}{\includegraphics{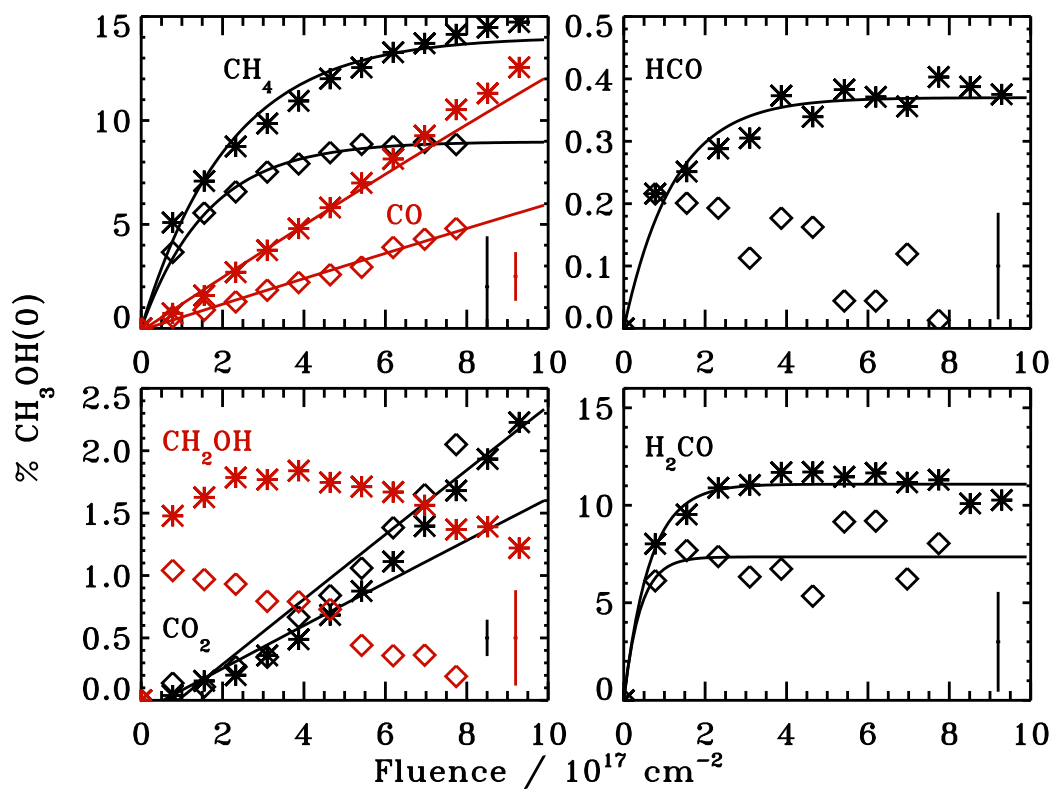}}
\caption{The evolution of small CH$_3$OH photo-products with respect to UV fluence in \% of the initial CH$_3$OH ice abundance (CH$_3$OH(0)) in the two high flux/fluence experiments at 20~K (stars) and 50~K (diamonds). The relative uncertainty for each abundance is indicated in the bottom right corner. The lines are exponential fits to the abundance growths.}
\label{fig:app_quant_simp_hf}
\end{figure}

\begin{figure}[ftp]
\resizebox{\hsize}{!}{\includegraphics{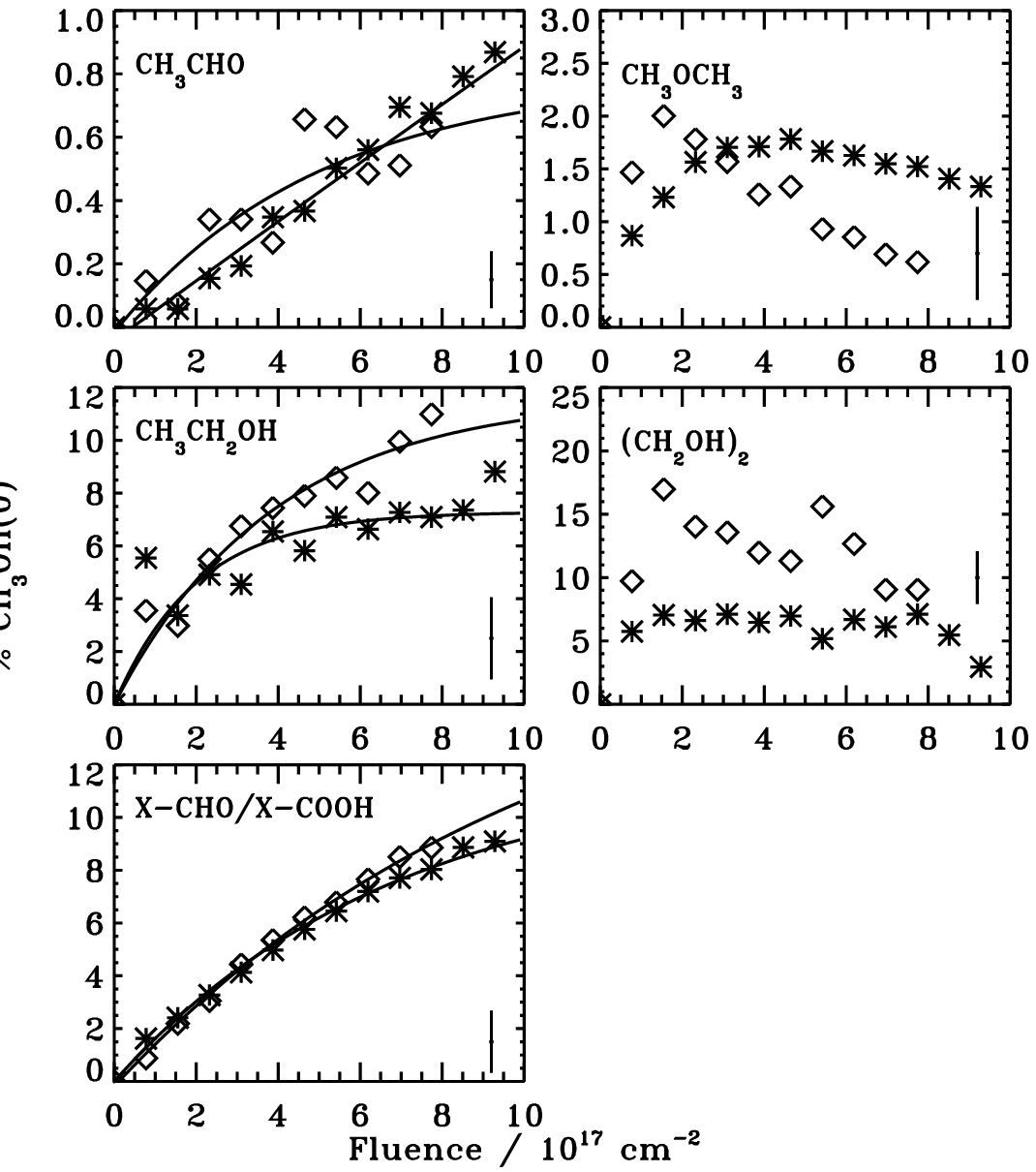}}
\caption{The evolution of complex CH$_3$OH photo-products with respect to UV fluence in \% of the initial CH$_3$OH ice abundance. Otherwise as Fig. \ref{fig:app_quant_simp_hf}.}
\label{fig:app_quant_comp_hf}
\end{figure}

\newpage

\subsection{CH$_4$ and CO mixtures}

\begin{figure}
\resizebox{\hsize}{!}{\includegraphics{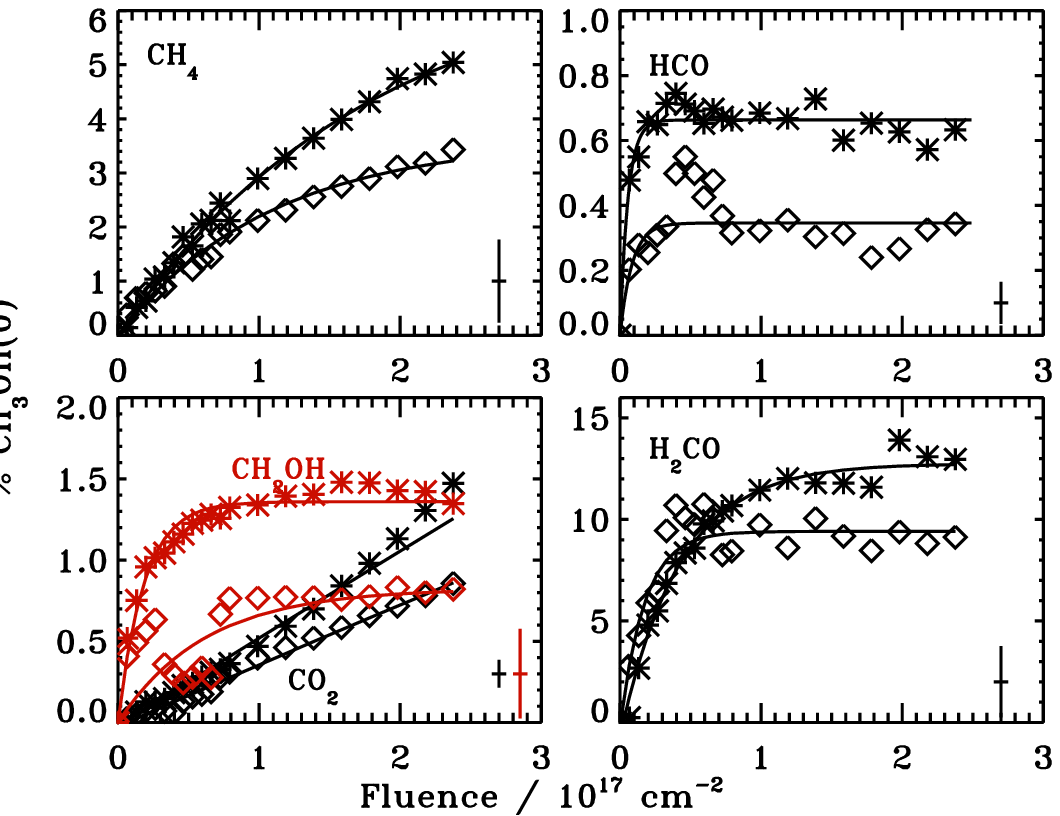}}
\caption{The evolution of small photo-products with respect to UV fluence in \% of the initial CH$_3$OH ice abundance in CH$_3$OH:CO 1:1 ice mixture experiments at 30~K (stars) and 50~K  (diamonds).}
\label{fig:app_quant_simp_co}
\end{figure}

\begin{figure}
\resizebox{\hsize}{!}{\includegraphics{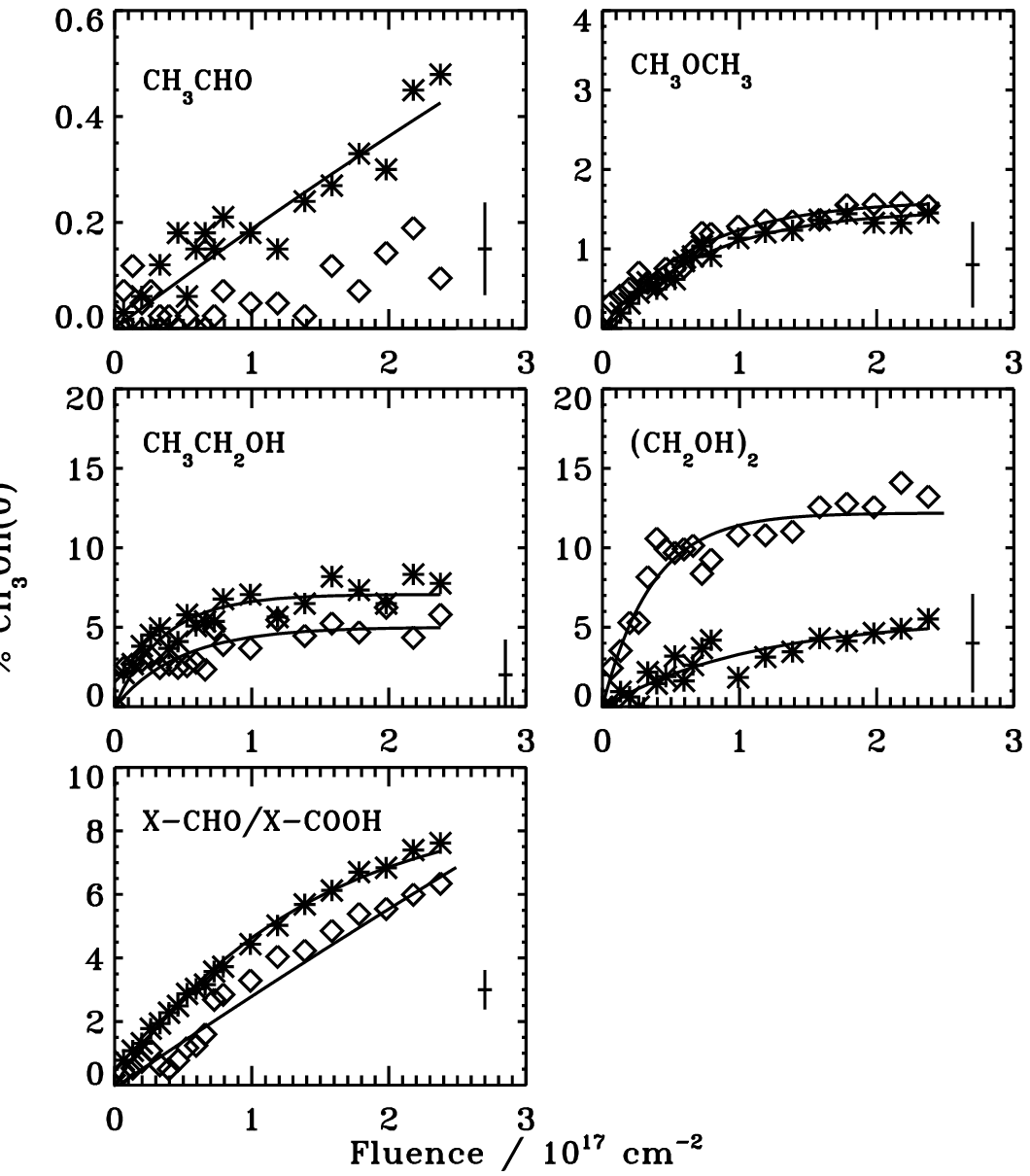}}
\caption{The evolution of complex photo-products with respect to UV fluence in \% of the initial CH$_3$OH ice abundance in CH$_3$OH:CO 1:1 ice mixture experiments at 30~K (stars) and 50~K  (diamonds). }
\label{fig:app_quant_comp_co}
\end{figure}

\begin{figure}
\resizebox{\hsize}{!}{\includegraphics{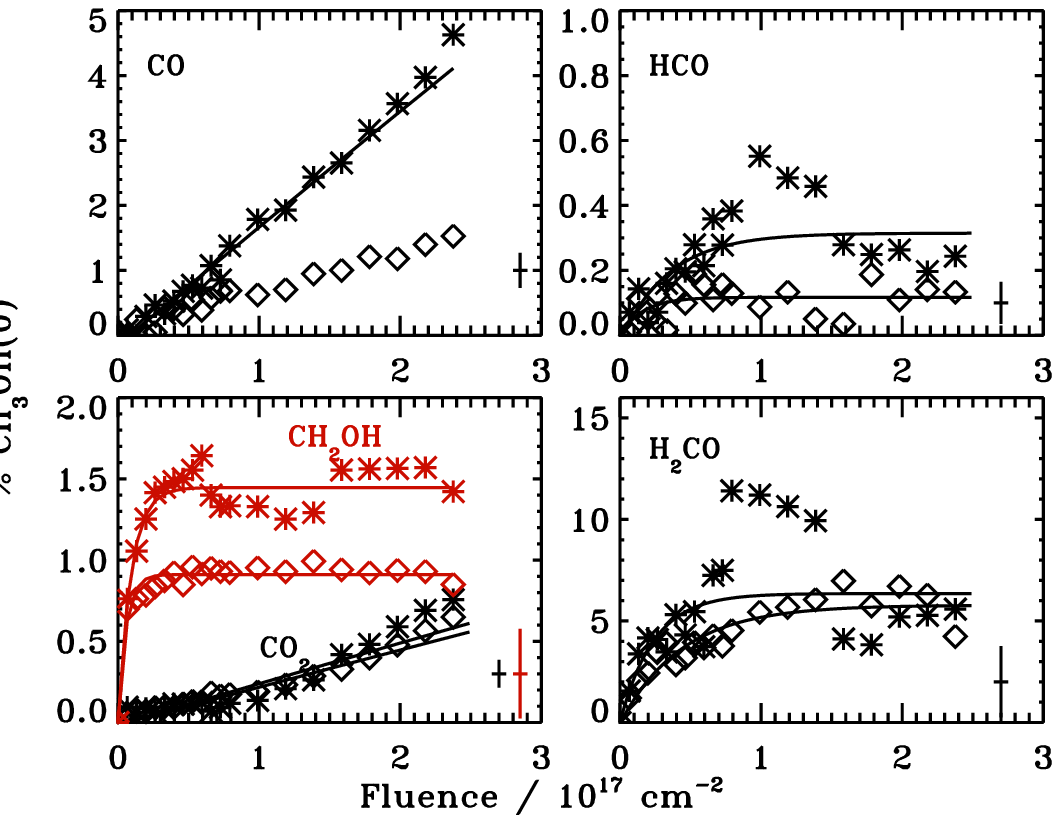}}
\caption{The evolution of small photo-products with respect to UV fluence in \% of the initial CH$_3$OH ice abundance in CH$_3$OH:CH$_4$ 1:2 ice mixture experiments at 30~K (stars) and 50~K  (diamonds).}
\label{fig:app_quant_simp_ch4}
\end{figure}

\begin{figure}
\resizebox{\hsize}{!}{\includegraphics{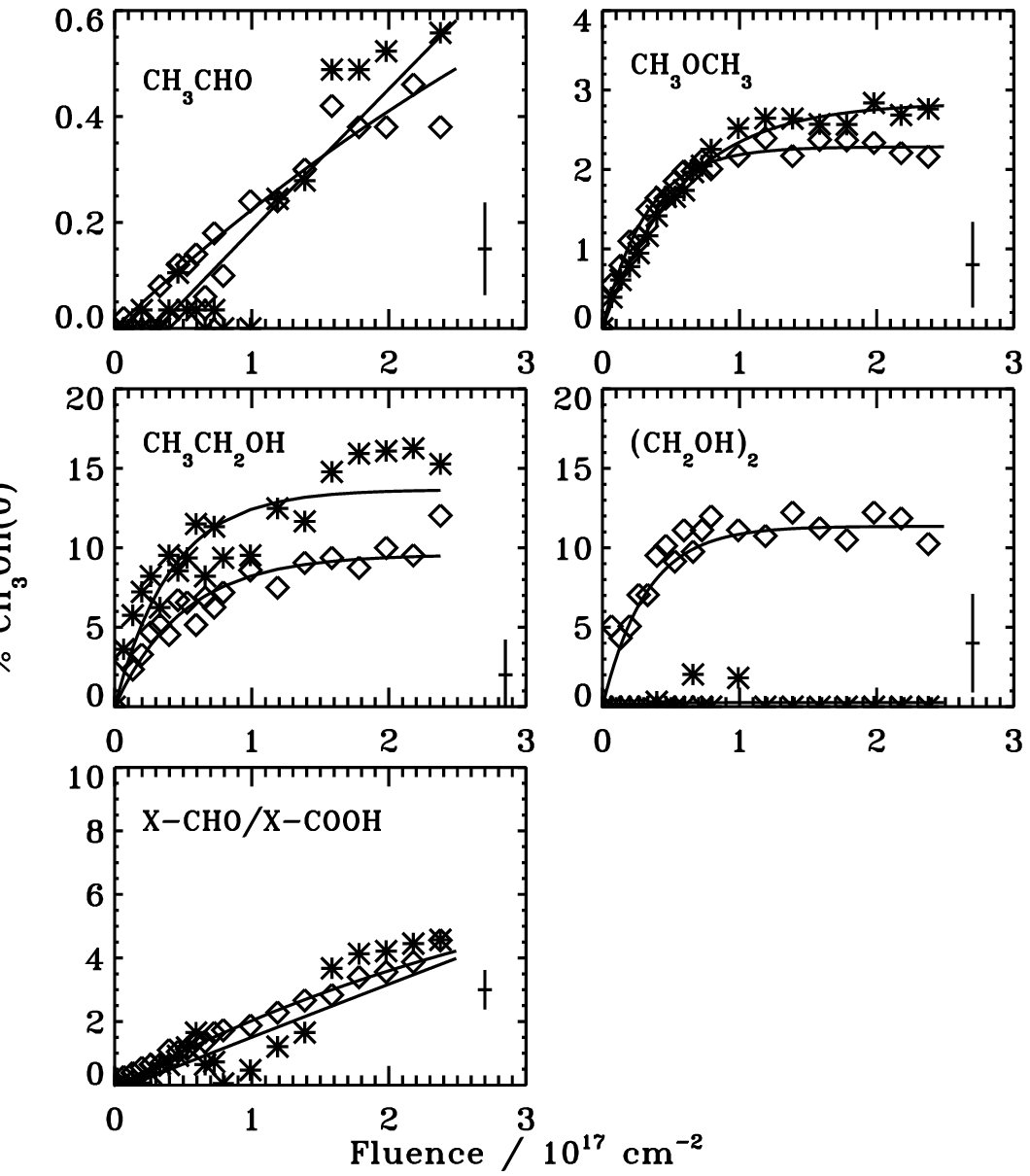}}
\caption{The evolution of small photo-products with respect to UV fluence in \% of the initial CH$_3$OH ice abundance in CH$_3$OH:CH$_4$ 1:2 ice mixture experiments at 30~K (stars) and 50~K  (diamonds). }
\label{fig:app_quant_comp_ch4}
\end{figure}

Figures \ref{fig:app_quant_simp_co}--\ref{fig:app_quant_comp_ch4} show the increasing abundances of photoproducts during irradiation of CH$_3$OH:CH$_4$ 1:2 and CH$_3$OH:CO 1:1 mixtures at 30~K and 50~K together with fitted growth curves. The temperature trends are similar to what is seen for pure CH$_3$OH ice.  As suggested from the spectra in \S\ref{sec:res_depend_comp}, all HCO containing species are increased in abundance in the CO mixtures, with the exception of CH$_3$CHO, which is mainly enhanced in the CH$_4$ ice mixture. The other two CH$_3$ containing species, CH$_3$CH$_2$OH and CH$_3$OCH$_3$, are also enhanced in the CH$_4$ mixture compared both to pure CH$_3$OH and the CO ice mixture -- in the CO mixtures the abundance points represent upper limits. In contrast the (CH$_2$OH)$_2$ production is suppressed in both ice mixtures at 30~K. The fit coefficients of these experiments and CO:CH$_3$OH 1:1 mixture irradiated at 20~K are reported in Table \ref{tab:fits_2}.

\section{Formation and destruction curves during warm-up\label{sec:app_warm}}

\subsection{Pure CH$_3$OH at different irradiation temperatures\label{sec:app_warm_temp}}

Figures \ref{fig:app_quant_simp_tpd} and \ref{fig:app_quant_comp_tpd} show the evolution of photoproduct abundances during warm-up following irradiation of pure CH$_3$OH ices at 30 and 50~K (experiments 2 and 3). The abundances follow the warm-up trends suggested by the 20 and 70~K ices (experiments 1 and 4). The CH$_3$CH$_2$OH desorption starts at a lower temperature than in pure CH$_3$CH$_2$OH, suggesting that similarly to the 20~K experiment, a substantial part of the CH$_3$CH$_2$OH desorbs together with CH$_3$OH.

\begin{figure}
\resizebox{\hsize}{!}{\includegraphics{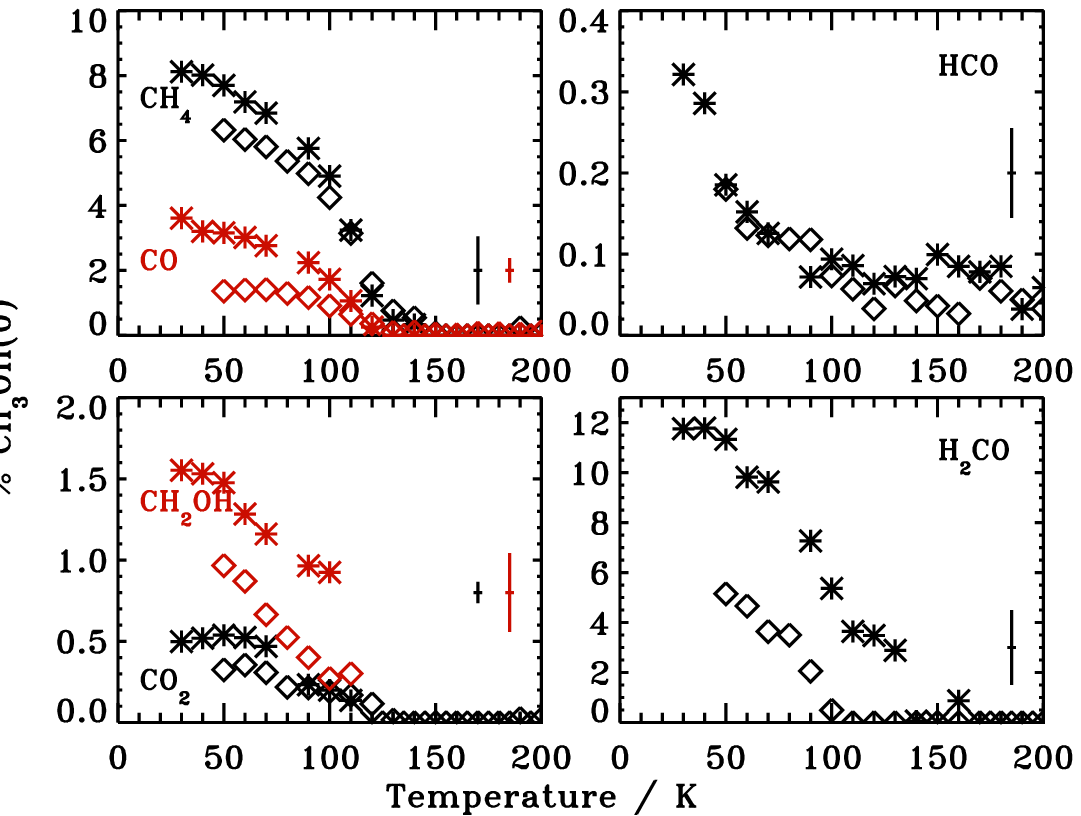}}
\caption{The evolution of small CH$_3$OH photo-products with respect to UV fluence in \% of the initial CH$_3$OH ice abundance in pure CH$_3$OH irradiation experiments at 30~K (stars) and 50~K (diamonds). The average uncertainties are indicated by the error bar to the right in each panel.}
\label{fig:app_quant_simp_tpd}
\end{figure}

\begin{figure}
\resizebox{\hsize}{!}{\includegraphics{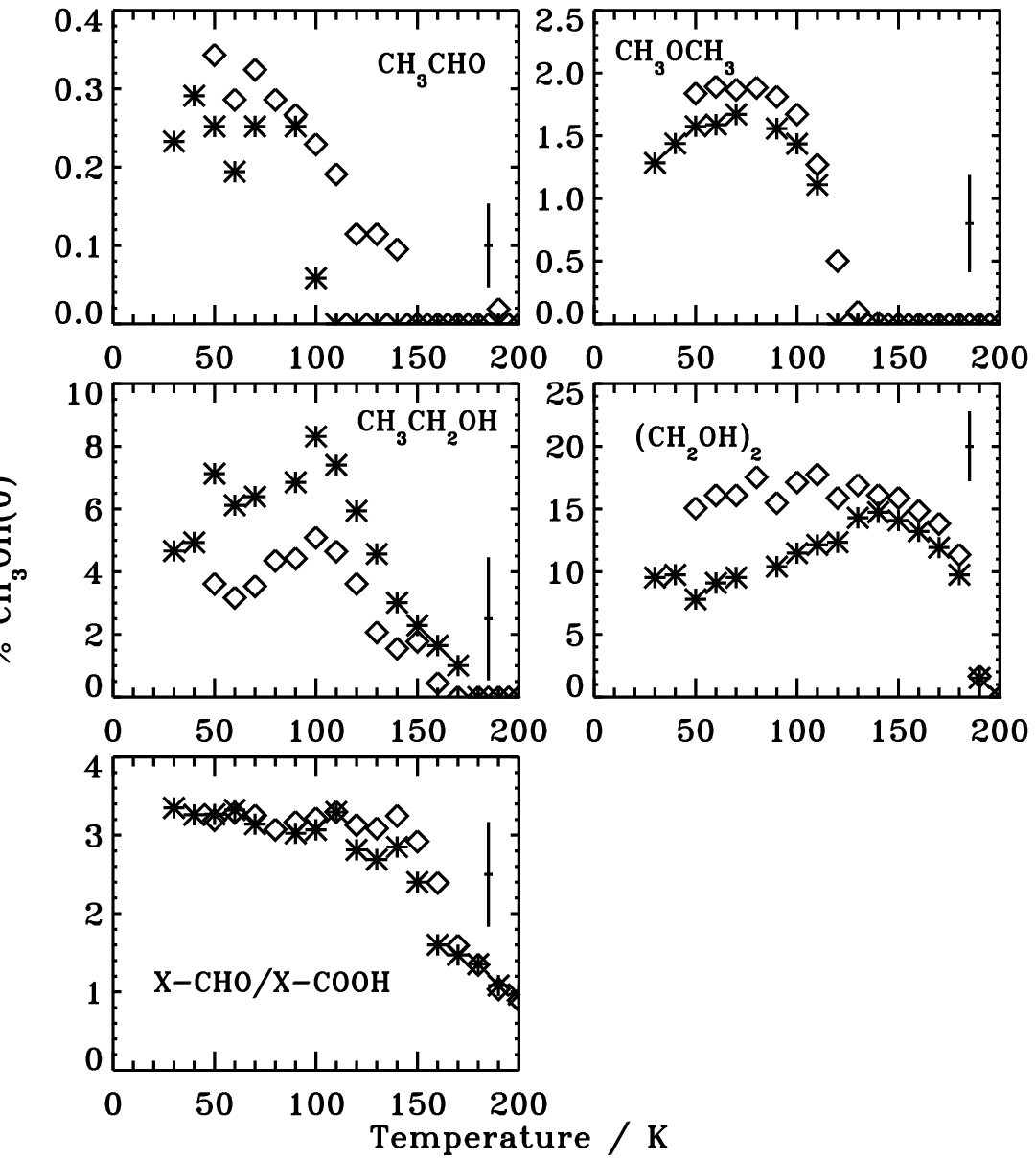}}
\caption{The evolution of complex CH$_3$OH photo-products with respect to UV fluence in \% of the initial CH$_3$OH ice abundance in pure CH$_3$OH irradiation experiments at 30~K (stars) and 50~K (diamonds). The average uncertainties are indicated by the error bar to the right in each panel.}
\label{fig:app_quant_comp_tpd}
\end{figure}

\subsection{CH$_4$ and CO mixtures}

The warm-up trends are similar in the experiments where CH$_4$ and CO are mixed with the CH$_3$OH ice, at a 1(2):1 ratio, compared to the pure CH$_3$OH experiments (Figs. \ref{fig:app_quant_simp_tpd_co}--\ref{fig:app_quant_comp_tpd_ch4}). The CH$_3$-containing molecules CH$_3$CHO and CH$_3$CH$_2$OH show a remarkable growth between 30 and 50~K, which is only hinted at in the pure CH$_3$OH experiments, suggesting a significant build-up of CH$_3$ radicals in these ice mixture experiments. In contrast the (CH$_2$OH)$_2$ formation rate is low during warm-up of the 30~K experiments. 

\begin{figure}
\resizebox{\hsize}{!}{\includegraphics{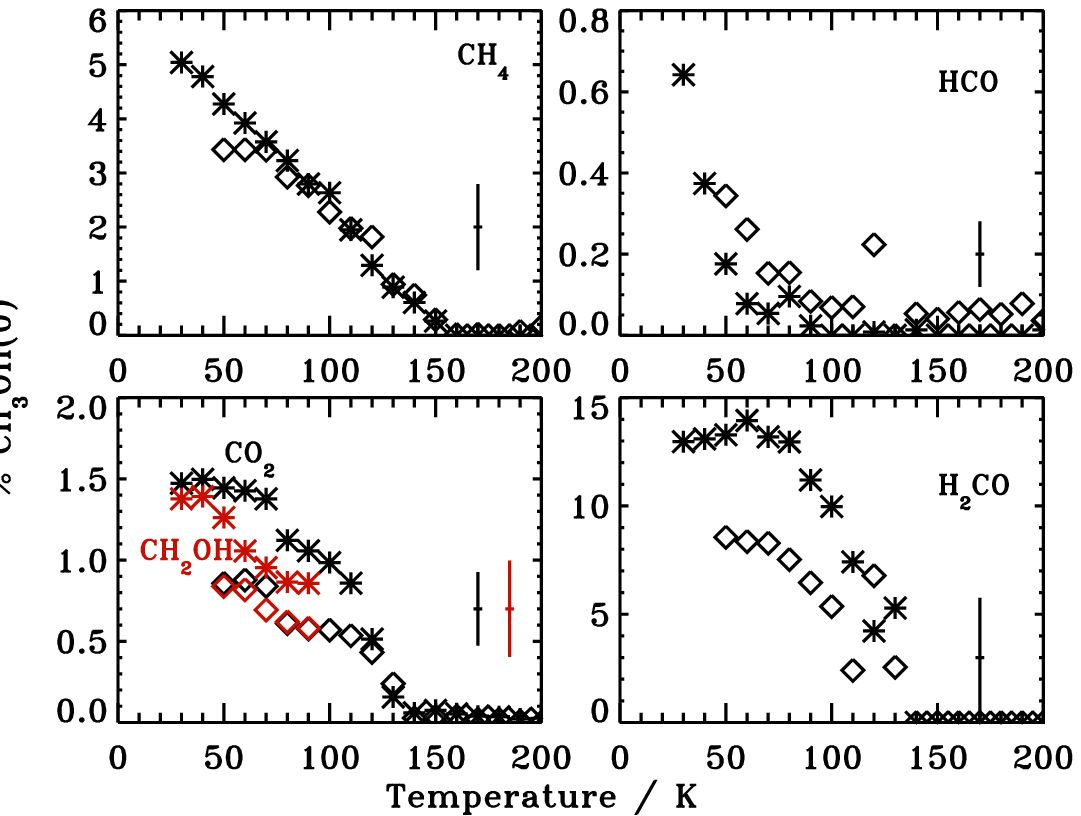}}
\caption{The evolution of small photo-products,  in \% of the initial CH$_3$OH ice abundance, with respect to temperature following irradiation of CH$_3$OH:CO 1:1 ice mixtures at 30 (stars) and 50~K (diamonds). The average uncertainties are indicated by the error bar to the right in each panel.}
\label{fig:app_quant_simp_tpd_co}
\end{figure}

\begin{figure}
\resizebox{\hsize}{!}{\includegraphics{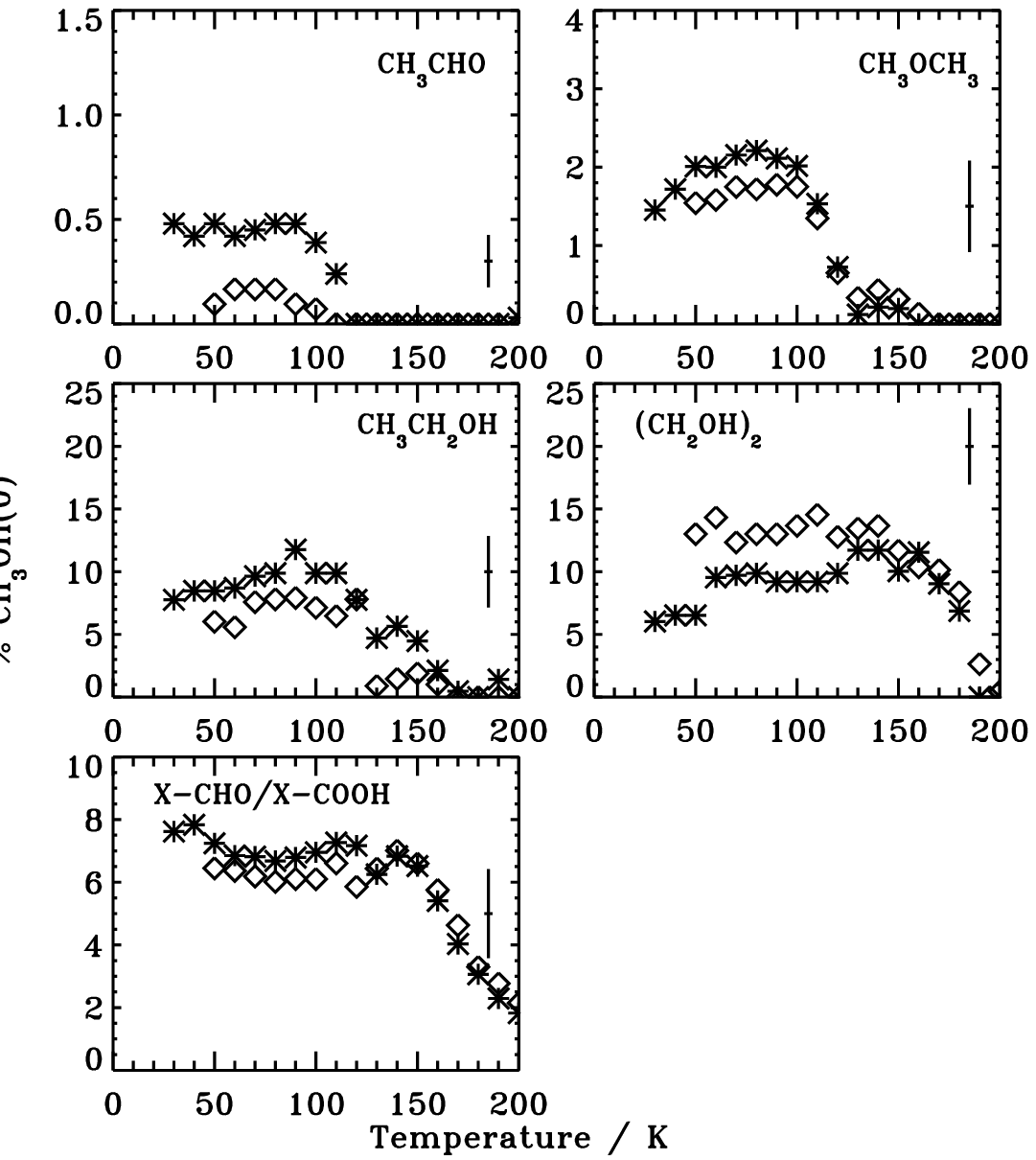}}
\caption{The evolution of small photo-products,  in \% of the initial CH$_3$OH ice abundance, with respect to temperature following irradiation of CH$_3$OH:CH$_4$ 1:2 ice mixtures at 30 (stars) and 50~K (diamonds). The average uncertainties are indicated by the error bar to the right in each panel.}
\label{fig:app_quant_comp_tpd_co}
\end{figure}

\begin{figure}
\resizebox{\hsize}{!}{\includegraphics{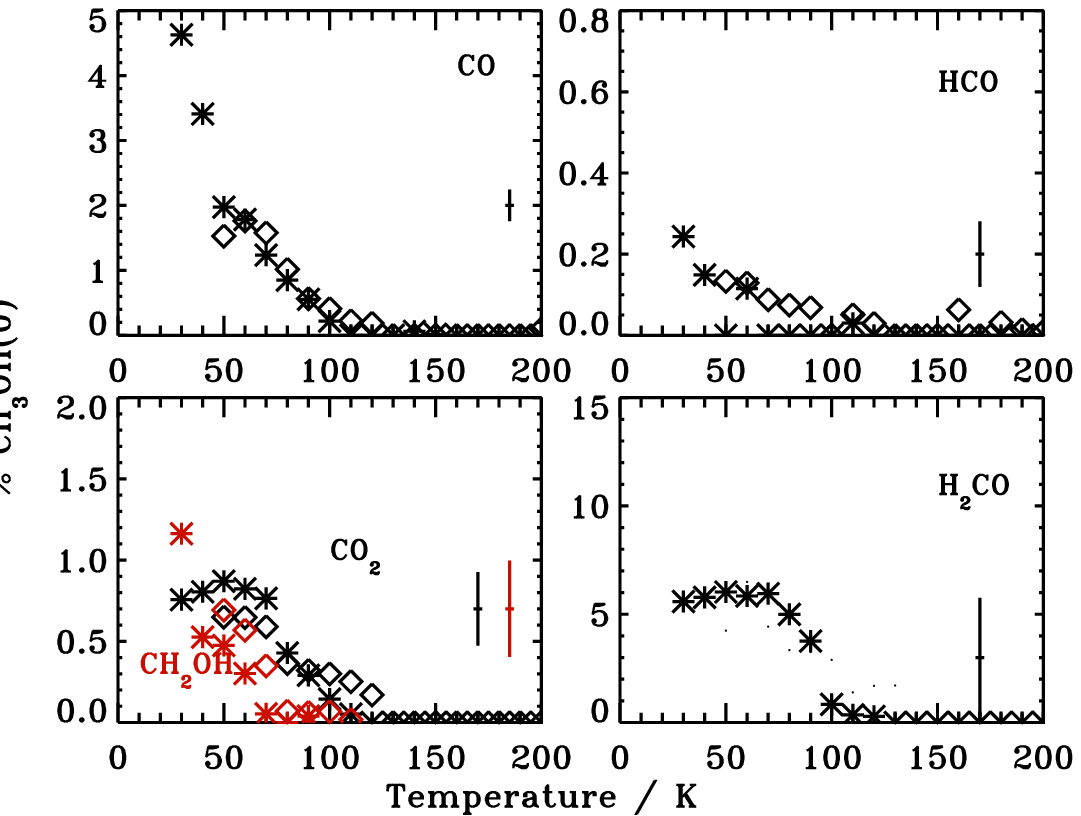}}
\caption{The evolution of small photo-products,  in \% of the initial CH$_3$OH ice abundance, with respect to temperature following irradiation of CH$_3$OH:CO 1:1 ice mixtures at 30 (stars) and 50~K (diamonds). The average uncertainties are indicated by the error bar to the right in each panel.}
\label{fig:app_quant_simp_tpd_ch4}
\end{figure}

\begin{figure}
\resizebox{\hsize}{!}{\includegraphics{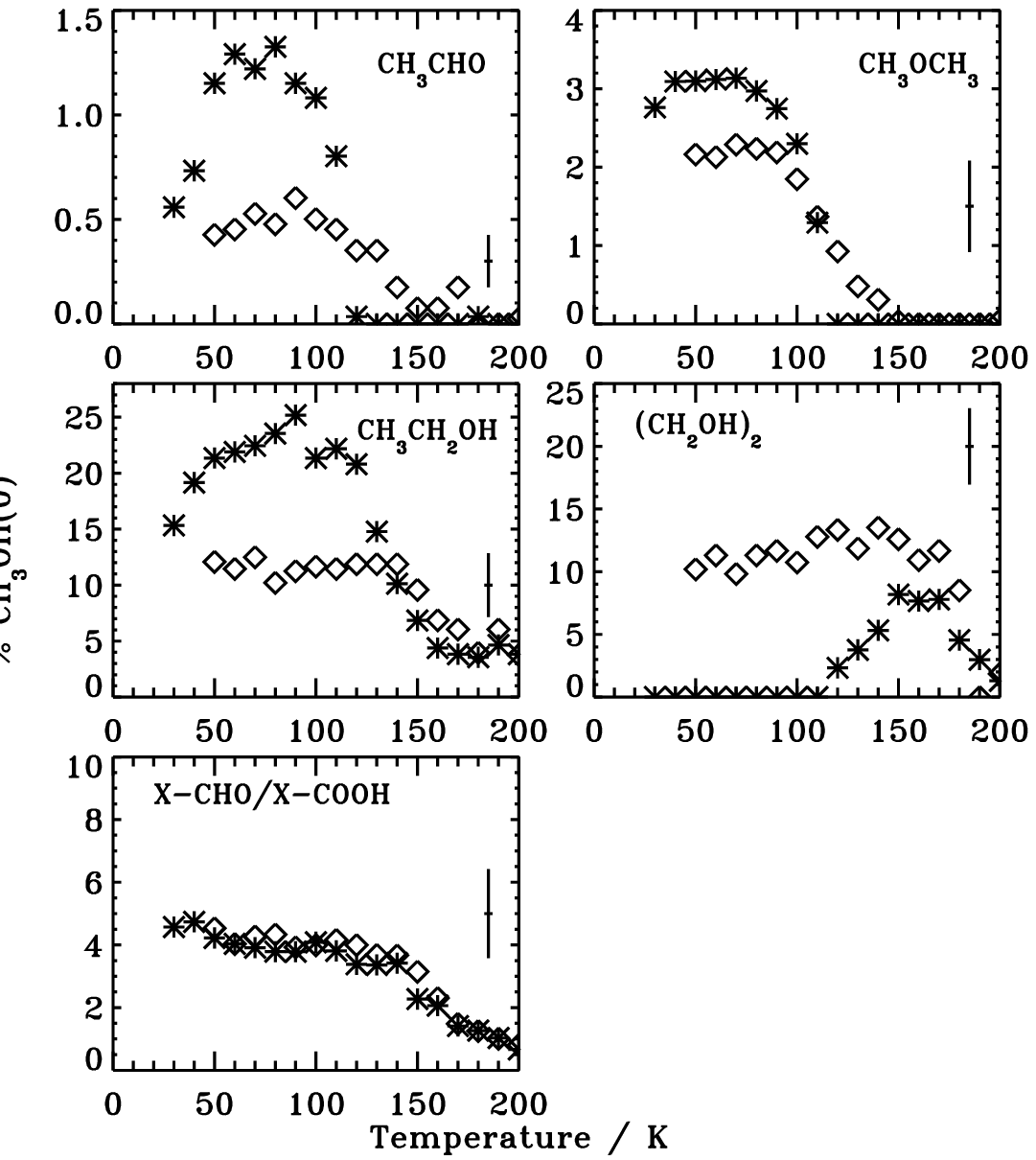}}
\caption{The evolution of small photo-products,  in \% of the initial CH$_3$OH ice abundance, with respect to temperature following irradiation of CH$_3$OH:CH$_4$ 1:2 ice mixtures at 30 (stars) and 50~K (diamonds). The average uncertainties are indicated by the error bar to the right in each panel.}
\label{fig:app_quant_comp_tpd_ch4}
\end{figure}

\newpage
\section{Formation rate parameters\label{sec:app_coeff}}

The production of simple and complex molecules during UV irradiation is parameterized with $A_1(1-e^{-A_2\times(\phi-A_3)})$ for 13 of the experiments, where $A_1$ correspond to the equilibrium abundance in \% of the initial CH$_3$OH abundance, $A_2$ is the fluence offset in 10$^{17}$ cm$^{-2}$ before production starts and $A_3$ describes the formation and destruction rates in fluence space in 10$^{-17}$ cm$^2$ (Tables \ref{tab:fits_1} and \ref{tab:fits_2}). The experiments include all pure 6--20~ML CH$_3$OH experiments irradiated with the normal flux setting, two 20~ML pure CH$_3$OH experiments irradiated with the higher flux setting, and all ice mixture experiments except for the CH$_3$OH:CO 1:10 experiment. The 1:10 experiment is excluded since the formation of any molecules is too low to detect during irradiation itself. The fitted experiments are numbered as in Table \ref{tab:ph_exps}. 

It is important to note  that the fitted formation `cross sections' cannot be used directly in astrophysical models, but rather should be used to compare models of the experiments with the experimental results. The quoted uncertainties in the tables are the fit error and do not include systematic uncertainties, in particular the conversion error between integrated band strength and abundances of $\sim$20\% and the fluence uncertainty of $\sim$ 30\%.

\newpage
\begin{table*}
\begin{center}
\caption{Photoproduct crosssection coefficents for pure CH$_3$OH experiments 1--6 as listed in Table 1.}             
\label{tab:fits_1}      
\centering                          
\begin{tabular}{l l |c c c c c c }        
\hline\hline                 
& & 1 & 2 & 3 & 4 & 5 & 6 \\
\hline                 
         CH$_4$&$A_1$&11.1[$   0.5$]& 9.6[$   0.6$]& 7.2[$   0.5$]& 6.1[$   0.3$]&14.7[$   0.3$]& 9.0[$   0.2$]\\
               &$A_2$&0.69[$  0.06$]&0.78[$  0.08$]&0.80[$  0.10$]&0.98[$  0.11$]&0.39[$  0.02$]&0.61[$  0.06$]\\
               &$A_3$& 0.0& 0.0& 0.0& 0.0& 0.0& 0.0\\
\hline
             CO&$A_1$&$<$99&$<$99&$<$99& 1.81[$   6.45$]&$<$99&$<$99\\
               &$A_2$&0.021[$  0.001$]&0.016[$  0.001$]&0.02[$  0.16$]&0.12[$  0.49$]&0.015[$  0.001$]&0.006[$  0.001$]\\
               &$A_3$& 0.16[$   0.02$]& 0.14[$   0.03$]& 0.07[$   0.09$]& 0.0& 0.36[$   0.05$]& 0.17[$   0.17$]\\
\hline
            HCO&$A_1$& 0.32[$   0.04$]& 0.29[$   0.07$]& 0.16[$   0.03$]& 0.13[$   0.14$]& 0.37[$   0.02$]& --\\
               &$A_2$&0.99[$  0.29$]&0.88[$  0.37$]&1.78[$  0.81$]&0.53[$  0.87$]&0.76[$  0.21$]&--\\
               &$A_3$& 0.0& 0.0& 0.0& 0.0& 0.0& --\\
\hline
        H$_2$CO&$A_1$& 9.1[$   0.4$]&10.3[$   0.5$]& 4.6[$   0.5$]& 1.1[$   0.3$]&11.1[$   0.3$]& 7.5[$   0.4$]\\
               &$A_2$&2.2[$  0.3$]&2.0[$  0.3$]&1.9[$  0.6$]&3.5[$  3.9$]&1.5[$  0.4$]&2.3[$  1.4$]\\
               &$A_3$& 0.0& 0.0&  0.0&  0.0&  0.0&  0.0\\
\hline
       CH$_2$OH&$A_1$& 1.52[$   0.04$]& 1.17[$   0.04$]& 1.05[$   0.04$]& 0.77[$   0.03$]& --&--\\
               &$A_2$&9.0[$  1.5$]&13.6[$  3.8$]&12.9[$  3.7$]&20.8[$  9.9$]& --&--\\
               &$A_3$& 0.0& 0.0& 0.0&0.0&--& --\\
\hline
         CO$_2$&$A_1$&$<$99&$<$99&$<$99&$<$99&$<$99&$<$99\\
               &$A_2$&0.008[$  0.025$]&0.014[$  0.099$]&0.015[$  0.117$]&0.008[$  0.095$]&0.002[$  0.001$]&0.003[$  0.006$]\\
               &$A_3$& 0.17[$   0.05$]& 0.17[$   0.08$]& 0.08[$   0.11$]& 0.22[$   0.18$]& 1.13[$   0.09$]& 1.00[$   0.10$]\\
\hline
      CH$_3$CHO&$A_1$&$<$99& 0.9[$   3.5$]& $<$99& 0.47[$   0.18$]&$<$99& 0.82[$   0.18$]\\
               &$A_2$&0.009[$  0.075$]&0.10[$  0.46$]&0.014[$  0.153$]&0.51[$  0.32$]&0.004[$  0.026$]&0.19[$  0.08$]\\
               &$A_3$& 0.23[$   0.10$]& 0.0& 0.30[$   0.11$]& 0.11[$   0.08$]& 0.54[$   0.25$]& 0.09[$   0.33$]\\
\hline
  CH$_3$OCH$_3$&$A_1$& 1.5[$   0.3$]& 1.6[$   0.3$]& 1.9[$   0.2$]& 1.40[$   0.09$]& --& --\\
               &$A_2$&0.86[$  0.38$]&1.04[$  0.46$]&1.50[$  0.38$]&3.01[$  0.82$]&--&--\\
               &$A_3$& 0.10[$   0.08$]& 0.05[$   0.08$]& 0.03[$   0.05$]& 0.00[$   0.04$]& --& --\\
\hline
 CH$_3$CH$_2$OH&$A_1$& 4.7[$   0.3$]& 5.4[$   0.3$]& 4.8[$   2.1$]& 6.9[$   3.5$]& 7.4[$   0.6$]&12.9[$   2.4$]\\
               &$A_2$&4.4[$  1.3$]&10.2[$  3.6$]&0.63[$  0.47$]&0.46[$  0.34$]&0.50[$  0.14$]&0.22[$  0.08$]\\
               &$A_3$& 0.0& 0.0& 0.0& 0.0& 0.0& 0.0\\
\hline
 (CH$_2$OH)$_2$&$A_1$& 4.9[$   0.5$]&10.5[$   2.0$]&14.6[$   0.7$]&18.8[$   0.6$]& 6.2[$   0.5$]&12.0[$   0.6$]\\
               &$A_2$&3.0[$  1.1$]&0.88[$  0.31$]&2.1[$  0.3$]&2.6[$  0.3$]&3.9[$  6.9$]&3.0[$  2.3$]\\
               &$A_3$&0.0& 0.0& 0.0&0.0& 0.0&0.0\\
\hline
   X-CHO/&$A_1$&12.5[$   7.8$]& 6.1[$   2.0$]& 4.5[$   0.8$]& 8.4[$3.7$]&13.5[$   1.1$]&16.8[$   3.2$]\\
X-COOH      &$A_2$&0.16[$  0.12$]&0.37[$  0.17$]&0.50[$  0.14$]&0.21[$  0.11$]&0.12[$  0.02$]&0.10[$  0.03$]\\
               &$A_3$& 0.03[$   0.06$]& 0.04[$   0.06$]& 0.0& 0.0& 0.0& 0.12[$   0.14$]\\
\hline
\end{tabular}
\end{center}
\end{table*}

\begin{table*}
\begin{center}
\caption{Photoproduct crosssection coefficents for ice mixture experiments 7--11, 13, 14 as listed in Table 1.}             
\label{tab:fits_2}      
\centering                          
\begin{tabular}{l l |c c c c c c c}        
\hline\hline                 
& & 7 & 8 & 9 & 10 & 11 &13 &14\\
\hline                 
         CH$_4$&$A_1$& 8.7[$   0.9$]& 7.0[$   1.6$]& 3.7[$   0.6$]&--&--&10.7[$   0.8$]& 5.6[$   0.8$]\\
               &$A_2$& 0.79[$   0.14$]& 0.54[$   0.19$]& 0.89[$   0.25$]&--&--& 1.27[$   0.22$]& 1.34[$   0.47$]\\
               &$A_3$& 0.0& 0.0&0.0& --&--& 0.0& 0.0\\
\hline
             CO&$A_1$&--&--&--&$<$99& 2.9[$   2.4$]&$<$99& 1.2[$   0.3$]\\
               &$A_2$&--&--& --& 0.02[$   0.01$]& 0.29[$   0.31$]& 0.03[$   0.01$]& 1.6[$   1.2$]\\
               &$A_3$&--&--& --& 0.11[$   0.05$]& 0.0&0.0& 0.0\\
\hline
            HCO&$A_1$& 0.98[$   0.02$]& 0.67[$   0.02$]& --& 0.34[$   0.04$]& 0.12[$   0.03$]& --&--\\
               &$A_2$&18.2[$   3.7$]&17.2[$   4.9$]& --& 2.9[$   1.2$]& 6.2[$   7.8$]& --& --\\
               &$A_3$& 0.0& 0.0&--& 0.03[$   0.06$]& 0.01[$   0.10$]& --&--\\
\hline
        H$_2$CO&$A_1$&15.2[$   0.6$]&12.8[$   0.7$]& 9.5[$   0.4$]& 7.1[$   0.7$]& 6.0[$   0.8$]& 5.4[$   0.9$]& --\\
               &$A_2$& 3.3[$   0.5$]& 2.4[$   0.5$]& 6.4[$   1.5$]& 3.7[$   1.6$]& 2.1[$   0.9$]& 7.3[$   7.8$]&--\\
               &$A_3$& 0.03[$   0.02$]& 0.02[$   0.03$]& 0.01[$   0.02$]& 0.0& 0.0& 0.0& --\\
\hline
       CH$_2$OH&$A_1$& 2.08[$   0.08$]& 1.37[$   0.08$]& 0.81[$   0.10$]& 1.46[$   0.07$]& 0.91[$   0.06$]& --& 0.82[$   0.13$]\\
               &$A_2$& 4.4[$   0.7$]& 5.1[$   1.3$]& 1.8[$   0.6$]&10.8[$   4.4$]&17[$  12$]& --&12[$  15$]\\
               &$A_3$& 0.0& 0.0&0.0& 0.0& 0.0& --& 0.0\\
\hline
     C$_2$H$_6$&$A_1$& --& --&--&13.8[$   7.4$]& --&--&--\\
               &$A_2$&--& --&--& 1.2[$   1.5$]&--& --&--\\
               &$A_3$&--& --&--& 0.0 &--& --&--\\
\hline
         CO$_2$&$A_1$&$<$99&$<$99& 8[$  23$]&$<$99&$<$99&$<$99&$<$99\\
               &$A_2$& 0.01[$   0.02$]& 0.01[$   0.05$]& 0.05[$   0.17$]& 0.01[$   0.06$]& 0.01[$   0.08$]& 0.01[$   0.11$]& 0.01[$   0.13$]\\
               &$A_3$& 0.10[$   0.03$]& 0.10[$   0.04$]& 0.07[$   0.07$]& 0.13[$   0.10$]& 0.0& 0.26[$   0.07$]& 0.16[$   0.09$]\\
\hline
      CH$_3$CHO&$A_1$&$<$99& $<$99& $<$99&$<$99& 0.90[$   0.93$]& $<$99]& 0.33[$   0.05$]\\
               &$A_2$& 0.01[$   0.16$]& 0.03[$   0.30$]& 0.01[$   0.37$]& 0.01[$   0.05$]& 0.31[$   0.43$]& 0.04[$   0.59$]& 6.4[$   5.0$]\\
               &$A_3$& 0.36[$   0.14$]& 0.0& 0.0& 0.32[$   0.09$]& 0.11[$   0.13$]& 0.0& 0.0\\
\hline
  CH$_3$OCH$_3$&$A_1$& 1.2[$   0.3$]& 1.5[$   0.3$]& 1.6[$   0.2$]& 2.8[$   0.3$]& 2.3[$   0.2$]& 2.1[$   0.6$]& 0.8[$   0.3$]\\
&$A_2$& 1.4[$   0.9$]& 1.4[$   0.8$]& 1.5[$   0.5$]& 1.8[$   0.6$]& 3.1[$   0.8$]& 1.4[$   1.0$]& 6[$  12$]\\
&$A_3$& 0.09[$   0.12$]& 0.03[$   0.11$]& 0.0&0.0& 0.0& 0.06[$   0.14$]& 0.01[$   0.13$]\\
\hline
 CH$_3$CH$_2$OH&$A_1$&10.0[$   0.4$]& 7.3[$   0.7$]& 5.2[$   0.6$]&14.9[$   0.9$]&10.0[$   0.9$]& 5.7[$   0.8$]& 8.1[$   6.2$]\\
               &$A_2$& 8.4[$   2.3$]& 2.6[$   0.8$]& 2.0[$   0.7$]& 1.9[$   0.3$]& 1.8[$   0.5$]&12[$  14$]& 0.8[$   1.1$]\\
               &$A_3$& 0.0& 0.0& 0.0& 0.0& 0.0& 0.01[$   0.04$]& 0.10[$   0.23$]\\
\hline
 (CH$_2$OH)$_2$&$A_1$& 5.4[$   0.8$]& 5.5[$   3.3$]&12.4[$   0.8$]& --&11.4[$   0.9$]& 6.4[$   1.7$]&14.6[$   2.8$]\\
               &$A_2$& 5.6[$   4.1$]& 0.9[$   1.3$]& 2.7[$   0.6$]& --& 3.8[$   1.2$]& 3.7[$   3.7$]& 1.9[$   1.1$]\\
               &$A_3$& 0.02[$   0.06$]& 0.01[$   0.28$]& 0.0& --& 0.0& 0.0& 0.02[$   0.12$]\\
\hline
   X-CHO/&$A_1$&12.1[$   0.7$]& 9.7[$   0.8$]&17.2[$   7.3$]&$<$99&11.3[$   8.3$]&11.6[$   4.7$]& 8.3[$   3.1$]\\
 X-COOH &$A_2$& 0.75[$   0.07$]& 0.64[$   0.09$]& 0.21[$   0.11$]& 0.02[$   0.01$]& 0.20[$   0.17$]& 0.41[$   0.25$]& 0.47[$   0.26$]\\
        &$A_3$& 0.0& 0.0& 0.05[$   0.04$]& 0.16[$   0.07$]& 0.0& 0.0& 0.0\\
\hline

\end{tabular}
\end{center}
\end{table*}

\end{appendix}

\end{document}